%% file: paper.tex
\documentclass[journal]{IEEEtran}

\usepackage[T1]{fontenc}
\usepackage{cite}

\usepackage{amsmath}
\usepackage{newtxmath}
\usepackage{graphicx}
\usepackage{booktabs}
\usepackage{array}
\usepackage{tabularx}
\usepackage{ragged2e}
\usepackage{multirow}
\usepackage{tikz}
\usetikzlibrary{positioning,arrows.meta,fit,backgrounds,calc,decorations.pathreplacing}
\usepackage{xurl}
\usepackage[hidelinks,bookmarksnumbered=true,pdftitle={Intelligence as a Founding Principle: The O-RAN Architecture for Artificial Intelligence and Its Reference Implementations},pdfauthor={Chih-Lin I, Hao Xu, Tao Chen, Frank H. P. Fitzek, Muhammad Ali Imran, Sen Wang}]{hyperref}
\usepackage{bookmark}

\input{figs/figure_style}

\begin{document}
\raggedbottom

\title{Intelligence as a Founding Principle: The O-RAN Architecture
for Artificial Intelligence and Its Reference Implementations}

\author{Chih-Lin~I,~\IEEEmembership{Life~Fellow,~IEEE,}
Hao~Xu,~\IEEEmembership{Member,~IEEE,}
Tao~Chen,~\IEEEmembership{Senior~Member,~IEEE,}
Frank~H.~P.~Fitzek,~\IEEEmembership{Fellow,~IEEE,}
Muhammad~Ali~Imran,~\IEEEmembership{Fellow,~IEEE,}
and~Sen~Wang%
\thanks{Chih-Lin I and Sen Wang are with China Mobile Research Institute, Beijing, China (e-mail: icl@chinamobile.com; wangsenyjy@chinamobile.com).}%
\thanks{Hao Xu is with the College of Electronic and Information Engineering, Tongji University, Shanghai, China (e-mail: hxu@tongji.edu.cn).}%
\thanks{Tao Chen is with VTT Technical Research Centre of Finland, Espoo, Finland (e-mail: tao.chen@vtt.fi).}%
\thanks{Frank H. P. Fitzek is with the Deutsche Telekom Chair of Communication Networks, Technische Universit\"at Dresden, Dresden, Germany (e-mail: frank.fitzek@tu-dresden.de).}%
\thanks{Muhammad Ali Imran is with the James Watt School of Engineering, University of Glasgow, Glasgow, U.K. (e-mail: Muhammad.Imran@glasgow.ac.uk).}}

\markboth{IEEE Journal on Selected Areas in Communications}%
{I \MakeLowercase{\textit{et al.}}: Intelligence as a Founding Principle}

\maketitle

\input{sec_abstract}
\input{sec1_intro}
\input{sec2_beginning}
\input{sec3_today}
\input{sec4_industry}
\input{sec5_academia}
\input{sec6_airan}
\input{sec7_6g}
\input{sec8_conclusion}

\bibliographystyle{IEEEtran}
\bibliography{oran,academic}

\end{document}

%% file: figs/figure_style.tex
\definecolor{FigInk}{HTML}{000000}
\definecolor{FigBlue}{HTML}{0055B3}
\definecolor{FigNavy}{HTML}{003366}
\definecolor{FigTeal}{HTML}{006B38}
\definecolor{FigAmber}{HTML}{B82514}
\definecolor{FigLine}{HTML}{262626}
\definecolor{FigHostLine}{HTML}{1F4E79}
\definecolor{FigGuide}{HTML}{BFBFBF}
\definecolor{FigBlueFill}{HTML}{BFD9F2}
\definecolor{FigTealFill}{HTML}{B3E2C7}
\definecolor{FigAmberFill}{HTML}{F8CBB8}
\definecolor{FigGreyFill}{HTML}{DCDCDC}

\colorlet{ink}{FigInk}
\colorlet{c3gpp}{FigInk}
\colorlet{coran}{FigBlue}
\colorlet{cimpl}{FigTeal}
\colorlet{cfill}{FigBlueFill}
\colorlet{cfillb}{FigTealFill}
\colorlet{cfillg}{FigGreyFill}
\colorlet{cgrey}{FigLine}

\newcommand{\FigFont}{\rmfamily\fontsize{8}{9.4}\selectfont}
\newcommand{\FigSmall}{\rmfamily\fontsize{7.4}{8.6}\selectfont}
\newcommand{\FigHead}{\rmfamily\fontsize{8.5}{9.8}\selectfont\bfseries}
\newcommand{\FigGroup}{\rmfamily\fontsize{8}{9.4}\selectfont\bfseries}

\tikzset{
  fig/base/.style={x=1mm,y=1mm,font=\FigFont,text=FigInk,
    line cap=butt,line join=miter},
  fig/node/.style={draw=FigLine,fill=white,line width=.55pt,
    sharp corners,align=center,inner sep=1.2mm,outer sep=0pt,
    font=\FigFont,text=FigInk},
  fig/host/.style={fig/node,draw=FigHostLine,fill=FigBlueFill},
  fig/app/.style={fig/node,draw=FigTeal,fill=FigTealFill,line width=.65pt},
  fig/prop/.style={fig/node,draw=FigAmber,fill=FigAmberFill,
    dash pattern=on 3.2pt off 1.6pt,line width=.65pt},
  fig/group/.style={draw=FigLine,fill=white,line width=.55pt,
    sharp corners},
  fig/host/group/.style={fig/group,draw=FigHostLine,fill=FigBlueFill},
  fig/prop/group/.style={fig/group,draw=FigAmber,fill=FigAmberFill,
    dash pattern=on 3.2pt off 1.6pt},
  fig/smo/.style={fig/host},
  fig/near/.style={fig/host},
  fig/cu/.style={fig/node},
  fig/du/.style={fig/node},
  fig/ru/.style={fig/node},
  fig/cloud/.style={fig/node,fill=FigGreyFill},
  fig/smo/group/.style={fig/host/group},
  fig/near/group/.style={fig/host/group},
  fig/cu/group/.style={fig/group},
  fig/du/group/.style={fig/group},
  fig/ru/group/.style={fig/group},
  fig/cloud/group/.style={fig/group,fill=FigGreyFill},
  fig/interface/.style={draw=FigBlue,line width=.8pt},
  fig/management/.style={fig/interface,dash pattern=on 3.2pt off 1.6pt},
  fig/inherited/.style={draw=FigInk,line width=.55pt},
  fig/arrow/.style={fig/interface,-{Latex[length=1.6mm,width=1.1mm]}},
  fig/bidir/.style={fig/interface,<->,>={Latex[length=1.6mm,width=1.1mm]}},
  fig/return/.style={draw=FigLine,line width=.65pt,
    -{Latex[length=1.6mm,width=1.1mm]}},
  fig/guide/.style={draw=FigGuide,line width=.35pt},
  fig/label/.style={font=\FigSmall,text=FigInk,fill=white,
    inner sep=.65mm,outer sep=0pt,align=center},
  fig/title/.style={font=\FigHead,text=FigNavy,inner sep=0pt,
    outer sep=0pt,anchor=west,align=left},
  fig/heading/.style={font=\FigGroup,text=FigNavy,inner sep=0pt,
    outer sep=0pt,anchor=west,align=left},
  blk/.style={fig/node,minimum height=5.5mm},
  blo/.style={fig/host,minimum height=5.5mm},
  bli/.style={fig/app,minimum height=5.5mm},
  blg/.style={fig/node,minimum height=5.5mm},
  ar/.style={fig/inherited,-{Latex[length=1.6mm,width=1.1mm]}},
  aro/.style={fig/arrow},
  ari/.style={fig/arrow,draw=FigTeal},
  lbl/.style={fig/label},
  ttl/.style={fig/heading},
  yr/.style={font=\FigSmall,text=FigLine,inner sep=1pt}
}

%% file: sec_abstract.tex
\begin{abstract}
Since its inception, O-RAN has established openness and intelligence as
founding principles of the RAN, transforming closed base stations into
disaggregated, multi-vendor systems with a hierarchical RIC. This article
traces AI/ML in O-RAN through open specifications and reference
implementations to 6G design. We examine how open fronthaul, RIC interfaces,
cloud orchestration, and Acceleration Abstraction Layer (AAL) profiles support control loops across multiple
timescales, model management, and hardware acceleration. Commercial
deployments and OTIC testing demonstrate multi-vendor interoperability.
Open-source stacks and testbeds support pre-standardization studies of
in-node dApps. Within this architecture, GPU-based AI-RAN platforms
realize the inline High-PHY split on shared RAN and AI compute, providing
one reference implementation path for the O-RAN AAL alongside other
implementations on heterogeneous hardware. Vector instruction sets and
analog in-memory computing offer longer-term options for improving energy
efficiency. For 6G, extending intelligence to time-critical radio processing and mobile
agent services calls for deadline-bounded third-party inference,
energy-efficient RAN/Edge-AI silicon co-design, and agent session
continuity across handovers. We propose a RAN Agentic Controller (RAC) that coordinates agents
across tiers through local shells and a shared exposure gateway, with
in-node operating bounds delivered asynchronously. We also propose
Recursive Standardization, in which results from federated open labs
inform candidate 6G requirements and standards review guides the next
round of implementation and testing.
\end{abstract}

\begin{IEEEkeywords}
O-RAN, RAN Intelligent Controller (RIC), artificial intelligence, hardware acceleration, AI-RAN, agentic AI, 6G.
\end{IEEEkeywords}

%% file: sec1_intro.tex
\section{Introduction}

In 2018, five mobile operators formed the O-RAN ALLIANCE through the
merger of the xRAN Forum and the C-RAN Alliance to build a more open and
intelligent radio access network (RAN) \cite{oran18launch}.
The alliance's white paper that year called for embedding
intelligence across every RAN layer through an artificial intelligence (AI)-enabled hierarchical RAN
Intelligent Controller (RIC) open to third-party applications
\cite{oran18whitepaper}. Open interfaces allow a RIC application from
one supplier to use measurements from another supplier's radio node and
request changes as network conditions evolve.

\input{figs/fig_hero}

In traditional vertically integrated base stations, closed-loop optimization
remained confined to vendor-proprietary software accessing internal state
without external exposure. Disaggregating the RAN across suppliers alters that
constraint. The placement of a control function depends on whether it can
obtain measurements at the required granularity, access sufficient computing
resources, and return a decision before the radio deadline. Subsequent
O-RAN specifications translated this principle into standardized application
hosts, lifecycle services, and open interfaces \cite{oran.wg1.arch}. The
non-real-time RIC (Non-RT RIC) hosts rApps for fleet-wide analytics, model
training, and policy generation; the near-real-time RIC (Near-RT RIC) hosts
xApps for faster inference and radio control. Standardized interfaces connect
model onboarding and policy guidance to actuation at disaggregated radio nodes,
while cloud and hardware acceleration abstractions decouple virtualized
baseband software from silicon.

Commercial deployments, shared research testbeds, and accelerated radio
platforms show how operators, equipment and silicon vendors, and academic
groups implement the O-RAN architecture.
Two related areas of work inform our analysis.

\begin{itemize}
\item \textbf{AI and machine learning (ML) in the O-RAN architecture.}
\emph{AI for RAN} refers to the use of AI/ML in radio operation. O-RAN's early RIC design supported this
use.

\item \textbf{The AI-RAN Alliance framework as an O-RAN Acceleration Abstraction Layer (AAL) reference implementation.}
The framework covers AI for RAN, shared computing for RAN and AI
workloads, and connectivity for distributed AI applications
\cite{airan26.architecture}.
We use inline physical-layer (PHY) acceleration on graphics processing
units (GPUs) in AI-RAN platforms as a reference implementation of the
host--accelerator division of functions described by the O-RAN AAL
inline High-PHY profile
\cite{oran.wg6.aalhiphy}.
\end{itemize}

We examine reference implementations that use selected O-RAN interfaces for
AI for RAN, as well as platforms that run RAN and AI workloads on shared
computing resources. For example, X5G runs the OpenAirInterface (OAI) upper
protocol layers on the host, executes High-PHY on the GPU using NVIDIA Aerial,
and integrates the O-RAN Software Community (OSC) Near-RT RIC
\cite{villa25x5g}. Comparing this path with
field-programmable gate array (FPGA), digital signal processor (DSP),
application-specific integrated circuit (ASIC), and processor-integrated
offload implementations identifies which interfaces are standardized and
where integration depends on the hardware and software platform.

Moving toward 6G, openness and intelligence must advance together as wireless AI
extends into time-critical radio processing and connects with physical-world agent
tasks. The O-RAN ALLIANCE established the next Generation Research Group (nGRG)
to study how O-RAN should evolve for 6G \cite{oran22.ngrg}. Distributed
applications (dApps) illustrate how academic research contributes to this work,
having progressed from university proposals to nGRG studies. O-RAN RICs are
already in use, hosting rApps and xApps in operator labs and research testbeds
\cite{dt23.nonrtric,villa25x5g}. Extending third-party control into radio
nodes raises the question of how applications on different hosts can
coordinate while the radio node meets its execution deadlines. We envision a
RAN Agentic Controller (RAC) whose coordination functions are distributed
across the Non-RT RIC, Near-RT RIC, and proposed in-node hosts. The RAC
exposes selected application interfaces and connects future apps that embed rAgents, xAgents,
and dAgents. For user-facing agents that rely on edge inference, we examine
the radio, compute, and memory demands of a session and how it could
continue during handover. We propose recursive standardization in which results from
federated open labs inform candidate requirements and revisions to those
requirements guide the next round of implementation and experimentation.

The remainder of this article follows the progression summarized in
Fig.~\ref{fig:hero}. Section~\ref{sec:beginning} traces three design principles
to early O-RAN work. These are hierarchical RAN control, four-stratum
disaggregation, and open application hosting.
Section~\ref{sec:today} details the resulting O-RAN AI/ML architecture,
including control timescales, interfaces for model management, control, and
analytics, O-Cloud and AAL, and 3GPP support for AI/ML in the RAN. Turning from specification to practice,
Section~\ref{sec:industry} examines multi-vendor testing, operator deployments,
and regional adoption.
Section~\ref{sec:academia} examines how open-source stacks, closed-loop
testbeds, and dApp research contribute to pre-standardization.
Section~\ref{sec:airan} then distinguishes
general AI for RAN from AI-RAN Alliance platforms and examines look-aside,
inline, reconfigurable, and analog acceleration, multi-supplier integration,
and system-level benchmarking. Finally, Section~\ref{sec:6g} examines
RAN control and edge inference, proposes the RAC for cross-tier
coordination, and outlines recursive
standardization across federated open labs.

%% file: figs/fig_hero.tex
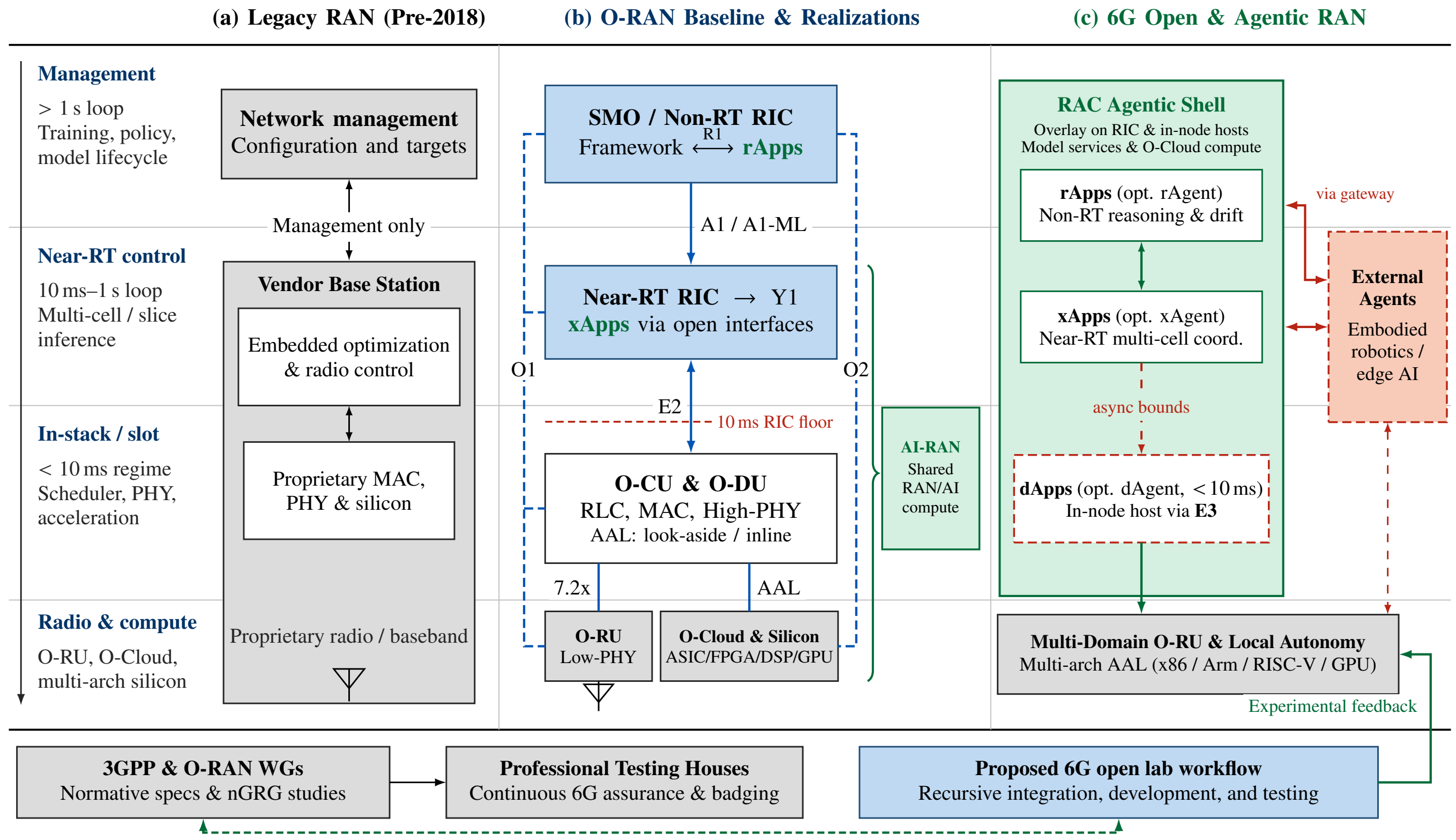
\begin{figure*}[!t]
\centering
\begin{tikzpicture}[fig/base,
  layer/.style={font=\FigSmall,text=FigLine,align=left,anchor=north west,inner sep=0pt},
  lhead/.style={font=\FigSmall\bfseries,text=FigNavy,align=left,anchor=north west,inner sep=0pt},
  era/.style={font=\FigGroup,text=FigNavy,anchor=center,inner sep=0pt},
  brace/.style={draw=FigTeal,line width=.75pt,decorate,decoration={brace,amplitude=1.3mm}}
]
\path[use as bounding box] (0,4.2) rectangle (178,-100.5);

\node[era,text=FigInk] at (42,1.2) {(a) Legacy RAN (Pre-2018)};
\node[era] at (90.5,1.2) {(b) O-RAN Baseline \& Realizations};
\node[era,text=FigTeal] at (149.5,1.2) {(c) 6G Open \& Agentic RAN};
\draw[fig/inherited,line width=.7pt] (0,-2) -- (178,-2);

\draw[fig/guide] (0,-24.5) -- (178,-24.5);
\draw[fig/guide] (0,-46.5) -- (178,-46.5);
\draw[fig/guide] (0,-70.5) -- (178,-70.5);
\draw[fig/inherited,line width=.7pt] (0,-86.5) -- (178,-86.5);

\draw[fig/inherited,draw=FigLine,-{Latex[length=1.4mm,width=1mm]}] (1.5,-4) -- (1.5,-83.5);
\node[lhead] at (3.5,-4.5) {Management};
\node[layer] at (3.5,-8.8) {$>1$\,s loop\\Training, policy,\\model lifecycle};

\node[lhead] at (3.5,-27) {Near-RT control};
\node[layer] at (3.5,-31.3) {10\,ms--1\,s loop\\Multi-cell / slice\\inference};

\node[lhead] at (3.5,-49) {In-stack / slot};
\node[layer] at (3.5,-53.3) {$<10$\,ms regime\\Scheduler, PHY,\\acceleration};

\node[lhead] at (3.5,-72.2) {Radio \& compute};
\node[layer] at (3.5,-76.5) {O-RU, O-Cloud,\\multi-arch silicon};

\draw[fig/guide] (60.5,-2) -- (60.5,-86.5);
\draw[fig/guide] (121.2,-2) -- (121.2,-86.5);

\node[fig/node,fill=FigGreyFill,minimum width=31mm,minimum height=11mm] (a_oss) at (42,-13)
  {\textbf{Network management}\\Configuration and targets};

\node[fig/group,fill=FigGreyFill,minimum width=31mm,minimum height=54.5mm] (a_bs) at (42,-56) {};
\node[font=\FigSmall\bfseries,text=FigInk] at (42,-31.5) {Vendor Base Station};
\node[fig/node,fill=white,minimum width=26mm,minimum height=12mm,font=\FigSmall] (a_son) at (42,-40.5)
  {Embedded optimization\\\& radio control};
\node[fig/node,fill=white,minimum width=26mm,minimum height=12mm,font=\FigSmall] (a_phy) at (42,-57)
  {Proprietary MAC,\\PHY \& silicon};
\draw[fig/inherited,<->,>={Latex[length=1.2mm,width=.9mm]}] (a_son.south) -- (a_phy.north);

\draw[fig/inherited,<->,>={Latex[length=1.3mm,width=.9mm]}] (a_oss.south) -- (a_bs.north);
\node[fig/label,fill=white,font=\FigSmall] at (42,-24.5) {Management only};
\node[font=\FigSmall,text=FigLine] at (42,-75.2) {Proprietary radio / baseband};
\draw[fig/inherited] (42,-79.0) -- (42,-83.0);
\draw[fig/inherited] (40.1,-79.0) -- (43.9,-79.0) -- (42,-81.4) -- cycle;

\node[fig/host,minimum width=36mm,minimum height=12mm] (b_smo) at (84.2,-13)
  {\textbf{SMO / Non-RT RIC}\\Framework $\stackrel{\text{R1}}{\longleftrightarrow}$ \textbf{\textcolor{FigTeal}{rApps}}};

\node[fig/host,minimum width=36mm,minimum height=11.5mm] (b_near) at (84.2,-35)
  {\textbf{Near-RT RIC} $\;\to\;$ Y1\\\textbf{\textcolor{FigTeal}{xApps}} via open interfaces};

\node[fig/node,minimum width=36mm,minimum height=13.5mm] (b_du) at (84.2,-59.2)
  {\textbf{O-CU \& O-DU}\\RLC, MAC, High-PHY\\[-1pt]{\scriptsize AAL: look-aside / inline}};

\draw[draw=FigAmber,dash pattern=on 2.5pt off 1.5pt,line width=.65pt] (66.2,-48.5) -- (102.2,-48.5);
\node[font=\fontsize{6.4}{7.2}\selectfont\rmfamily,text=FigAmber,fill=white,inner sep=.5mm,anchor=east]
  at (102.3,-48.5) {10\,ms RIC floor};

\node[fig/node,fill=FigGreyFill,minimum width=13.2mm,minimum height=8.6mm,
  font=\fontsize{6.5}{7.4}\selectfont\rmfamily,inner sep=.6mm] (b_ru) at (72.8,-76.2)
  {\textbf{O-RU}\\Low-PHY};
\node[fig/cloud,minimum width=21.6mm,minimum height=8.6mm,
  font=\fontsize{6.2}{7.1}\selectfont\rmfamily,inner sep=.6mm] (b_cld) at (91.4,-76.2)
  {\textbf{O-Cloud \& Silicon}\\ASIC/FPGA/DSP/GPU};
\draw[fig/inherited] (72.8,-80.5) -- (72.8,-84.2);
\draw[fig/inherited] (71.0,-80.9) -- (74.6,-80.9) -- (72.8,-83.1) -- cycle;

\draw[fig/arrow] (b_smo.south) -- node[fig/label,right=.5mm]{A1 / A1-ML} (b_near.north);
\draw[fig/bidir] (b_near.south) -- node[fig/label,left=.5mm]{E2} (b_du.north);
\draw[fig/interface] (b_du.south -| 72.8,-66) -- node[fig/label,left=.3mm]{7.2x} (b_ru.north);
\draw[fig/interface] (b_du.south -| 91.4,-66) -- node[fig/label,right=.3mm]{AAL} (b_cld.north);
\draw[fig/management] (b_smo.west) -- (63.6,-13) -- (63.6,-76.2) -- (b_ru.west);
\draw[fig/management] (63.6,-35) -- (b_near.west);
\draw[fig/management] (63.6,-59.2) -- (b_du.west);
\node[fig/label,font=\FigSmall,inner sep=.45mm] at (63.6,-42.0) {O1};
\draw[fig/management] (b_smo.east) -- (104.6,-13) -- (104.6,-76.2) -- (b_cld.east);
\node[fig/label,font=\FigSmall,inner sep=.45mm] at (104.6,-42.0) {O2};

\draw[brace] (106.0,-29.2) -- (106.0,-80.5);
\node[fig/app,fill=FigTealFill!55!white,minimum width=12.0mm,minimum height=17.5mm,inner sep=.7mm,
  font=\fontsize{5.8}{6.7}\selectfont\rmfamily] at (113.8,-55.5)
  {\textbf{\textcolor{FigTeal}{AI-RAN}}\\[1pt]Shared\\RAN/AI\\compute};

\node[fig/group,draw=FigTeal,fill=FigTealFill!55!white,line width=.75pt,
  minimum width=35.0mm,minimum height=63.6mm] (c_rac) at (139.8,-38.2) {};
\node[font=\FigSmall\bfseries,text=FigTeal,anchor=north] at (139.8,-7.2)
  {RAC Agentic Shell};
\node[font=\fontsize{5.8}{6.6}\selectfont\rmfamily,text=FigInk,align=center,anchor=north] at (139.8,-10.5)
  {Overlay on RIC \& in-node hosts\\[-0.5pt]Model services \& O-Cloud compute};

\node[fig/node,draw=FigTeal,fill=white,minimum width=29.8mm,minimum height=8.8mm,
  font=\fontsize{6.9}{7.9}\selectfont\rmfamily,inner sep=.7mm] (c_ra) at (139.8,-21.8)
  {\textbf{rApps} {\scriptsize (opt.\ rAgent)}\\Non-RT reasoning \& drift};
\node[fig/node,draw=FigTeal,fill=white,minimum width=29.8mm,minimum height=8.8mm,
  font=\fontsize{6.9}{7.9}\selectfont\rmfamily,inner sep=.7mm] (c_xa) at (139.8,-36.8)
  {\textbf{xApps} {\scriptsize (opt.\ xAgent)}\\Near-RT multi-cell coord.};
\node[fig/prop,fill=white,minimum width=29.8mm,minimum height=10.8mm,
  font=\fontsize{6.7}{7.7}\selectfont\rmfamily,inner sep=.7mm] (c_da) at (139.8,-58.0)
  {\textbf{dApps} {\scriptsize (opt.\ dAgent, $<\!10$\,ms)}\\In-node host via \textbf{E3}};

\draw[ari,<->,>={Latex[length=1.3mm,width=.9mm]}] (c_ra.south) -- (c_xa.north);
\draw[draw=FigAmber,line width=.75pt,dashed,-{Latex[length=1.3mm,width=.9mm]}] (c_xa.south) --
  node[fig/label,text=FigAmber,font=\fontsize{6.3}{7.1}\selectfont\rmfamily,fill=FigTealFill!55!white,inner sep=1.4pt]{async bounds} (c_da.north);

\node[fig/prop,minimum width=14.6mm,minimum height=23.5mm,font=\fontsize{6.8}{7.8}\selectfont\rmfamily,inner sep=.6mm] (c_ext) at (170.2,-36.8)
  {\textbf{External}\\\textbf{Agents}\\[1mm]Embodied\\robotics /\\edge AI};
\draw[fig/bidir,draw=FigAmber] (c_rac.east |- c_ra.east) -- (160,-21.8) -- (160,-31) -- (c_ext.west |- 162,-31);
\draw[fig/bidir,draw=FigAmber] (c_rac.east |- c_xa.east) -- (c_ext.west |- c_xa.east);
\node[font=\fontsize{5.8}{6.6}\selectfont\rmfamily,text=FigAmber,fill=white,inner sep=0.7pt,align=center] at (166.2,-20.5)
  {via gateway};

\node[fig/node,fill=FigGreyFill,minimum width=49.5mm,minimum height=9.8mm,font=\fontsize{6.8}{7.8}\selectfont\rmfamily] (c_sub) at (146.8,-77.2)
  {\textbf{Multi-Domain O-RU \& Local Autonomy}\\Multi-arch AAL (x86 / Arm / RISC-V / GPU)};
\draw[ari] (c_da.south) -- (c_da.south |- c_sub.north);
\draw[fig/inherited,draw=FigAmber,dashed,<->,>={Latex[length=1.2mm,width=.9mm]}]
  (c_ext.south) -- (170.2,-72.3);

\node[fig/node,fill=FigGreyFill,minimum width=46mm,minimum height=8.8mm,font=\FigSmall] (st_sdo) at (24,-93)
  {\textbf{3GPP \& O-RAN WGs}\\Normative specs \& nGRG studies};
\node[fig/node,fill=FigGreyFill,minimum width=44mm,minimum height=8.8mm,font=\FigSmall] (st_otic) at (76,-93)
  {\textbf{Professional Testing Houses}\\Continuous 6G assurance \& badging};
\node[fig/host,minimum width=64mm,minimum height=8.8mm,font=\FigSmall] (st_fed) at (137,-93)
  {\textbf{Proposed 6G open lab workflow}\\Recursive integration, development, and testing};

\draw[fig/inherited,-{Latex[length=1.4mm,width=1mm]}] (st_sdo.east) -- (st_otic.west);
\draw[fig/bidir,draw=FigTeal,dash pattern=on 2.5pt off 1.3pt] (st_sdo.south) -- (24,-99.2) -- (137,-99.2) -- (st_fed.south);
\draw[ari] (st_fed.east) -- (175.5,-93) -- (175.5,-77.2) -- (c_sub.east);
\node[fig/label,font=\fontsize{6.4}{7.2}\selectfont\rmfamily,text=FigTeal,anchor=east] at (174.5,-83.8)
  {Experimental feedback};
\end{tikzpicture}
\caption{Evolution of openness and intelligence across RAN architectures.
(a)~Vendor-integrated self-organizing network (SON) functions.
(b)~O-RAN hosts, Service Management and Orchestration (SMO), and
Acceleration Abstraction Layer (AAL) profiles. (c)~Proposed RAN Agentic
Controller (RAC) shells and exposure gateway. Federated open labs test
candidate requirements and inform their revision.}
\label{fig:hero}
\end{figure*}

%% file: sec2_beginning.tex
\section{Architectural Origins and Design Principles}
\label{sec:beginning}

Early Open RAN work laid the groundwork for the current O-RAN architecture.
Section~\ref{sec:open-fronthaul} traces its technical origins.
We organize the analysis around three design principles.

\begin{enumerate}
\item \textbf{Hierarchical RAN control.}
The Non-RT RIC supports model training and policy generation on longer
timescales. The Near-RT RIC hosts applications for low-latency inference
and radio control (Section~\ref{sec:lineage}).

\item \textbf{Four-stratum disaggregation.}
The architecture separates management, programmable control, disaggregated
3GPP RAN nodes, and compute and acceleration infrastructure
(Section~\ref{sec:baseline}).

\item \textbf{Open application hosting.}
Standardized hosts and application programming interfaces (APIs)
let independent suppliers deploy their optimization algorithms
(Section~\ref{sec:hosting}).
\end{enumerate}

\subsection{Disaggregation, Open Fronthaul, and Programmable Control}
\label{sec:open-fronthaul}

Founded in June 2016, the xRAN Forum sought to
decouple RAN control and user planes, run modular base-station software on
commodity servers, and standardize open northbound and southbound interfaces
\cite{xran18merger}: its northbound vision anticipated external applications on
a programmable controller, while its southbound work defined the functional split
between baseband processing and radio units.
Table~\ref{tab:lineage} places this work in the context of early Open RAN
and network data analytics.

Packet fronthaul made that split viable at scale. Building on the packet
transport and Radio-over-Ethernet encapsulations of the IEEE~1914 Next
Generation Fronthaul Interface (NGFI) working group \cite{ieee1914.1,ieee1914.3},
the xRAN Forum and later the O-RAN ALLIANCE specified the Option~7.2x functional split
across control, user, and synchronization planes
\cite{oran.wg4.cus}. In parallel, SoftRAN (2013) and FlexRAN
(2016) demonstrated software-defined networking control over softwarized
base stations \cite{softran13,flexran16}, while ETSI's Experiential Networked
Intelligence (ENI) group, the Telecom Infra Project (TIP), and wireless-AI
study groups examined open interfaces and automation \cite{etsi17.eni,vodafone17.tip,ieee5gtf18.levels,wwrf.wgwai}.
The xRAN Forum's early agenda thus paired an open fronthaul split between
baseband and radio units with open interfaces for programmable control.

\begin{table}[!t]
\centering
\caption{Milestones in Early Open RAN and \mbox{Network Data Analytics (2016--2019)}}
\label{tab:lineage}
\footnotesize
\setlength{\tabcolsep}{2.2pt}
\renewcommand{\arraystretch}{1.15}
\begin{tabularx}{\columnwidth}{@{}>{\raggedright\arraybackslash}p{1.15cm}>{\raggedright\arraybackslash}p{1.65cm}>{\raggedright\arraybackslash}X@{}}
\toprule
\textbf{Date} & \textbf{Venue} & \textbf{Milestone} \\
\midrule
2016-02 & IEEE 1914 & Packet fronthaul project authorized \\
2016-06 & xRAN Forum & Open northbound and southbound interfaces \\
2017-02 & ETSI ENI & Cognitive management group formed \\
2017-06 & 3GPP SA2 & Core network analytics adopted \\
2017-11 & ITU-T SG13 & ML5G focus group established \\
2017-11 & TIP & Open RAN project group launched \\
2017-12 & IEEE & Data-driven RAN architecture \\
2018-02 & O-RAN & Launched by AT\&T, China Mobile, Deutsche Telekom,
NTT~DOCOMO, and Orange \\
2018-03 & IEEE 5G TF & Data analytics in the CU and DU \\
2018-04 & xRAN Forum & Southbound 7.2x fronthaul spec \\
2018-06 & 3GPP RAN3 & RAN data-collection study approved \\
2018-10 & O-RAN & White paper specifying two-tier RIC \\
2018-12 & IEEE & Management, core, and RAN analytics \\
2019-01 & ITU-T ML5G & Unified ML architecture for 5G \\
\bottomrule
\end{tabularx}
\end{table}

\subsection{RAN Data Analytics and the Origins of the RIC}
\label{sec:lineage}
\label{sec:whitepaper}

\input{figs/fig_archmap}

Applying machine learning to programmable RAN control raises the question of
where to train models and where to make radio decisions. Model training can
draw on measurements collected over time. Radio scheduling uses recent
measurements and must meet radio deadlines. In 2017, 3GPP adopted the Network
Data Analytics Function (NWDAF) for the 5G core \cite{3gpp.ts23501}. An early RAN architecture in \emph{IEEE Vehicular Technology Magazine}
described how network data could guide radio control
\cite{i17bigdata}. A subsequent design assigned measurement aggregation and
model training to the central unit (CU) and radio control at transmission time
interval (TTI) granularity to the distributed unit (DU)
\cite{ieee5gtf18.cudaduda}.

The 2018 O-RAN white paper proposed a two-tier RIC to support model training
and radio control on different timescales \cite{oran18whitepaper}. In this
design, the Non-RT RIC trains models and sends policy guidance over A1 to
the Near-RT RIC. The Near-RT RIC runs inference and uses E2 to obtain
measurements from radio nodes and issue control actions. The white paper also
proposed hosting third-party control applications on the Near-RT RIC.
Subsequent O-RAN specifications defined R1 for rApps to access Non-RT RIC
services (Section~\ref{sec:lifecycle}).

Related work on machine learning in mobile networks also distinguished model
training from its use in radio control
\cite{ieee5gtf18.levels,itu19.ml5garc,itu19.y3172}. The CU in the earlier
proposal served a training role comparable to that of the Non-RT RIC
\cite{ieee5gtf18.cudaduda}. The DU, however, made radio decisions at TTI
granularity, on a shorter timescale than the Near-RT RIC control loop.
In O-RAN, the Near-RT RIC guides radio operation over E2, and the DU retains
responsibility for per-TTI scheduling. This timing boundary motivates research
on hosting applications inside radio nodes
(Sections~\ref{sec:timescale} and~\ref{sec:6g-theses}).

\subsection{Disaggregated Radio Nodes and the Architectural Baseline}
\label{sec:baseline}

\input{figs/fig_socket}

3GPP studied the CU/DU split in TR~38.801 \cite{3gpp.tr38801}. The O-RAN
central unit (O-CU) has separate control-plane and user-plane entities
\cite{oran.wg1.arch}. O-RAN further divides physical-layer
processing between the O-RAN distributed unit (O-DU) and the O-RAN radio unit (O-RU) at the
Option~7.2x Open Fronthaul
interface \cite{oran.wg4.cus}.

\begin{itemize}
\item \textbf{O-CU-CP (control plane).} Hosts Radio Resource Control (RRC)
and control-plane Packet Data Convergence Protocol (PDCP) functions.
\item \textbf{O-CU-UP (user plane).} Hosts the Service Data Adaptation
Protocol (SDAP) and user-plane PDCP functions.
\item \textbf{O-DU.} Hosts Radio Link Control (RLC) and Medium Access
Control (MAC), including scheduling, and performs High-PHY processing.
\item \textbf{O-RU.} Performs Low-PHY and radio-frequency (RF) processing.
\end{itemize}

The architecture integrates these nodes with the functions that manage them,
control their operation, and provide computing resources.

\begin{enumerate}
\item \textbf{Management and AI lifecycle.}
The Service Management and Orchestration (SMO) framework manages the network,
and its Non-RT RIC supports model training and policy generation
\cite{oran.wg1.arch}.
\item \textbf{Programmable control.}
The Near-RT RIC hosts applications that use radio measurements to guide
near-real-time control.
\item \textbf{Disaggregated RAN and core.}
The O-CU, O-DU, and O-RU implement the radio protocol functions described
above, and the O-CU connects the RAN to the 5G core.
\item \textbf{Compute and acceleration.}
O-Cloud provides compute resources \cite{oran.wg6.o2}, and the AAL gives
O-DU software access to hardware accelerators.
\end{enumerate}

This division allows an independently developed control application to use
the measurements and actions exposed by another supplier's radio node,
as the following traffic-steering example illustrates.

\begin{samepage}
Figure~\ref{fig:archmap} shows this four-stratum architecture. The teal and
amber elements preview how the 6G proposals in Section~\ref{sec:6g},
including the RAC in Section~\ref{sec:rac-concept},
build on the existing O-RAN architecture.
\par\end{samepage}

\subsection{Hosting Third-Party RAN Control Applications}
\label{sec:hosting}

For an operator introducing a new radio optimization algorithm, a practical
question is whether it can be deployed independently of the base-station
software. Earlier cellular networks already supported closed-loop optimization
through the self-organizing network (SON) functions specified by 3GPP
\cite{3gpp.ts32500,3gpp.ts28313}. In the distributed SON
deployment model \cite{3gpp.tr36902}, the operator configures performance
targets and management parameters \cite{3gpp.ts32500,3gpp.ts28313}. The
algorithm that uses radio measurements to adjust the base station's operation
typically runs within the equipment supplier's software.

Configuring a performance target gives the operator limited control over how
the base station meets it. O-RAN introduced third-party application hosting so
that operators could also choose the algorithms used to pursue their
objectives \cite{oran18whitepaper,oran.wg1.arch}.
The roles of rApps and xApps reflect the existing division between network
management and radio resource management.
For traffic steering, an rApp on the Non-RT RIC can use performance trends
across cells to formulate a load-balancing policy. An xApp on the Near-RT RIC
can apply that policy to current radio measurements when selecting a handover
target for an individual user \cite{oran.uc036}.
When the radio nodes expose the measurements and control actions an algorithm
requires, the operator can deploy or update the algorithm as a RIC application
independently of the base-station software.

%% file: figs/fig_archmap.tex
\begin{figure*}[!t]
\centering
\begin{tikzpicture}[fig/base,
  loopstep/.style={draw=FigNavy,fill=white,circle,minimum size=3.4mm,inner sep=0pt,
    font=\fontsize{6.2}{6.2}\selectfont\rmfamily\bfseries,text=FigNavy},
  extarrow/.style={draw=FigAmber,line width=.75pt,dashed,{Latex[length=1.35mm,width=.95mm]}-{Latex[length=1.35mm,width=.95mm]}},
  extdown/.style={draw=FigAmber,line width=.75pt,dashed,-{Latex[length=1.35mm,width=.95mm]}},
  racspine/.style={draw=FigTeal,line width=1.5pt},
  ractab/.style={draw=FigTeal,line width=.8pt,fill=FigTealFill,text=FigInk,
    font=\fontsize{5.8}{6.4}\selectfont\rmfamily\bfseries,
    sharp corners,inner sep=1pt,minimum width=10.2mm,minimum height=4.4mm,align=center},
  racgw/.style={draw=FigTeal,line width=.9pt,fill=FigTealFill!55!white,text=FigInk,
    font=\fontsize{6.1}{6.8}\selectfont\rmfamily\bfseries,sharp corners,inner sep=1.4pt,align=center}]
\path[use as bounding box] (0,1.2) rectangle (178,-130.5);

\foreach \y in {-35.5,-61,-105.5} \draw[fig/guide,dashed] (1,\y) -- (22.5,\y);
\node[fig/title] at (1,-7.5) {1. Management};
\node[fig/title] at (1,-12.0) {\& AI lifecycle};
\node[font=\FigSmall,text=FigLine,anchor=west,inner sep=0pt] at (1,-17.0) {$>1$\,s training, policy};
\node[fig/title] at (1,-45.0) {2. Programmable};
\node[fig/title] at (1,-49.5) {control};
\node[font=\FigSmall,text=FigLine,anchor=west,inner sep=0pt] at (1,-54.5) {10\,ms--1\,s loop};
\node[fig/title] at (1,-75.5) {3. Disaggregated};
\node[fig/title] at (1,-80.0) {RAN \& core};
\node[font=\FigSmall,text=FigLine,anchor=west,inner sep=0pt] at (1,-85.0) {3GPP + $<10$\,ms loops};
\node[fig/title] at (1,-111.5) {4. Compute \&};
\node[fig/title] at (1,-116.0) {AAL substrate};

\node[fig/smo/group,minimum width=143mm,minimum height=31mm,inner sep=0pt,
  outer sep=0pt] (smo) at (101.5,-15.0) {};
\node[fig/heading] at (31.5,-2.6) {SMO (Service Management \& Orchestration)};

\node[fig/group,fill=white,minimum width=73mm,minimum height=18.2mm,
  inner sep=0pt,outer sep=0pt] (nrtbox) at (68.5,-19.9) {};
\node[fig/heading] at (33.5,-13.0) {Non-RT RIC};
\node[fig/node,minimum width=28.5mm,minimum height=9.4mm,font=\fontsize{6.0}{6.8}\selectfont\rmfamily]
  (nrtfw) at (48.0,-22.5) {\textbf{SMO AI/ML Model Services}\\Train, catalog, drift, infer.};

\node[fig/app,minimum width=30.5mm,minimum height=9.4mm,font=\fontsize{6.2}{7.0}\selectfont\rmfamily]
  (rapp) at (86.0,-22.5) {\textbf{rApps} {\scriptsize (opt.\ rAgent)}\\Train, policy \& drift};
\draw[fig/bidir] (nrtfw.east) -- node[fig/label,above=.35mm,font=\fontsize{6.2}{7.0}\selectfont\rmfamily] {R1} (rapp.west);

\node[fig/node,fill=white,minimum width=40mm,minimum height=8.8mm,font=\fontsize{6.1}{6.9}\selectfont\rmfamily]
  (smoorc) at (151.0,-12.6) {\textbf{SMO Orchestration}\\Model and network management};
\node[fig/prop,fill=FigAmberFill,minimum width=40mm,minimum height=8.8mm,
  font=\fontsize{5.9}{6.7}\selectfont\rmfamily] (trust) at (151.0,-23.8)
  {\textbf{\textcolor{FigAmber}{Cross-domain trust}}\\[-1pt]Attestation \& multi-domain policy};
\draw[fig/arrow] (105,-14.2) -- (smoorc.west |- 131,-14.2);
\node[loopstep] at (115.5,-14.2) {1};

\draw[fig/arrow] (60.5,-17.8) -- (60.5,-7.2) -- (122.5,-7.2) -- (122.5,-32.2);
\node[fig/label,fill=FigBlueFill,font=\fontsize{5.9}{6.7}\selectfont\rmfamily,inner sep=.65pt]
  at (91.5,-7.2) {Model serving (R1/A1-ML; O2 compute)};

\node[fig/near/group,minimum width=73mm,minimum height=18.0mm,inner sep=0pt,
  outer sep=0pt] (near) at (68.5,-49.2) {};
\node[fig/heading] at (33.5,-42.2) {Near-RT RIC};
\node[fig/node,minimum width=28.5mm,minimum height=9.4mm,font=\fontsize{6.2}{7.0}\selectfont\rmfamily]
  (ricapi) at (48.0,-51.2) {RIC APIs \&\\[-1pt]conflict mgmt.};

\node[fig/app,minimum width=30.5mm,minimum height=9.4mm,font=\fontsize{6.2}{7.0}\selectfont\rmfamily]
  (xapp) at (86.0,-51.2) {\textbf{xApps} {\scriptsize (opt.\ xAgent)}\\Near-RT radio control};
\draw[fig/bidir] (ricapi.east) -- (xapp.west);

\draw[fig/arrow] (48.0,-29.0) -- node[fig/label,right=2.2mm,font=\fontsize{6.2}{7.0}\selectfont\rmfamily,align=left]
  {A1 / A1-ML\\[-1pt]policy \& cross-RIC ML} (48.0,-40.2);
\node[loopstep] at (48.0,-34.8) {2};

\node[fig/node,minimum width=36.5mm,minimum height=7.8mm,font=\fontsize{6.0}{6.8}\selectfont\rmfamily]
  (consumer) at (154.2,-43.8) {\textbf{Authorized Y1 consumers}\\{\scriptsize Read-only RAN analytics}};
\node[fig/prop,fill=FigAmberFill,minimum width=36.5mm,minimum height=8.8mm,
  font=\fontsize{6.0}{6.8}\selectfont\rmfamily] (edgeagent) at (154.2,-54.4)
  {\textbf{\textcolor{FigAmber}{External / Edge AI Agents}}\\[-1pt]Edge compute \& partner domains};

\node[fig/cu,minimum width=24.5mm,minimum height=9.4mm,font=\fontsize{6.4}{7.3}\selectfont\rmfamily]
  (cp) at (44.25,-72.5) {\textbf{O-CU-CP}\\RRC, PDCP-C};
\node[fig/cu,minimum width=24.5mm,minimum height=9.4mm,font=\fontsize{6.4}{7.3}\selectfont\rmfamily]
  (up) at (74.5,-72.5) {\textbf{O-CU-UP}\\SDAP, PDCP-U};
\node[fig/node,fill=FigGreyFill,minimum width=24mm,minimum height=9.4mm,font=\fontsize{6.4}{7.3}\selectfont\rmfamily]
  (core) at (151,-72.5) {\textbf{5GC}\\User \& ctrl};

\node[fig/du,minimum width=73mm,minimum height=17.4mm,inner sep=0pt] (du) at (68.5,-92.9) {};
\node[fig/heading] at (33.5,-86.3) {O-DU};
\node[fig/node,fill=white,minimum width=25.5mm,minimum height=10.6mm,inner sep=.6mm,
  font=\fontsize{5.9}{6.7}\selectfont\rmfamily]
  (dustack) at (46.2,-95.0) {\textbf{O-DU Stack}\\RLC, MAC, High-PHY\\[-1pt]{\fontsize{5.5}{6.2}\selectfont Look-aside / inline AAL}};

\node[fig/prop/group,minimum width=33.0mm,minimum height=12.4mm,
  inner sep=0pt,outer sep=0pt] (dhost) at (86.0,-94.4) {};
\node[font=\fontsize{5.9}{6.6}\selectfont\rmfamily,text=FigAmber,anchor=north] at (86.0,-88.7)
  {\textbf{In-node host}};
\node[fig/app,minimum width=27.5mm,minimum height=5.5mm,inner sep=.5mm,
  font=\fontsize{6.0}{6.8}\selectfont\rmfamily] (dapp) at (86.0,-96.8)
  {\textbf{dApps} {\scriptsize (opt.\ dAgent)}};

\draw[extarrow] (dustack.east |- 59,-95.5) -- (dhost.west |- 69.5,-95.5);
\node[fig/label,text=FigAmber,font=\fontsize{5.8}{6.5}\selectfont\rmfamily,fill=white,inner sep=0.6pt,above=.5mm]
  at (64.2,-95.2) {\textbf{E3} (samples)};

\node[fig/ru,fill=FigGreyFill,minimum width=24mm,minimum height=15.4mm,font=\fontsize{6.4}{7.3}\selectfont\rmfamily]
  (ru) at (151,-93.2) {\textbf{O-RU}\\Low-PHY, RF \&\\[-1pt]{\scriptsize antenna arrays}};

\draw[fig/inherited] (cp.east) -- node[fig/label,above=.3mm,font=\fontsize{6.3}{7.2}\selectfont\rmfamily]{E1} (up.west);
\draw[fig/inherited,preaction={draw=white,line width=2pt}] (up.east) -- (core.west);
\node[fig/label,above=.2mm,font=\fontsize{6.3}{7.2}\selectfont\rmfamily] at (133,-72.3) {NG-U};
\draw[fig/inherited,preaction={draw=white,line width=2pt}]
  ([xshift=-2mm]cp.north east) -- (54.5,-65.6) -- (151,-65.6) -- (core.north);
\node[fig/label,font=\fontsize{6.3}{7.2}\selectfont\rmfamily] at (133,-65.6) {NG-C};
\draw[fig/inherited] (cp.south) -- (44.25,-81.2) -- (44.25,-84.2);
\node[fig/label,font=\fontsize{6.3}{7.2}\selectfont\rmfamily] at (44.25,-81.0) {F1-C};
\draw[fig/inherited] (up.south) -- (74.5,-79.6) -- (55.5,-79.6) -- (55.5,-84.2);
\node[fig/label,font=\fontsize{6.3}{7.2}\selectfont\rmfamily] at (65,-79.6) {F1-U};

\draw[fig/bidir] (61.5,-58.2) -- (61.5,-62.2);
\draw[fig/interface] (44.25,-62.2) -- (109.5,-62.2) -- (109.5,-82.6) -- (61.5,-82.6) -- (61.5,-84.2);
\draw[fig/bidir] (44.25,-62.2) -- (cp.north);
\draw[fig/bidir] (74.5,-62.2) -- (up.north);
\foreach \x in {44.25,61.5,74.5} \fill[FigBlue] (\x,-62.2) circle (.55mm);
\node[fig/label,font=\fontsize{6.3}{7.2}\selectfont\rmfamily,inner sep=1pt] at (76.5,-60.1) {E2: live indications \& ctrl};
\node[loopstep] at (53,-62.2) {3};
\node[loopstep] at (109.5,-74.8) {4};

\draw[fig/bidir,preaction={draw=white,line width=2pt}] (105,-88.8) -- (139,-88.8);
\node[fig/label,font=\fontsize{6.2}{7.0}\selectfont\rmfamily] at (131.5,-86.3) {Open FH 7.2x};
\node[loopstep] at (131.5,-88.8) {5};
\draw[fig/management,<->,>={Latex[length=1.25mm,width=.85mm]},preaction={draw=white,line width=1.8pt}]
  (105,-100.2) -- (139,-100.2);
\node[fig/label,font=\fontsize{6.0}{6.8}\selectfont\rmfamily] at (123,-102.6) {Open FH M-plane};

\draw[fig/management,-{Latex[length=1.5mm,width=1mm]}]
  (du.west) -- (26.2,-92.9) -- (26.2,-15.0) -- (smo.west);
\draw[fig/management] (26.2,-49.2) -- (near.west);
\draw[fig/management] (26.2,-72.5) -- (cp.west);
\foreach \y in {-49.2,-72.5} \fill[FigBlue] (26.2,\y) circle (.55mm);
\node[fig/label,font=\fontsize{6.6}{7.5}\selectfont\rmfamily] at (26.2,-35.5) {O1};
\node[loopstep] at (26.2,-61) {6};

\node[fig/cloud/group,minimum width=143mm,minimum height=14.0mm,inner sep=0pt,
  outer sep=0pt] (cloud) at (101.5,-116.6) {};
\node[fig/heading,align=left] at (31.5,-116.6) {O-Cloud\\[-1pt]{\scriptsize \& Edge}};

\node[fig/node,fill=white,minimum width=48.5mm,minimum height=9.4mm,font=\FigSmall] (acc) at (69.2,-116.6)
  {\textbf{AAL Accelerators} (Today)\\[-1pt]ASIC / FPGA / DSP / GPU};
\node[fig/prop,fill=FigAmberFill,minimum width=72mm,minimum height=9.4mm,
  font=\fontsize{6.3}{7.2}\selectfont\rmfamily] (comp6g) at (133,-116.6)
  {\textbf{\textcolor{FigAmber}{6G Compute \& Accelerators}}\\[-1pt]GPU, vector processors, analog memory};

\draw[fig/bidir] (50,-101.6) -- node[fig/label,right=.7mm,font=\fontsize{6.2}{7.0}\selectfont\rmfamily]{AAL API} (50,-109.6);
\draw[extarrow] (91,-101.6) -- node[fig/label,left=.7mm,text=FigAmber,font=\fontsize{6.1}{6.9}\selectfont\rmfamily]{Fast-path RT offload} (91,-109.6);

\draw[extarrow] (166.5,-58.8) -- (166.5,-109.6);
\node[fig/label,text=FigAmber,font=\fontsize{6.1}{6.9}\selectfont\rmfamily,rotate=90] at (169.0,-83.5)
  {RAN--compute coordination};
\draw[fig/management,<->,>={Latex[length=1.4mm,width=.95mm]}] (smo.east) -- (175.2,-15.0) -- (175.2,-116.6) -- (cloud.east);
\node[fig/label,font=\fontsize{6.2}{7.0}\selectfont\rmfamily,align=center] at (175.2,-61)
  {\textbf{O2}\\[-1pt]{\scriptsize compute}};

\node[ractab,anchor=west] (tab1) at (105,-22.5) {RAC port};
\node[ractab,anchor=west] (tab2) at (105,-51.2) {RAC port};
\node[ractab,minimum width=11.4mm,anchor=west] (tab3) at (105,-96.8) {dShell guard};

\draw[draw=FigTeal,line width=.8pt] (rapp.east) -- (tab1.west);
\draw[draw=FigTeal,line width=.8pt] (xapp.east) -- (tab2.west);
\draw[draw=FigAmber,line width=.75pt,dashed] (dhost.east |- tab3) -- (tab3.west);

\node[racgw,minimum width=12.4mm,minimum height=6.2mm] (gw) at (122.5,-35.3)
  {Exposure\\[-1pt]GW};

\draw[racspine] (tab1.east) -- (122.5,-22.5) -- (gw.north);
\draw[racspine] (gw.south) -- (122.5,-51.2) -- (tab2.east);

\node[font=\fontsize{5.8}{6.5}\selectfont\rmfamily\bfseries,text=FigTeal,align=center,fill=white,inner sep=0.6pt]
  at (113.2,-56.8) {RAC shell};

\draw[fig/arrow,preaction={draw=white,line width=2pt}]
  (near.east |- consumer) -- node[fig/label,above=.3mm,pos=.70,font=\fontsize{6.2}{7.0}\selectfont\rmfamily] {Y1} (consumer.west);

\draw[extdown,preaction={draw=white,line width=2pt}] (122.5,-51.2) -- (122.5,-96.8) -- (tab3.east);
\node[fig/label,text=FigAmber,font=\fontsize{5.9}{6.7}\selectfont\rmfamily,align=left,fill=white,inner sep=0.8pt,anchor=west]
  at (123.8,-78.2) {\textbf{Async bounds}\\[-1pt]{\scriptsize no synchronous calls $<10$\,ms}};

\draw[extarrow] (trust.west) -- (130.6,-23.8) -- (130.6,-34.0) -- (gw.east |- 130.6,-34.0);
\draw[draw=FigTeal,line width=.8pt,<->,>={Latex[length=1.4mm,width=.95mm]}]
  (gw.east |- 131.8,-36.6) -- (131.8,-36.6) -- (131.8,-54.4) -- (edgeagent.west);
\node[fig/label,text=FigTeal,font=\fontsize{5.8}{6.6}\selectfont\rmfamily,align=center,fill=white,inner sep=0.6pt]
  at (131.8,-49.2) {Capability\\[-1pt]calls};

\draw[fig/guide] (1,-125.4) -- (177,-125.4);
\draw[fig/interface] (1.5,-129.0) -- (6.5,-129.0);
\node[anchor=west,inner sep=0pt,font=\FigSmall] at (7.3,-129.0) {O-RAN IF};
\draw[fig/management] (23,-129.0) -- (28,-129.0);
\node[anchor=west,inner sep=0pt,font=\FigSmall] at (28.8,-129.0) {Mgmt.\ (O1/O2/M)};
\draw[fig/inherited] (54,-129.0) -- (59,-129.0);
\node[anchor=west,inner sep=0pt,font=\FigSmall] at (59.8,-129.0) {3GPP link};
\node[fig/app,minimum width=3.8mm,minimum height=2.5mm,inner sep=0pt] at (76.5,-129.0) {};
\node[anchor=west,inner sep=0pt,font=\FigSmall] at (79.2,-129.0) {Hosted app};
\node[ractab,minimum width=3.8mm,minimum height=2.5mm,inner sep=0pt] at (97.5,-129.0) {};
\node[anchor=west,inner sep=0pt,font=\FigSmall,text=FigTeal] at (100.2,-129.0) {\textbf{RAC overlay}};
\node[fig/prop,minimum width=3.8mm,minimum height=2.5mm,inner sep=0pt] at (120.5,-129.0) {};
\node[anchor=west,inner sep=0pt,font=\FigSmall] at (123.2,-129.0) {Proposed 6G extensions};
\node[loopstep] at (150.5,-129.0) {1};
\node[anchor=west,inner sep=0pt,font=\FigSmall] at (153.2,-129.0) {AI/ML loop (Sec.~III-B)};
\end{tikzpicture}
\caption{The four-stratum O-RAN architecture and its AI/ML workflow (steps~\textbf{1}--\textbf{6}). Amber marks proposed 6G extensions, and teal marks the RAC overlay. In-node hosting through E3 is under study for both the O-DU and O-CU. The RAC (Section~\ref{sec:rac-concept}) coordinates applications across hosts. The exposure gateway connects external agents to the overlay. Edge service continuity during handover is discussed in Section~\ref{sec:6g-theses}.}
\label{fig:archmap}
\end{figure*}
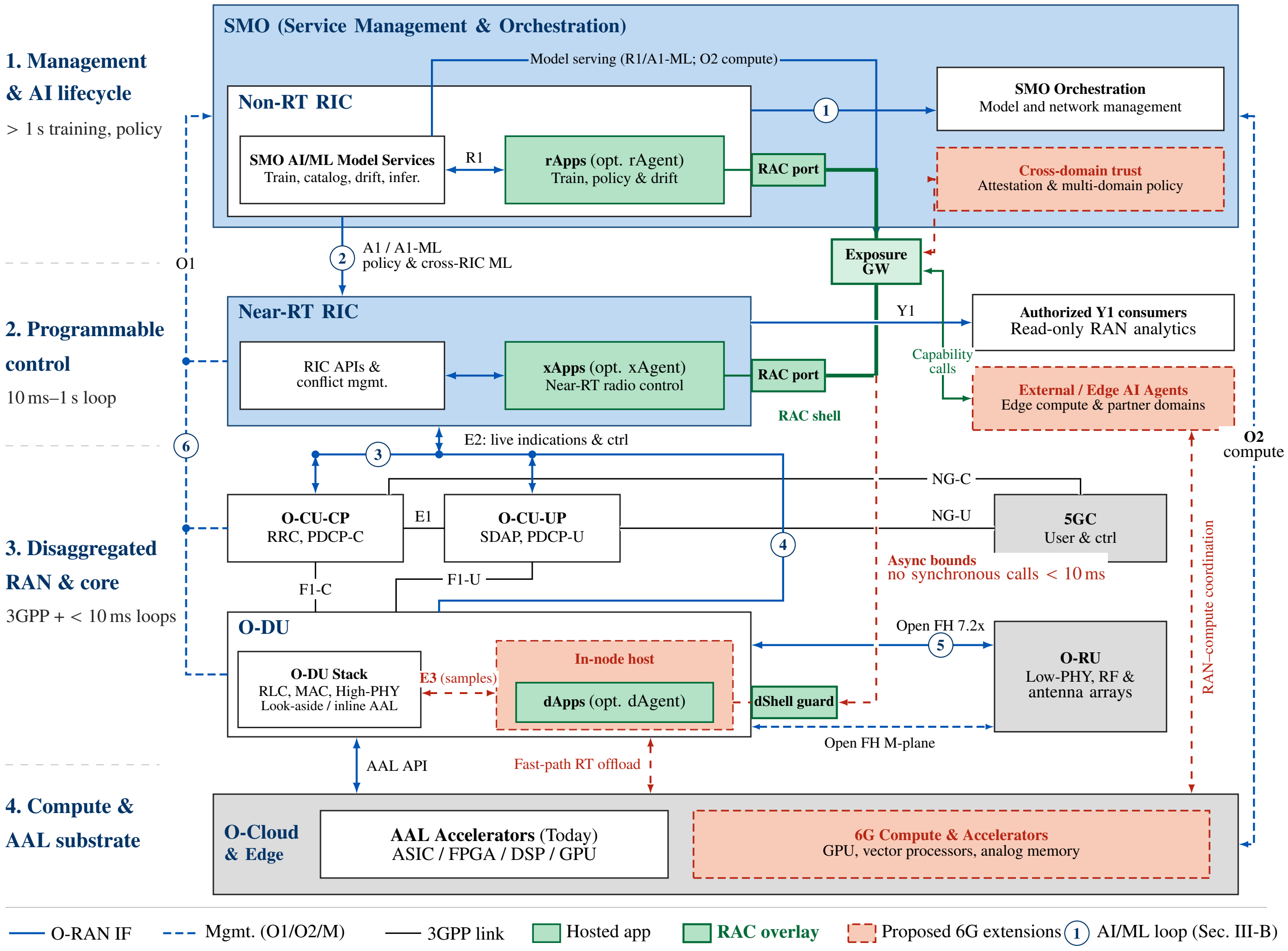

%% file: figs/fig_socket.tex
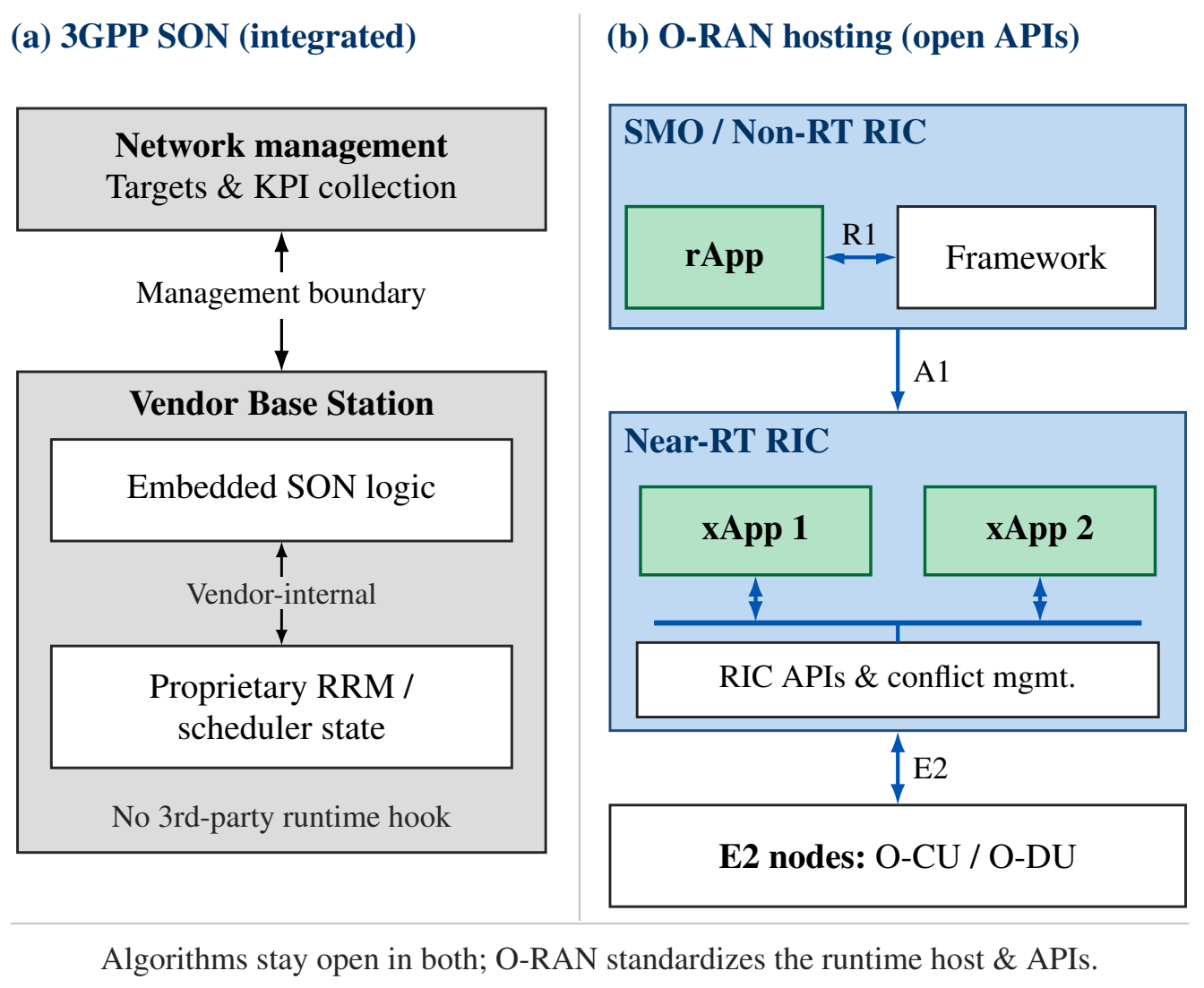
\begin{figure}[!t]
\centering
\begin{tikzpicture}[
  fig/base,
  every node/.style={align=center,inner sep=0pt,outer sep=0pt},
  sframe/.style={fig/group,inner sep=0pt,outer sep=0pt},
  bus/.style={draw=FigBlue,line width=1.1pt}
]
\path[use as bounding box] (0,0) rectangle (88,73);

\draw[fig/guide] (42.5,5) -- (42.5,72);
\node[fig/title,anchor=base west,font=\FigSmall\bfseries] at (0.5,69.5) {(a) 3GPP SON (integrated)};
\node[fig/title,anchor=base west,font=\FigSmall\bfseries] at (44.5,69.5) {(b) O-RAN hosting (open APIs)};

\node[fig/cloud,minimum width=39mm,minimum height=9mm,font=\FigSmall] (smgmt) at (20.5,60.5)
  {\textbf{Network management}\\Targets \& KPI collection};

\draw[fig/bidir,draw=FigInk,line width=.6pt] (smgmt.south) -- (20.5,45.5);
\node[fig/label,font=\fontsize{6.6}{7.6}\selectfont\rmfamily] at (20.5,51.2)
  {Management boundary};

\node[sframe,draw=FigLine,fill=FigGreyFill,minimum width=39mm,minimum height=35.5mm] (sbase) at (20.5,27.8) {};
\node[font=\FigSmall\bfseries,text=FigInk,anchor=north] at (20.5,44.2) {Vendor Base Station};

\node[fig/node,minimum width=34mm,minimum height=7.5mm,font=\FigSmall] (sson) at (20.5,36.8)
  {Embedded SON logic};
\node[fig/node,minimum width=34mm,minimum height=9mm,font=\FigSmall] (srrm) at (20.5,20.8)
  {Proprietary RRM /\\scheduler state};

\draw[fig/inherited,<->,>={Latex[length=1.2mm,width=.9mm]}] (sson.south) -- (srrm.north);
\node[font=\fontsize{6.4}{7.4}\selectfont\rmfamily,text=FigLine,fill=FigGreyFill,inner sep=.4mm]
  at (20.5,29.2) {Vendor-internal};
\node[font=\fontsize{6.6}{7.6}\selectfont\rmfamily,text=FigLine] at (20.5,12.5)
  {No 3rd-party runtime hook};

\node[sframe,fig/smo/group,minimum width=42.5mm,minimum height=16.5mm] (ssmo) at (66,57) {};
\node[font=\FigSmall\bfseries,text=FigNavy,anchor=north west] at (45.8,64.2) {SMO / Non-RT RIC};
\node[fig/app,minimum width=14.5mm,minimum height=7.5mm,font=\FigSmall] (srapp) at (53.2,54) {\textbf{rApp}};
\node[fig/node,minimum width=19mm,minimum height=7.5mm,font=\FigSmall] (snonrt) at (75.5,54) {Framework};
\draw[fig/bidir] (srapp.east) -- node[fig/label,fill=FigBlueFill,above=.3mm,font=\fontsize{6.6}{7.6}\selectfont\rmfamily]{R1} (snonrt.west);

\draw[fig/arrow] (ssmo.south) -- node[fig/label,right=.5mm,font=\fontsize{6.6}{7.6}\selectfont\rmfamily]{A1} (66,42.5);

\node[sframe,fig/near/group,minimum width=42.5mm,minimum height=23.5mm] (snear) at (66,30.8) {};
\node[font=\FigSmall\bfseries,text=FigNavy,anchor=north west] at (45.8,41.4) {Near-RT RIC};

\node[fig/app,minimum width=17mm,minimum height=6.5mm,font=\FigSmall] (sx1) at (55.5,33.8) {\textbf{xApp 1}};
\node[fig/app,minimum width=17mm,minimum height=6.5mm,font=\FigSmall] (sx2) at (76.5,33.8) {\textbf{xApp 2}};

\draw[bus] (48,27) -- (84,27);
\draw[fig/bidir] (sx1.south) -- (55.5,27);
\draw[fig/bidir] (sx2.south) -- (76.5,27);
\node[fig/node,minimum width=38.5mm,minimum height=5.5mm,font=\fontsize{6.8}{7.8}\selectfont\rmfamily] (sserv) at (66,22.8)
  {RIC APIs \& conflict mgmt.};
\draw[fig/interface] (66,27) -- (sserv.north);

\draw[fig/bidir] (snear.south) -- node[fig/label,right=.5mm,font=\fontsize{6.6}{7.6}\selectfont\rmfamily]{E2} (66,13.5);
\node[fig/node,minimum width=42.5mm,minimum height=7.5mm,font=\FigSmall] (senodes) at (66,9.8)
  {\textbf{E2 nodes:} O-CU / O-DU};

\draw[fig/guide] (0.5,4.8) -- (87.5,4.8);
\node[font=\fontsize{6.8}{7.8}\selectfont\rmfamily,text=FigLine] at (44,2)
  {Algorithms stay open in both; O-RAN standardizes the runtime host \& APIs.};
\end{tikzpicture}
\caption{Standardization boundaries in (a)~distributed 3GPP SON
(vendor-internal logic) and (b)~O-RAN hosting (standardized R1 and Near-RT RIC
APIs with A1/E2 control). In (a) the equipment supplier implements the
algorithm within the base station. In (b) an independent supplier can
deploy its algorithm as a RIC application.}
\label{fig:socket}
\end{figure}

%% file: sec3_today.tex
\section{The O-RAN Architecture for AI/ML}
\label{sec:today}

Building on the open hosts introduced in Section~\ref{sec:beginning},
this section examines how AI/ML applications obtain measurements, use models,
and issue radio control actions within their timing and compute budgets.

\begin{enumerate}
\item \textbf{O-RAN control loops.}
The time available for a control action, the measurements it requires, and
its computational cost guide the placement of control functions in the
Non-RT RIC, the Near-RT RIC, or the radio node (Section~\ref{sec:timescale}).

\item \textbf{Model lifecycle and control interfaces.}
A radio energy-saving example traces how model registration leads to policy
guidance and radio control, and how operational measurements feed back into
model updates
(Section~\ref{sec:lifecycle}).

\item \textbf{Cloud infrastructure and acceleration.}
The SMO provisions O-Cloud resources over O2. The AAL defines how O-DU software
accesses hardware-accelerated functions (Section~\ref{sec:substrate}).

\item \textbf{RAN AI/ML in 3GPP and O-RAN.}
Section~\ref{sec:milestones} reviews 3GPP work on AI/ML for the New Radio (NR)
air interface, the next-generation RAN (NG-RAN),
and model management, alongside O-RAN's work on RIC applications.
\end{enumerate}

\subsection{O-RAN Control Loops and Application Placement}
\label{sec:timescale}

The choice of where to run a RAN control function depends on how quickly it must act,
which measurements it needs, and how much computation it requires. An rApp on the
Non-RT RIC can draw on historical measurements from multiple cells to update a policy
over seconds or longer. An xApp on the Near-RT RIC can use recent cell and user
measurements to guide radio resource management (RRM), typically on the order of 10\,ms to
1\,s. The O-DU makes scheduling decisions for each slot using locally available radio
and buffer information \cite{oran.wg1.arch}. Figures~\ref{fig:socket}
and~\ref{fig:loops} show the application hosts and the time scales of their
control loops. Table~\ref{tab:son} relates representative 3GPP SON functions
to the applications and radio nodes that can support them.

\input{figs/fig_loops}
\input{figs/fig_aal}

\begin{table}[!tb]
\centering
\caption{Mapping 3GPP SON Functions to O-RAN Loops
  \mbox{(\textup{\textbullet}~primary, \textup{\textopenbullet}~secondary)}}
\label{tab:son}
\footnotesize
\setlength{\tabcolsep}{2pt}
\renewcommand{\arraystretch}{1.05}
\begin{tabular}{@{}>{\raggedright\arraybackslash}p{1.85cm}>{\raggedright\arraybackslash}p{2.05cm}>{\centering\arraybackslash}p{0.68cm}>{\centering\arraybackslash}p{0.68cm}>{\centering\arraybackslash}p{0.68cm}>{\raggedright\arraybackslash}p{1.25cm}@{}}
\toprule
\textbf{Use case} & \textbf{3GPP scope} & \textbf{rApp} & \textbf{xApp} & \textbf{Node} & \textbf{Interfaces} \\
\midrule
ANR & Neighbor KPIs & \textopenbullet & \textbullet & \textopenbullet & O1, E2 \\
Mobility robustness & Handover targets & \textopenbullet & \textbullet & -- & O1, A1, E2 \\
Load balancing & Load thresholds & \textopenbullet & \textbullet & -- & A1, E2 \\
RACH opt. & Enable \& targets & \textopenbullet & \textbullet & -- & O1, E2 \\
PCI assign. & PCI conflict lists & \textbullet & \textopenbullet & -- & O1, E2 \\
Cov.\ \& cap. & Cell targets/KPIs & \textopenbullet & \textbullet & -- & O1, A1, E2 \\
Energy saving & Cell/RF sleep & \textbullet & \textbullet & -- & O1, A1, E2 \\
Interference & RRM parameters & -- & \textbullet & -- & E2 \\
Traffic steering & Load-balancing policy & -- & \textbullet & -- & A1, E2 \\
\bottomrule
\end{tabular}
\par\smallskip
\begin{minipage}{\columnwidth}\raggedright
ANR: automatic neighbor relation. KPI: key performance indicator. RACH:
random access channel. PCI: physical cell identity.
\end{minipage}
\end{table}

Slot-level control poses a further question for third-party algorithms. A round trip
between the O-DU and an external controller can take longer than the time available
for a scheduling decision. Some algorithms also require physical-layer data, such as
in-phase and quadrature (I/Q) samples, that the Near-RT RIC cannot obtain through E2
\cite{ngrg2024.10}. These constraints motivated the proposal for
dApps that run alongside the relevant O-DU or O-CU functions
\cite{doro22dapps,lacava25dapps}. The O-RAN ALLIANCE's nGRG subsequently studied dApp
requirements \cite{ngrg2024.10} and proposed E3 as an interface for access to local
measurements and control functions \cite{ngrg2025.05}. This placement can shorten the
control loop, provided the application completes within the radio deadline using the
node's available compute resources. Section~\ref{sec:6g-theses} examines
what happens when a result arrives too late, and Section~\ref{sec:rac-concept}
describes how applications on different hosts could interact while the node
retains responsibility for local execution.

\subsection{Model Lifecycle and Control Interfaces}
\label{sec:lifecycle}

\input{figs/fig_timeline}

Consider a radio unit using massive multiple-input multiple-output (MIMO).
The operator aims to save energy while maintaining service quality.
Figure~\ref{fig:archmap}
(steps~1--6) illustrates how a trained model can guide RF channel
reconfiguration \cite{oran25.eswp,oran.ts104226}. An rApp registers the model
through R1 \cite{oran.wg2.r1ap} and uses it to determine how
far the antenna array can be reduced while meeting the service requirements.
The Non-RT RIC conveys these limits to the Near-RT RIC as an A1 policy.
Table~\ref{tab:ifaces} summarizes the interfaces used in this workflow.

\begin{table}[!tb]
\centering
\caption{Interfaces for Model Management, Radio Control, and Monitoring}
\label{tab:ifaces}
\footnotesize
\setlength{\tabcolsep}{3pt}
\renewcommand{\arraystretch}{1.05}
\begin{tabular}{@{}l >{\raggedright\arraybackslash}p{2.9cm} p{3.9cm}@{}}
\toprule
\textbf{Interface} & \textbf{Endpoints} & \textbf{Role} \\
\midrule
R1 & rApp and \mbox{Non-RT RIC} framework & Access to model lifecycle services \\
A1 & \mbox{Non-RT RIC} and \mbox{Near-RT RIC} & Policy guidance and model management \\
E2 & \mbox{Near-RT RIC} and E2 nodes & Radio measurements and control \\
O1 & SMO and managed functions & Performance monitoring \\
Y1 & \mbox{Near-RT RIC} and authorized consumers & Read-only access to RAN analytics \cite{oran.wg3.y1gap} \\
\bottomrule
\end{tabular}
\end{table}

Within those limits, an xApp uses current radio measurements to select an
array configuration and requests the change over E2 \cite{oran.wg3.e2gap}.
The O-DU applies the change to the O-RU over Open Fronthaul. Energy and
service-quality measurements collected through O1 then allow the operator to
assess the savings and determine whether the model needs updating.
When the model, host, and radio node come from different suppliers, the
operator also needs to verify the execution environment and identify which
model version the node is using after an update or a rollback
\cite{ngrg2024.05,oran.security.aiml}.

\subsection{Cloud Infrastructure and Hardware Acceleration}
\label{sec:substrate}

When RAN and AI workloads share a server, the SMO allocates compute resources
through O2. Within the O-DU, the AAL defines how baseband software uses
those resources for hardware acceleration.

\subsubsection{Shared O-Cloud infrastructure}
The SMO uses O2 to provision O-Cloud resources for virtualized RAN functions
and controllers as traffic and compute demand change
\cite{oran.wg6.o2}. O-RAN's 2024 report on
Communication and Computing Integrated Networks (CCIN) described how radio
and computing workloads could share this infrastructure. Subsequent work
addressed edge application orchestration \cite{oran.wg1.ccin} and distributed
AI training and inference alongside radio processing \cite{oran26.specupdate}.

AI applications can also draw on measurements analyzed by the RAN. Through
RAN Analytics Information Exposure (RAIE), the Near-RT RIC provides these
analytics to authorized applications over Y1 to support RAN for AI services
\cite{oran.wg3.y1gap}.

\subsubsection{Hardware acceleration in the O-DU}
Some physical-layer computations must finish in less than a millisecond.
The AAL lets O-DU software invoke accelerated functions through a common
interface across host processor architectures. It presents ASICs, FPGAs,
DSPs, and GPUs as logical processing units (LPUs) and keeps hardware-specific
drivers separate from the baseband code. A companion interface makes the
available accelerator resources visible to O-Cloud orchestration through O2
(Fig.~\ref{fig:aal}(a)) \cite{oran.wg6.aalganp,oran.wg6.aalapi}.

Figure~\ref{fig:aal}(b) shows how the data path changes with the AAL profile.
\begin{itemize}
\item \textbf{Look-aside acceleration.} The host central processing unit (CPU) runs the physical-layer
pipeline. It sends an individual task, such as forward error correction (FEC), to
the accelerator and uses the result to continue processing \cite{oran.wg6.aalfec}.
\item \textbf{Inline acceleration.} The accelerator sits between the upper
protocol layers and Open Fronthaul. It executes the full High-PHY pipeline
without returning intermediate symbols to host memory \cite{oran.wg6.aalhiphy}.
O-RAN also defines a profile for multi-user MIMO beamforming
\cite{oran.wg6.aalmumimo}.
\end{itemize}

Moving an O-DU implementation to a different accelerator requires compatible
AAL profiles and APIs, as well as compatible memory layouts and real-time
transport queues. Section~\ref{sec:airan} examines these requirements in
reference implementations.

\subsection{RAN AI/ML in 3GPP and O-RAN}
\label{sec:milestones}

Figure~\ref{fig:timeline} traces the development of RAN AI/ML in 3GPP and
O-RAN beyond the early proposals in Table~\ref{tab:lineage}.

\begin{itemize}
\item \textbf{AI/ML in 3GPP radio and management specifications.}
3GPP studied AI/ML for the NR air interface and mobility
\cite{3gpp.38843,3gpp.38744}. Its NG-RAN specifications define
information exchange procedures that support several AI/ML use cases,
including network energy saving \cite{3gpp.ts38300}. Its management
specifications also address the AI/ML model lifecycle \cite{3gpp.ts28105}.

\item \textbf{RIC applications in O-RAN.}
RIC applications use the interfaces described in Section~\ref{sec:lifecycle}
to manage models and participate in radio control.
In the energy-saving use case in Section~\ref{sec:lifecycle}, an A1 policy
guides the Near-RT RIC's choice of RF channel configuration
\cite{oran.ts104226}.
\end{itemize}

%% file: figs/fig_loops.tex
\begin{figure}[!t]
\centering
\begin{tikzpicture}[fig/base,
  tier/.style={font=\FigSmall\bfseries,text=FigNavy,align=left,anchor=north west,inner sep=0pt},
  tmeta/.style={font=\fontsize{6.3}{7.2}\selectfont\rmfamily,text=FigLine,align=left,anchor=north west,inner sep=0pt}]
\path[use as bounding box] (0,0) rectangle (88,-73);

\draw[fig/inherited,<->,>={Latex[length=1.4mm,width=1mm]}] (1.8,-1.5) -- (1.8,-66);
\node[tier] at (3.6,-2.0) {$>1$\,s};
\node[tmeta] at (3.6,-5.8) {$10^2$--$10^4$ cells\\$\sim$kbps KPIs};
\node[tier] at (3.6,-22.5) {10\,ms--1\,s};
\node[tmeta] at (3.6,-26.3) {$1$--$10^2$ cells\\$\sim$Mbps KPMs};
\node[tier,text=FigAmber] at (3.6,-44) {$<10$\,ms};
\node[tmeta,text=FigAmber] at (3.6,-47.8) {1 cell / beam\\$\sim$Gbps I/Q};
\node[font=\fontsize{6.6}{7.5}\selectfont\rmfamily,text=FigLine] at (44,-70.5)
  {Tighter latency $\leftrightarrow$ finer state granularity \& smaller spatial span.};
\draw[fig/guide] (0.5,-68.0) -- (87.5,-68.0);

\node[fig/smo/group,minimum width=68.5mm,minimum height=64.5mm] (out) at (53.2,-33.75) {};
\node[fig/heading,anchor=north west,font=\FigSmall\bfseries] at (20.2,-2.6) {SMO / Non-RT RIC (Training \& policy)};
\node[fig/app,minimum width=13.2mm,minimum height=6.2mm,font=\FigSmall] (rapp) at (28.5,-10.8) {\textbf{rApp}};
\node[fig/node,fill=white,minimum width=28.0mm,minimum height=6.2mm,font=\FigSmall] (o1store) at (61.2,-10.8)
  {O1 KPIs \& model store};
\draw[fig/bidir] (rapp.east) -- node[fig/label,fill=FigBlueFill,above=.25mm,font=\fontsize{6.4}{7.3}\selectfont\rmfamily] {R1} (o1store.west);

\node[fig/group,draw=FigHostLine,fill=white,line width=.65pt,minimum width=58.5mm,minimum height=43mm] (mid) at (51.25,-42.8) {};
\draw[fig/arrow] (28.5,-13.9) -- node[fig/label,fill=FigBlueFill,right=.5mm,font=\fontsize{6.4}{7.3}\selectfont\rmfamily] {A1 / A1-ML} (28.5,-21.3);
\node[fig/heading,anchor=north west,font=\FigSmall\bfseries] at (23.4,-22.5) {Near-RT RIC (Multi-cell / slice RRM)};
\node[fig/app,minimum width=13.2mm,minimum height=6.2mm,font=\FigSmall] (xapp) at (31.2,-30.5) {\textbf{xApp}};
\node[fig/node,fill=FigBlueFill!55!white,minimum width=26.5mm,minimum height=6.2mm,font=\FigSmall] (kpm) at (61.8,-30.5)
  {Cell / UE KPM state};
\draw[fig/bidir] (xapp.east) -- node[fig/label,above=.25mm,font=\fontsize{6.3}{7.2}\selectfont\rmfamily] {APIs} (kpm.west);

\node[fig/prop/group,minimum width=51.5mm,minimum height=21mm] (inn) at (50.75,-51.8) {};
\draw[fig/bidir] (31.2,-33.6) -- node[fig/label,right=.5mm,font=\fontsize{6.4}{7.3}\selectfont\rmfamily] {E2 ctrl / ind.} (31.2,-41.3);
\node[fig/heading,text=FigAmber,anchor=north west,font=\FigSmall\bfseries] at (26.4,-42.5) {O-DU / O-CU Stack (Per-TTI $<10$\,ms)};
\node[fig/prop,fill=white,minimum width=13.0mm,minimum height=6.5mm,font=\FigSmall] (dapp) at (34.0,-53.5) {\textbf{dApp}};
\node[fig/node,fill=white,minimum width=24.5mm,minimum height=6.5mm,font=\FigSmall] (sched) at (61.5,-53.5)
  {I/Q \& MAC scheduler};
\draw[fig/bidir,draw=FigAmber] (dapp.east) -- node[fig/label,fill=FigAmberFill,above=.25mm,font=\fontsize{6.4}{7.3}\selectfont\rmfamily] {E3} (sched.west);

\draw[fig/management,-{Latex[length=1.4mm,width=1mm]}]
  (sched.east) -- (84.8,-53.5) -- (84.8,-10.8) -- (o1store.east);
\node[fig/label,fill=FigBlueFill,rotate=90,inner sep=.45mm,font=\fontsize{6.3}{7.2}\selectfont\rmfamily] at (84.8,-32.5) {O1 drift \& KPIs};
\end{tikzpicture}
\caption{RAN control loops at three timescales. The Non-RT RIC uses
longer-term key performance indicators (KPIs) across cells. The Near-RT RIC
uses recent cell and user equipment (UE) key performance measurements (KPMs)
for control on the order of 10\,ms to 1\,s.
The proposed dApp/E3 path would give in-node applications access to local
measurements for sub-10\,ms operation.}
\label{fig:loops}
\end{figure}
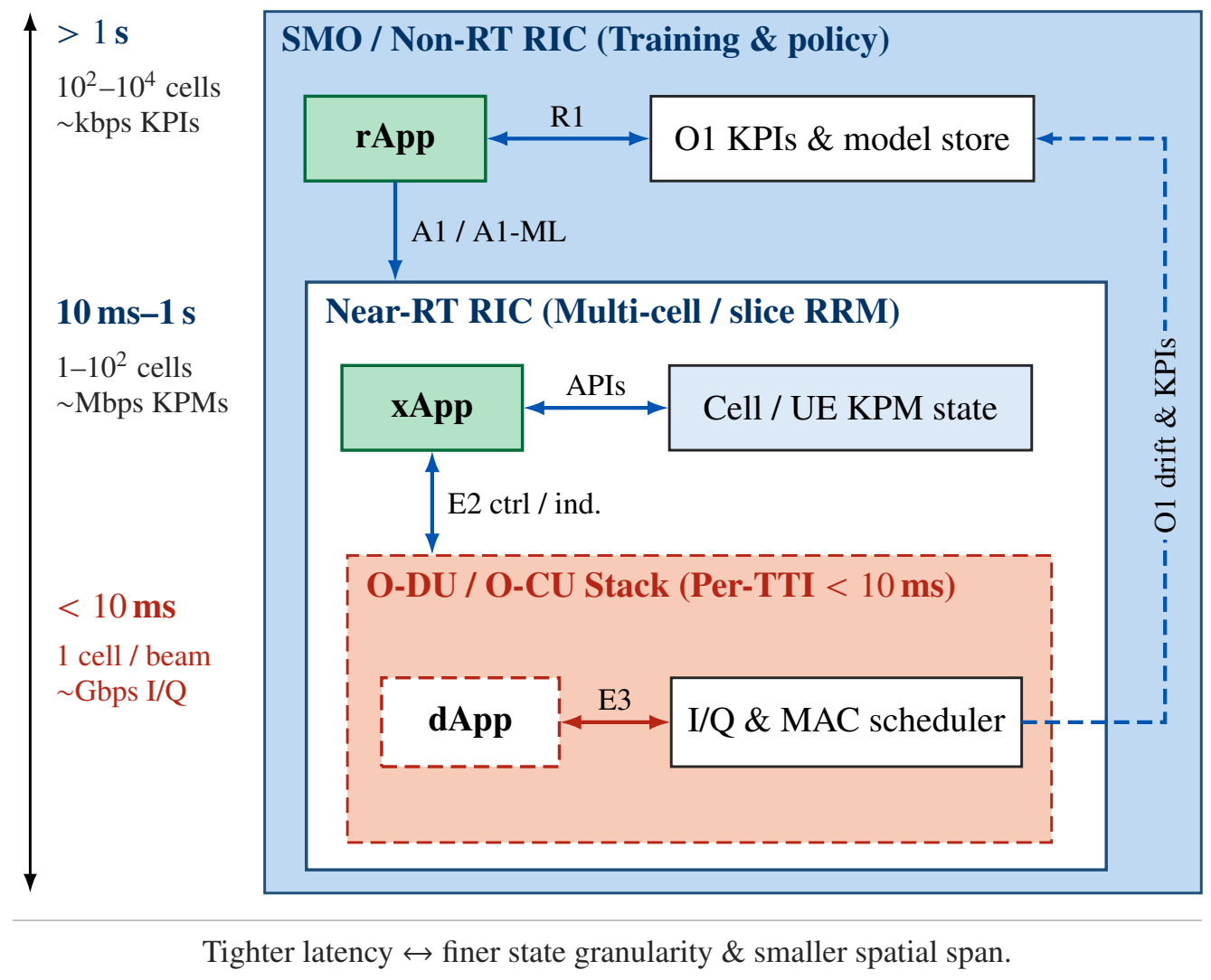

%% file: figs/fig_aal.tex
\begin{figure*}[!t]
\centering
\begin{tikzpicture}[
  fig/base,
  every node/.style={align=center,inner sep=1.05mm,outer sep=0pt},
  aal/data/.style={fig/arrow,draw=FigInk,line width=.75pt},
  aal/offload/.style={fig/bidir,draw=FigTeal,line width=.8pt},
  aal/label/.style={fig/label,font=\FigSmall},
  stage/.style={fig/node,minimum width=12.6mm,minimum height=8.2mm,font=\fontsize{6.8}{7.8}\selectfont\rmfamily,inner sep=.65mm}
]
\path[use as bounding box] (0,2) rectangle (178,76.5);
\draw[fig/guide] (90,3) -- (90,75);
\node[fig/title,anchor=base west] at (1,73.2)
  {(a) AAL multi-architecture LPU abstraction};
\node[fig/title,anchor=base west] at (93,73.2)
  {(b) High-PHY stage partition: look-aside vs.\ inline};

\node[fig/node,minimum width=38.5mm,minimum height=9.2mm] (aapphost) at (21.5,61)
  {\textbf{O-DU Software}\\(RLC / MAC / PHY)};
\node[fig/cloud,minimum width=38.5mm,minimum height=9.2mm] (amgmthost) at (66.5,61)
  {\textbf{O-Cloud}\\Orchestration};
\draw[fig/host/group] (2,32.5) rectangle (86.5,44.5);
\draw[fig/interface,line width=.65pt] (48.5,32.5) -- (48.5,44.5);
\node[font=\FigGroup,text=FigNavy] at (25.2,38.5) {Acceleration\\Abstraction Layer (AAL)};
\node[font=\FigGroup,text=FigNavy] at (67.5,38.5) {Accelerator\\Manager};
\draw[fig/bidir] (aapphost.south) -- node[aal/label,right=.5mm]{AAL API} (21.5,44.5);
\draw[fig/bidir,fig/management] (amgmthost.south) -- node[aal/label,left=.5mm]{O2 / mgmt} (66.5,44.5);

\node[fig/cloud,minimum width=84.5mm,minimum height=7.2mm] (alpu) at (44.25,24.2)
  {\textbf{Logical Processing Units (LPUs)}};
\draw[fig/interface] (25.2,32.5) -- (25.2,27.8);
\draw[fig/management] (66.5,32.5) -- (66.5,27.8);

\draw[fig/interface] (44.25,20.6) -- (44.25,16.2);
\draw[fig/interface] (16,16.2) -- (74.5,16.2);
\node[fig/node,minimum width=27.5mm,minimum height=9.2mm,font=\FigSmall] (hw1) at (16,8.5)
  {\textbf{ASIC / DSP / GPU}\\Deployed digital};
\node[fig/app,minimum width=27mm,minimum height=9.2mm,font=\FigSmall] (hw2) at (46,8.5)
  {\textbf{FPGA / RVV}\\Reconfigurable vector};
\node[fig/prop,minimum width=23mm,minimum height=9.2mm,font=\FigSmall] (hw3) at (74.5,8.5)
  {\textbf{Analog IMC}\\Matrix engines};
\foreach \x in {16,46,74.5} \draw[fig/interface] (\x,16.2) -- (\x,13.1);

\node[fig/heading,anchor=base west] at (93,65.5)
  {Look-aside offload (host CPU runs PHY chain)};
\node[stage] (lmac) at (99.8,55) {RLC /\\MAC};
\node[stage,draw=FigHostLine,fill=FigBlueFill] (lcod) at (118.2,55) {\textbf{Coding}\\(LDPC)};
\node[stage] (lmod) at (136.5,55) {Mod.\ /\\MIMO};
\node[stage] (lifft) at (153.5,55) {RE /\\iFFT};
\node[stage,fill=FigGreyFill,minimum width=12.2mm] (lfh) at (170.2,55) {Open\\FH 7.2x};
\draw[aal/data] (lmac.east) -- (lcod.west);
\draw[aal/data] (lcod.east) -- (lmod.west);
\draw[aal/data] (lmod.east) -- (lifft.west);
\draw[aal/data] (lifft.east) -- (lfh.west);

\node[fig/app,minimum width=31mm,minimum height=7.6mm,font=\FigSmall] (afec) at (118.2,39.8)
  {\textbf{FEC / Kernel Accel.}};
\draw[aal/offload] (lcod.south) -- node[aal/label,right=.7mm,font=\fontsize{6.6}{7.5}\selectfont\rmfamily]
  {Task offload and result return} (afec.north);

\draw[fig/guide] (93,33.2) -- (177,33.2);

\node[fig/heading,anchor=base west] at (93,27.2)
  {Inline High-PHY acceleration};
\node[stage,minimum width=12mm] (imac) at (99.2,14.8) {RLC /\\MAC};

\node[fig/app,minimum width=44.4mm,minimum height=14.2mm] (iaccel) at (139,14.8) {};
\node[font=\fontsize{6.6}{7.4}\selectfont\rmfamily\bfseries,text=FigTeal,anchor=north] at (139,21.6)
  {Inline High-PHY Accelerator};
\node[stage,fill=white,minimum width=12mm,minimum height=7mm] (icod) at (124.5,12.8) {Coding};
\node[stage,fill=white,minimum width=12mm,minimum height=7mm] (imod) at (139,12.8) {MIMO};
\node[stage,fill=white,minimum width=12mm,minimum height=7mm] (iifft) at (153.5,12.8) {iFFT};
\draw[aal/data] (icod.east) -- (imod.west);
\draw[aal/data] (imod.east) -- (iifft.west);

\node[stage,fill=FigGreyFill,minimum width=12.2mm] (ifh) at (170.2,14.8) {Open\\FH 7.2x};

\draw[draw=FigAmber,dashed,line width=.75pt] (110.5,4.2) -- (110.5,13.6);
\draw[draw=FigAmber,dashed,line width=.75pt] (110.5,19.4) -- (110.5,23.5);
\draw[aal/data] (imac.east) -- node[fig/label,above=.75mm,font=\fontsize{6.4}{7.2}\selectfont\rmfamily,inner sep=.4mm] {FAPI} (iaccel.west);
\draw[aal/data] (iaccel.east) -- (ifh.west);
\node[font=\fontsize{6.5}{7.4}\selectfont\rmfamily,text=FigAmber,anchor=west] at (111.8,4.8)
  {Host--accelerator split (no return to host memory)};
\end{tikzpicture}
\caption{O-RAN AAL and baseband profiles. (a)~Logical processing units (LPUs) abstract fixed hardware, FPGAs, processors
with the RISC-V Vector extension (RVV), and analog in-memory computing (IMC).
(b)~Look-aside and inline High-PHY paths. LDPC: low-density parity-check.
FAPI: functional application platform interface.}
\label{fig:aal}
\end{figure*}
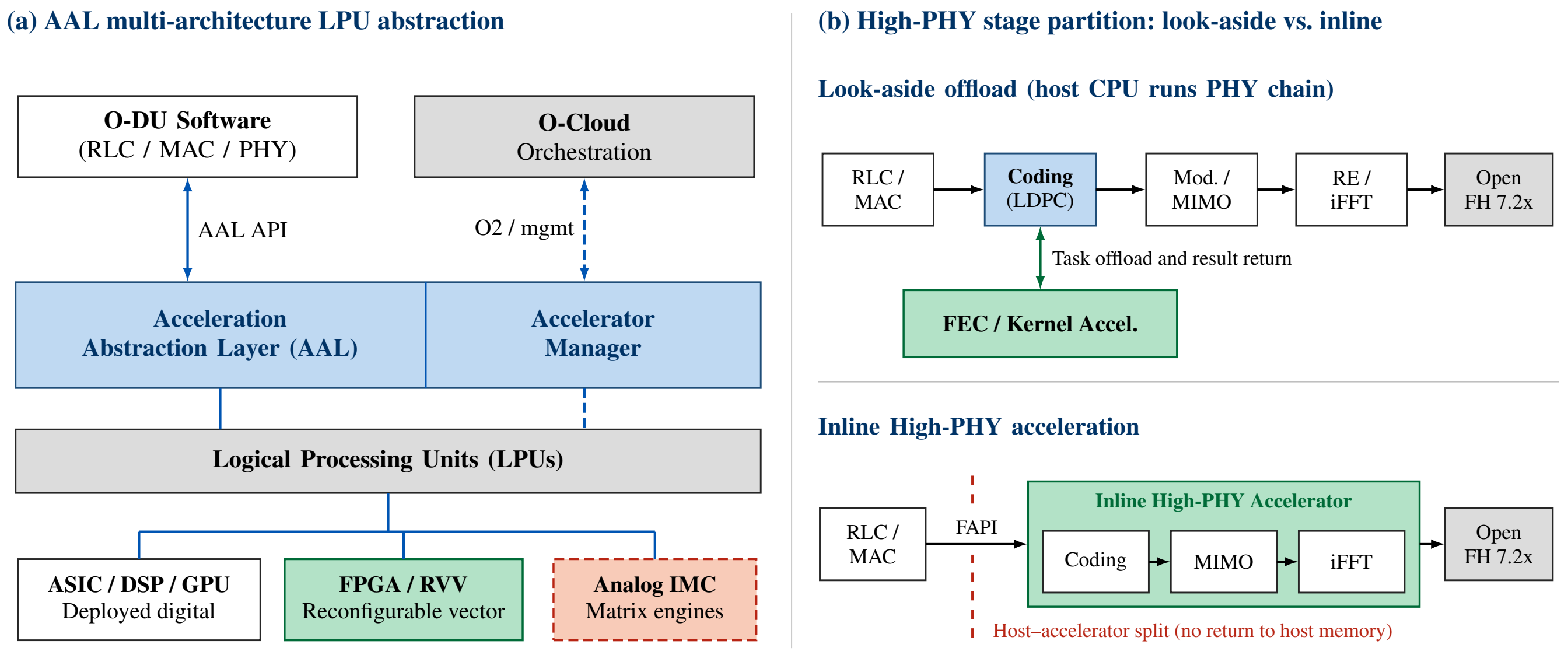

%% file: figs/fig_timeline.tex
\begin{figure*}[!t]
\centering
\begin{tikzpicture}[fig/base,y=.85mm,
  timeline/org/.style={anchor=east,align=right,
    inner sep=0pt,outer sep=0pt},
  timeline/entry/.style={anchor=north west,align=left,text width=44.5mm,
    inner sep=0pt,outer sep=0pt,
    execute at begin node={\hyphenpenalty=10000\exhyphenpenalty=10000\relax}},
  timeline/row/.style={fig/heading,anchor=west,align=left},
  timeline/note/.style={fig/label,font=\fontsize{6.6}{7.4}\selectfont\rmfamily,fill=white,
    anchor=west,align=left,inner sep=.85pt},
  timeline/life/.style={line width=2.6pt,draw=FigLine!45,line cap=butt},
  timeline/olife/.style={timeline/life,draw=FigBlue!65},
  timeline/ilife/.style={timeline/life,draw=FigTeal!70}]
\path[use as bounding box] (0,2.5) rectangle (178,-168);
\node[fig/title] at (0,0) {(a) Alliances, study groups, and open-source communities (2014--2026+)};
\foreach \x/\year in {48/2014,68/2016,88/2018,108/2020,128/2022,148/2024,168/2026}
 { \draw[fig/guide,line cap=butt] (\x,-10)--(\x,-69);
   \node[fig/label,font=\FigFont,anchor=south,inner sep=0pt]
     at (\x,-9) {\year}; }
\draw[fig/inherited,draw=FigLine,line cap=butt,
  -{Latex[length=1.55mm,width=1.05mm]}] (47,-10)--(176,-10);

\node[timeline/org] at (44,-17.5) {C-RAN Alliance};
\node[timeline/org] at (44,-23.5) {xRAN Forum};
\node[timeline/org,font=\FigGroup,text=FigNavy] at (44,-29.5) {O-RAN ALLIANCE};
\node[timeline/org] at (44,-36.0) {O-RAN Software Community};
\node[timeline/org] at (44,-42.0) {next Generation Research Group};
\node[timeline/org] at (44,-48.5) {IEEE 1914 NGFI working group};
\node[timeline/org] at (44,-54.5) {ITU-T FG ML5G};
\node[timeline/org] at (44,-60.5) {Wireless AI Alliance};
\node[timeline/org] at (44,-66.0) {AI-RAN Alliance};

\draw[timeline/life,dashed] (48,-17.5)--(89.5,-17.5);
\node[timeline/note,anchor=south west] at (49.2,-16.2) {Joined Feb.\ 2018 merger};
\draw[timeline/life] (72.5,-23.5)--(89.5,-23.5);
\draw[timeline/olife] (89.5,-29.5)--(175,-29.5);
\draw[timeline/olife] (101,-36.0)--(171,-36.0);
\draw[draw=FigBlue,fill=white,line width=.65pt,line join=miter]
  (170.2,-36.8) rectangle (171.8,-35.2);
\node[timeline/note,anchor=south east] at (172,-34.6) {2026 move to Linux Foundation Networking};
\draw[timeline/olife] (132.5,-42.0)--(175,-42.0);
\draw[timeline/life] (69,-48.5)--(175,-48.5);
\node[timeline/note,anchor=south west] at (70.2,-47.1) {Project approved Feb.\ 2016};
\draw[timeline/life] (87,-54.5)--(113.5,-54.5);
\node[timeline/note] at (115.5,-54.5) {Concluded 2020};
\draw[timeline/life] (78,-60.5)--(98,-60.5);
\draw[fig/inherited,draw=FigLine,line width=.75pt,
  dash pattern=on 3pt off 1.6pt] (98,-60.5)--(104,-60.5);
\node[timeline/note] at (105.5,-60.5) {2017--2018 papers};
\draw[timeline/ilife] (149.5,-66.0)--(175,-66.0);

\foreach \x/\y in {72.5/-23.5,69/-48.5,87/-54.5,78/-60.5}
  \fill[FigLine] (\x,\y) circle (.65mm);
\foreach \x/\y in {101/-36.0,132.5/-42.0}
  \fill[FigBlue] (\x,\y) circle (.65mm);
\fill[FigTeal] (149.5,-66.0) circle (.65mm);
\foreach \y/\trackcolor in {-29.5/FigBlue,-42.0/FigBlue,-48.5/FigLine,-66.0/FigTeal}
  \draw[fig/arrow,draw=\trackcolor,line cap=butt] (171,\y)--(176,\y);
\draw[fig/inherited,draw=FigLine,line width=.6pt,line join=miter]
  (89.5,-17.5)--(95,-17.5)--(95,-23.5)--(89.5,-23.5);
\draw[fig/arrow,draw=FigLine] (95,-23.5)--(95,-28.3);
\node[timeline/note] at (97.5,-20.5) {Merged Feb.\ 2018};

\node[fig/title] at (0,-75) {(b) O-RAN, 3GPP, and related milestones (2019--2026+)};
\fill[FigBlueFill] (0,-80) rectangle (178,-88);
\draw[fig/inherited,line width=.6pt] (0,-80)--(178,-80);
\draw[fig/inherited,line width=.6pt] (0,-88)--(178,-88);

\node[fig/heading,text=FigNavy] at (3,-84) {Track};
\node[fig/heading,text=FigNavy] at (34,-84) {2019--2021 (Normative v1)};
\node[fig/heading,text=FigNavy] at (83,-84) {2022--2024 (Lifecycle \& AAL)};
\node[fig/heading,text=FigNavy] at (132,-84) {2025--2026+ (Converged AI \& 6G)};
\foreach \y in {-100,-114,-127,-140,-153}
 \draw[fig/guide,line cap=butt] (0,\y)--(178,\y);
\draw[fig/inherited,line width=.6pt] (0,-166)--(178,-166);
\foreach \x in {31,80,129}
 \draw[fig/guide,line cap=butt] (\x,-80)--(\x,-166);

\node[timeline/row] at (3,-94) {O-RAN\\architecture};
\node[timeline/entry] at (34,-90) {\textbf{2020} Architecture v1; O1 \& O2\\\textbf{2020--21} First OTICs \& PlugFests};
\node[timeline/entry] at (83,-90) {\textbf{2022} nGRG launched (6G studies)\\\textbf{2024} Decoupled SMO; CCIN v1};
\node[timeline/entry] at (132,-90) {\textbf{2025} Joint 3GPP 6G workshop\\\textbf{2026} 21 OTICs; in-node dApp/E3};

\node[timeline/row] at (3,-107) {O-RAN\\AI/ML};
\node[timeline/entry] at (34,-102) {\textbf{2019} First AI/ML workflow report\\\textbf{2020--21} A1 policy \& E2SM v1};
\node[timeline/entry] at (83,-102) {\textbf{2022} Normative AI/ML \& R1\\\textbf{2023--24} RAN analytics exposure; Y1};
\node[timeline/entry] at (132,-102) {\textbf{2025} R1/RIC APIs; cross-RIC A1-ML\\\textbf{2026} Cross-tier AI workflow enh.};

\node[timeline/row] at (3,-120.5) {O-Cloud\\\& AAL};
\node[timeline/entry] at (34,-116) {\textbf{2020} O-Cloud \& AAL principles\\\textbf{2021} Look-aside FEC profile};
\node[timeline/entry] at (83,-116) {\textbf{2022} Inline High-PHY profile\\\textbf{2023--24} Common AAL API; CCIN};
\node[timeline/entry] at (132,-116) {\textbf{2025--26} MU-MIMO BF; CCIN v2\\\textbf{2026} O-Cloud AI enhancement};

\node[timeline/row] at (3,-133.5) {3GPP};
\node[timeline/entry] at (34,-129) {\textbf{2019} TR 37.816 RAN data study\\\textbf{2021} Rel-17 RAN intelligence};
\node[timeline/entry] at (83,-129) {\textbf{2022--24} TR 37.817 (NG-RAN AI);\\TR 38.843 (NR AI air interface)};
\node[timeline/entry] at (132,-129) {\textbf{2025--26} TS 28.105; TR 38.744;\\Rel-20 6G TR 38.914 / 38.960};

\node[timeline/row] at (3,-146.5) {ITU \& AI\\frameworks};
\node[timeline/entry] at (34,-142) {\textbf{2019} ITU-T ML5G; Y.3172\\(50\,\textmu s--10\,ms RAN loop classes)};
\node[timeline/entry] at (83,-142) {\textbf{2023} ITU-R M.2160 (IMT-2030\\native AI, sensing \& compute)};
\node[timeline/entry] at (132,-142) {\textbf{2024--26} AI-RAN Alliance founded;\\architecture \& 3-pillar reports};

\node[timeline/row] at (3,-159.5) {IEEE \&\\open stacks};
\node[timeline/entry] at (34,-155) {\textbf{2019} OSC Amber; FlexRIC/OAI\\\textbf{2020} IEEE 1914.1 packet FH};
\node[timeline/entry] at (83,-155) {\textbf{2022--24} OpenRAN Gym, X5G,\\split-7.2x OAI \& srsRAN};
\node[timeline/entry] at (132,-155) {\textbf{2025--26} Duranta; OCUDU;\\Sionna Research Kit \& testbeds};

\end{tikzpicture}
\caption{Milestones for (a)~alliances and open-source organizations
(2014--2026+) and (b)~O-RAN, 3GPP Releases~17--20, ITU, AI-RAN, and
open-source stacks (2019--2026+, extending Table~\ref{tab:lineage})
\cite{oran22.release002,oran26.specupdate}.}
\label{fig:timeline}
\end{figure*}

%% file: sec4_industry.tex
\section{Industrial Deployment, Testing, and Operator Economics}
\label{sec:industry}

Deploying Open RAN brings together radio units, baseband software, and cloud
infrastructure from multiple suppliers. Operators must integrate these
components and meet radio performance requirements on the available hardware.

\begin{enumerate}
\item \textbf{Testing and certification.}
Open Testing and Integration Centers (OTICs) test products against interface specifications and exercise
multi-vendor configurations before deployment
(Section~\ref{sec:otic-testing}).

\item \textbf{Operator economics and shared compute.}
Operator experience with modular deployment and shared computing illustrates
the potential benefits of pooling resources for RAN and AI workloads
(Section~\ref{sec:operator-economics}).

\item \textbf{Regional adoption and standards coordination.}
Regional standards bodies publish transpositions and establish cooperation
agreements. 3GPP and the O-RAN ALLIANCE coordinate their 6G studies
(Section~\ref{sec:6g-coordination}).
\end{enumerate}

\subsection{OTICs and Their Role in 6G Research}
\label{sec:otic-testing}

OTICs provide vendor-independent facilities for conformance, interoperability,
and end-to-end testing of O-RAN products \cite{oran.otic}.
For example, an OTIC can pair an O-DU and an O-RU from different suppliers
to test Open Fronthaul interoperability. OTICs also host PlugFests at which
participants integrate and exercise RIC applications for energy saving and
massive-MIMO optimization
\cite{oran25.mwcsummit,oran25.plugfestfall,oran26.plugfestspring}.

We distinguish two types of OTIC by institutional role. \emph{Professional
testing houses}, including centers supported by operators or national programs,
focus on testing products against adopted specifications and qualifying
multi-vendor configurations for deployment. \emph{Academic-led OTICs} combine
this testing capability with university research infrastructure, including
open-source radio stacks, over-the-air (OTA) testbeds, and heterogeneous computing
platforms \cite{oran.otic,ukgov22.sonic,powder.oran}.

Academic-led OTICs can play a central role in 6G pre-standardization by
testing a candidate mechanism while its requirements are still open to
revision. For example, centers can run the same radio-control workload on
different accelerator platforms to determine when a controller misses a
deadline and how the radio node responds. Comparing the results across
sites can help a working group specify the timing requirement and fallback
behavior. Once the specification is adopted, professional testing houses
can use those requirements to assess interoperability and repeat the tests
after software or model updates (Sections~\ref{sec:academia} and~\ref{sec:6g}).

\subsection{Operator Economics and Shared Compute}
\label{sec:operator-economics}

Early adopters targeted hardware consolidation \cite{rakuten20.launch}, modular
procurement \cite{mou24.priorities}, and controller-driven energy savings
\cite{oran25.mwcsummit}. Examples of deployment include DISH's greenfield
5G buildout and Vodafone's introduction of Open RAN into an existing network
\cite{dish23.buildout,vodafone22.bath}. OREX SAI focuses on integrating
multi-vendor equipment, while European operators identify accelerator
portability as a technical priority \cite{docomo24.orexsai,mou24.priorities}.
Trials of software-defined massive MIMO and shared RAN/AI infrastructure
are discussed in Section~\ref{sec:platform-benchmarking}.

Shared-server trials explore whether operators can improve utilization by
running communication and AI services on shared infrastructure.
O-Cloud and the AAL provide the resource and hardware abstractions
described in Section~\ref{sec:substrate}.
Sharing resources among baseband processing, AI for RAN control, and co-hosted
enterprise AI workloads also requires scheduling that preserves radio
deadlines (Sections~\ref{sec:airan} and~\ref{sec:6g}). Cross-supplier
deployments depend on software portability across certified hardware
platforms. The economic case for consolidation therefore depends on how
pooling affects utilization and cost for the operator's workload mix, at the
required radio performance and with the intended supplier mix.

\subsection{Regional Adoption and Coordination of 6G Standards}
\label{sec:6g-coordination}

\subsubsection{Regional adoption of O-RAN specifications}
ETSI and ATIS have published regional transpositions of O-RAN specifications
\cite{etsi.ts103859,oran26.atisdone}. Agreements with TTA and
TSDSI provide further channels for regional adoption
\cite{oran24.tta,oran21.etsitsdsi}. Public R\&D programs in North America,
Europe, and the United Kingdom support this work through open testbeds
\cite{ntia.pwscif.r1,bmvi21.openran,ukgov22.sonic}.

\subsubsection{The joint 3GPP--O-RAN workshop and 6G coordination}
The 2025 joint 3GPP--O-RAN workshop laid the foundation for coordinating
their 6G work. Both organizations subsequently endorsed the division of
responsibilities through liaison statements, providing a basis for aligning
their 6G research and standardization schedules \cite{oran25.6gworkshop}.

%% file: sec5_academia.tex
\section{Open-Source Ecosystems and Academic 6G Open Labs}
\label{sec:academia}

Open-source communities and academic testbeds support 6G research and
pre-standardization. Before the
introduction of the open hosting model described in Section~\ref{sec:beginning}
(Fig.~\ref{fig:socket}(b)), academic machine-learning research on cellular
networks was largely confined to link- or system-level simulations because
commercial base stations exposed neither real-time telemetry nor programmable
control hooks. By standardizing controller APIs and fronthaul splits, O-RAN
enabled university prototypes to inform pre-standardization studies, including
research on the sub-10\,ms control interval identified in
Section~\ref{sec:beginning}.

\begin{enumerate}
\item \textbf{Open-source software.}
Open controllers and protocol stacks let researchers implement and test RAN
intelligence on programmable systems (Section~\ref{sec:open-stacks}).

\item \textbf{Closed-loop experiments.}
Open toolchains and shared testbeds support repeatable evaluation of RAN
control under repeatable radio conditions
(Section~\ref{sec:experimental-testbeds}).

\item \textbf{Contributions to pre-standardization.}
The development of dApps illustrates how university research can inform
nGRG studies (Section~\ref{sec:prestandardization-research}).
\end{enumerate}

\subsection{Open-Source Controllers and Protocol Stacks}
\label{sec:open-stacks}

Open controller interfaces allow researchers to validate traffic steering, RAN
slicing, and inference logic on softwarized stacks before committing to
licensed spectrum. The O-RAN ALLIANCE and the Linux Foundation jointly
established the OSC, which moved under Linux Foundation Networking (LFN) in 2026
\cite{lfn26.oscmigration}. The OSC provides reference Non-RT/Near-RT RICs and
E2/A1/O1 components \cite{osc.releases}, complemented by
lightweight research controllers such as FlexRIC and POWDER's NexRAN
\cite{schmidt21flexric,johnson21nexran}.

Independently governed open-source projects provide implementations of RAN
nodes. OAI integrates OSC's \texttt{xran} library for Option~7.2x Open
Fronthaul \cite{oai.fhi72} and participates in LFN's Duranta project
\cite{lfn25.duranta}. The srsRAN stack
supports split 7.2 and E2 \cite{srsran.release}. The OCUDU Ecosystem
Foundation was launched in 2026 to advance open DU software
\cite{ocudu26}. The ability to combine these open-source stacks with
different processors, accelerators, and radio hardware gives research
institutes and universities greater autonomy in developing and validating
RAN functions. Researchers can modify a control algorithm in the stack they
select and compare its behavior on different platforms under the same radio
conditions.

Northeastern University uses Colosseum for repeatable hardware-in-the-loop (HIL)
experiments and X5G for end-to-end experiments on a
multi-vendor 5G network \cite{bonati21colosseum,villa25x5g}.
The Singapore University of Technology and Design and the Japan
OTIC obtained consistent results in Open Fronthaul conformance tests on
the same O-RU model at both sites \cite{nguyen24crosslab}.
OpenLab at the University of Texas at Dallas tests interoperability
among O-RUs, O-DUs, and O-CUs from different suppliers
\cite{oran.otic.utdallas}.

These practices provide a starting point for the proposed 6G open lab
initiative, which is gaining support from research institutes and
universities worldwide. Its proponents hope to carry out this work under
the governance of the O-RAN ALLIANCE's nGRG and within its research scope.
In response to the initiative, Tongji University plans to build an end-to-end
radio-frequency research testbed based on open-source RAN software.
Its hardware will include x86, Arm, and RISC-V
processors, as well as GPUs, FPGAs, and Intel virtualized RAN (vRAN) Boost accelerators.
The initiative envisages a shared experimental framework through which
participating labs could run the same RAN control experiments on different
hardware platforms and compare results before
standardization (Section~\ref{sec:6g}).

\subsection{Closed-Loop Testbeds and Experimental Toolchains}
\label{sec:experimental-testbeds}

Credible closed-loop evaluation requires five elements: an executable RAN
stack, a programmable RIC, real-time telemetry, a pipeline to onboard trained
policies or models, and repeatable propagation conditions.
Table~\ref{tab:research} summarizes how major open-source stacks, toolchains,
and shared test facilities fulfill these roles
\cite{powder.oran,bonati23openrangym,schmidt21flexric,lacava23nsoran,cammerer25sionnark,ukgov22.sonic,i14y.lab}.

\begin{table}[!t]
\centering
\caption{Open-Source Stacks, Toolchains, and Testbeds for O-RAN AI/ML}
\label{tab:research}
\footnotesize
\setlength{\tabcolsep}{2pt}
\renewcommand{\arraystretch}{1.05}
\begin{tabular}{@{}>{\raggedright\arraybackslash}p{2.05cm}>{\raggedright\arraybackslash}p{1.15cm}>{\raggedright\arraybackslash}p{1.65cm}>{\raggedright\arraybackslash}p{2.7cm}@{}}
\toprule
\textbf{Resource} & \textbf{Role} & \textbf{Interfaces} & \textbf{Primary capability} \\
\midrule
OSC & RIC stack & R1, A1, E2, O1 & Reference Non-/Near-RT RICs \\
OpenAirInterface & RAN stack & F1, 7.2x, E2 & 5G CU/DU with \texttt{xran} fronthaul \\
srsRAN & RAN stack & F1, 7.2x, E2 & Modular 5G CU/DU stack \\
Duranta / OCUDU & DU projects & 7.2x, E2, AAL & Open 5G/6G DU software \\
FlexRIC & RIC toolkit & E2 & Lightweight research RIC \\
ns-O-RAN & Simulator & E2 & Scale-out ns-3 + live E2 \\
Sionna / Research Kit & PHY--AI & Neural PHY & GPU-based differentiable PHY \\
PAWR / POWDER & OTA OTIC & 7.2x, E2, O1 & City-scale wireless testbeds \\
Colosseum & HIL emulator & E2, O1 & Repeatable RF emulation \\
SCOPE / Gym / ColO & Toolchain & E2 & Learning experiments and datasets \\
SONIC / i14y Lab & Cert. OTIC & 7.2x, O1, E2 & End-to-end testing \& badging \\
\bottomrule
\end{tabular}
\end{table}

Shared experimental platforms support the academic open labs discussed in
Section~\ref{sec:industry} by combining repeatable tests with OTA measurements. Examples include city-scale
testbeds from the U.S.\ Platforms for Advanced Wireless Research (PAWR) program
\cite{pawr.program,powder.oran} and European test facilities such as
SONIC Labs and i14y Lab \cite{ukgov22.sonic,i14y.lab}. OpenRAN Gym combines
SCOPE for data collection and control with ColO-RAN for deploying xApps
in repeatable closed-loop experiments \cite{bonati23openrangym}.
The tools in Table~\ref{tab:research} and multi-supplier platforms such as X5G
(Section~\ref{sec:airan}) let researchers combine telemetry and model
deployment with controlled RF conditions. Reproducing
those experiments at independent sites requires compatible adapters and
comparable test procedures, motivating the federation in Section~\ref{sec:6g}.

\subsection{Research Activity and Contributions to 6G Pre-Standardization}
\label{sec:prestandardization-research}

Prior work has examined how Open RAN could shape the 6G cellular architecture
\cite{polese24empowering}.

To channel academic research into future specifications, O-RAN established the
nGRG in June 2022 to coordinate pre-standardization 6G studies and transfer
findings to normative working groups \cite{oran22.ngrg,chen25oranrole}. The
development of dApps illustrates this process. Researchers proposed in-node
CU/DU inference in 2022 \cite{doro22dapps}. nGRG studied use cases and
requirements in 2024, and researchers tested a full-stack implementation
in 2025. In 2026, nGRG reported on the architecture and the proposed E3
interface \cite{ngrg2024.10,lacava25dapps,ngrg2025.05}. While
nGRG reports identify architectural gaps and candidate mechanisms, normative
standardization remains the responsibility of O-RAN working groups.

\input{figs/fig_coverage}

The path from a university prototype to an nGRG interface study shows how
academic research can shape a candidate mechanism. Section~\ref{sec:6g}
proposes how independent sites could verify the behavior of such a mechanism
and report comparable results to the groups drafting normative specifications.

%% file: figs/fig_coverage.tex
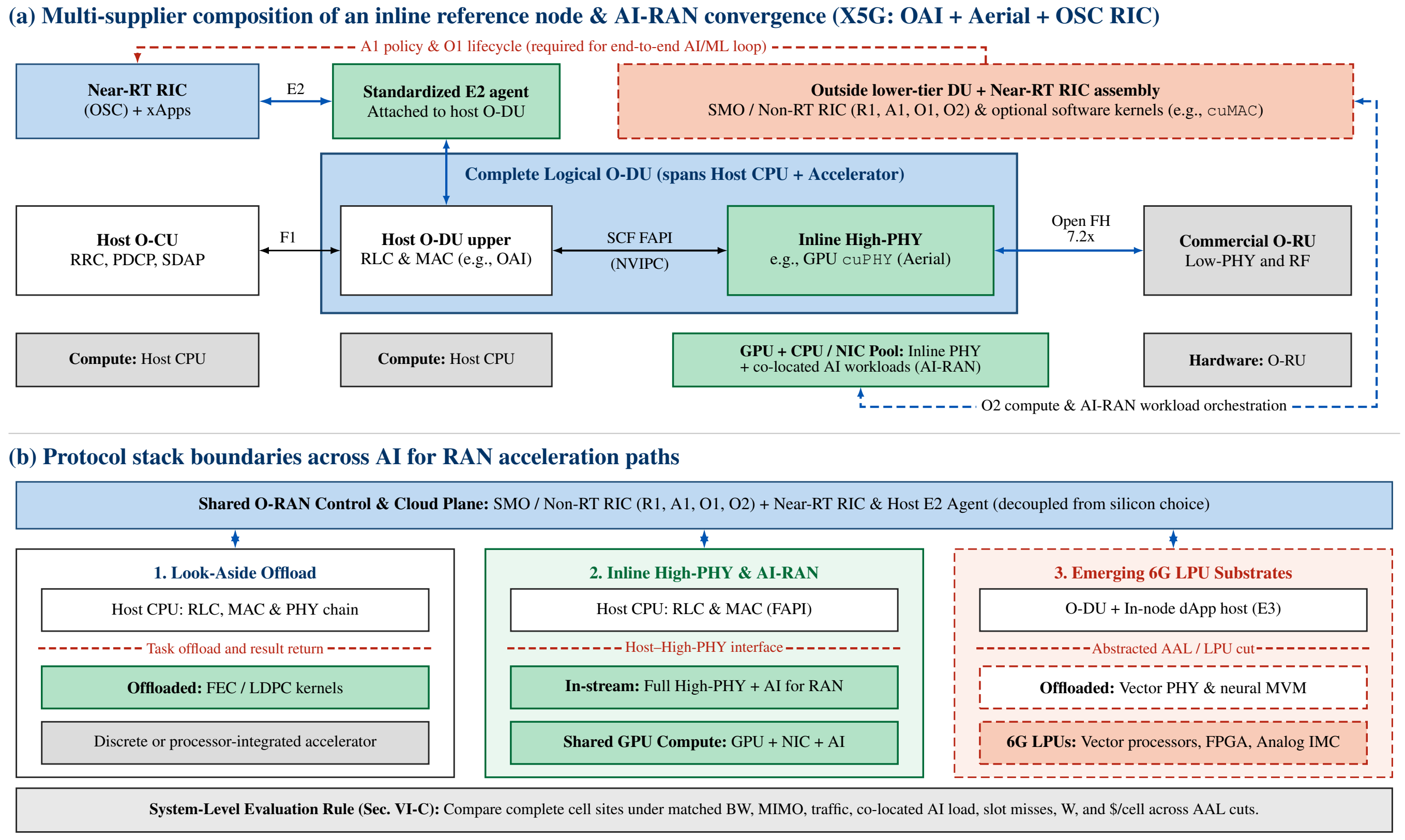
\begin{figure*}[!t]
\centering
\begin{tikzpicture}[fig/base,
  f62/.style={font=\fontsize{6.2}{7.1}\selectfont\rmfamily},
  f58/.style={font=\fontsize{5.8}{6.6}\selectfont\rmfamily},
  hostlayer/.style={fig/node,fill=white,minimum width=49.5mm,minimum height=5.4mm,inner sep=.45mm,font=\fontsize{6.0}{6.8}\selectfont\rmfamily},
  applayer/.style={fig/app,minimum width=49.5mm,minimum height=5.4mm,inner sep=.45mm,font=\fontsize{6.0}{6.8}\selectfont\rmfamily},
  proplayer/.style={fig/prop,minimum width=49.5mm,minimum height=5.4mm,inner sep=.45mm,font=\fontsize{6.0}{6.8}\selectfont\rmfamily},
  sublayer/.style={fig/node,fill=FigGreyFill,minimum width=49.5mm,minimum height=5.4mm,inner sep=.45mm,font=\fontsize{5.9}{6.7}\selectfont\rmfamily},
  cutline/.style={draw=FigAmber,line width=.8pt,dash pattern=on 2.6pt off 1.3pt}]
\path[use as bounding box] (0,12.2) rectangle (178,-93.0);

\node[fig/title,anchor=west] at (0,11.2)
  {(a) Multi-supplier composition of an inline reference node \& AI-RAN convergence (X5G: OAI + Aerial + OSC RIC)};

\node[fig/host,minimum width=31mm,minimum height=9.5mm,f62] (ric) at (16.5,0.5)
  {\textbf{Near-RT RIC}\\(OSC) + xApps};
\node[fig/app,minimum width=29mm,minimum height=9.5mm,f62] (agent) at (56,0.5)
  {\textbf{Standardized E2 agent}\\Attached to host O-DU};
\node[fig/prop,minimum width=94mm,minimum height=9.5mm,f62] (oob) at (125,0.5)
  {\textbf{Outside lower-tier DU + Near-RT RIC assembly}\\SMO / Non-RT RIC (R1, A1, O1, O2) \& optional software kernels (e.g., \texttt{cuMAC})};

\draw[fig/bidir] (ric.east) -- node[fig/label,above=.3mm,f58]{E2} (agent.west);
\draw[fig/management,draw=FigAmber,-{Latex[length=1.4mm,width=1mm]}]
  (oob.north) -- (125,7.4) -- (16.5,7.4) -- (ric.north);
\node[fig/label,text=FigAmber,f58] at (71,7.4) {A1 policy \& O1 lifecycle (required for end-to-end AI/ML loop)};

\draw[fig/host/group,line width=.8pt] (40,-6.2) rectangle (129,-26.6);
\node[fig/label,fill=FigBlueFill,font=\fontsize{6.6}{7.4}\selectfont\rmfamily\bfseries,text=FigNavy]
  at (86.5,-9.0) {Complete Logical O-DU (spans Host CPU + Accelerator)};

\node[fig/node,minimum width=31mm,minimum height=11.4mm,f62] (cu) at (16.5,-18.6)
  {\textbf{Host O-CU}\\RRC, PDCP, SDAP};
\node[fig/node,minimum width=27mm,minimum height=11.4mm,f62] (mac) at (56,-18.6)
  {\textbf{Host O-DU upper}\\RLC \& MAC (e.g., OAI)};
\node[fig/app,minimum width=34mm,minimum height=11.4mm,f62] (phy) at (109,-18.6)
  {\textbf{Inline High-PHY}\\e.g., GPU \texttt{cuPHY} (Aerial)};
\node[fig/node,fill=FigGreyFill,minimum width=26.5mm,minimum height=11.4mm,f62] (ru) at (158.5,-18.6)
  {\textbf{Commercial O-RU}\\Low-PHY and RF};

\draw[fig/inherited,<->,>={Latex[length=1.3mm,width=.95mm]}]
  (cu.east) -- (mac.west);
\node[fig/label,above=.45mm,f58,inner sep=.4mm] at (35.8,-18.4) {F1};

\draw[fig/inherited,line width=.7pt,<->,>={Latex[length=1.3mm,width=.95mm]}]
  (mac.east) -- node[fig/label,fill=FigBlueFill,above=.3mm,f58]{SCF FAPI}
  node[fig/label,fill=FigBlueFill,below=.3mm,f58]{(NVIPC)} (phy.west);

\draw[fig/interface,<->,>={Latex[length=1.3mm,width=.95mm]}]
  (phy.east) -- (ru.west);
\node[fig/label,above=.45mm,align=center,f58,inner sep=.5mm] at (137.2,-18.4) {Open FH\\[-1pt]7.2x};

\draw[fig/interface,<->,>={Latex[length=1.3mm,width=.95mm]}] (agent.south) -- (mac.north);

\node[fig/cloud,minimum width=31mm,minimum height=6.8mm,f58] (c_cu) at (16.5,-32.6)
  {\textbf{Compute:} Host CPU};
\node[fig/cloud,minimum width=27mm,minimum height=6.8mm,f58] (c_mac) at (56,-32.6)
  {\textbf{Compute:} Host CPU};
\node[fig/app,minimum width=48mm,minimum height=6.8mm,f58] (c_phy) at (109,-32.6)
  {\textbf{GPU + CPU / NIC Pool:} Inline PHY\\[-1pt]+ co-located AI workloads (AI-RAN)};
\node[fig/cloud,minimum width=26.5mm,minimum height=6.8mm,f58] (c_ru) at (158.5,-32.6)
  {\textbf{Hardware:} O-RU};

\draw[fig/management,<->,>={Latex[length=1.3mm,width=.95mm]}]
  (oob.east) -- (175,0.5) -- (175,-38.6) -- (109,-38.6) -- (c_phy.south);
\node[fig/label,f58,fill=white] at (144,-38.6) {O2 compute \& AI-RAN workload orchestration};

\draw[fig/guide] (0,-42.0) -- (178,-42.0);
\node[fig/title,anchor=west] at (0,-45.2)
  {(b) Protocol stack boundaries across AI for RAN acceleration paths};

\node[fig/host,minimum width=176mm,minimum height=6.0mm,f62] (uplane) at (89,-51.2)
  {\textbf{Shared O-RAN Control \& Cloud Plane:} SMO / Non-RT RIC (R1, A1, O1, O2) + Near-RT RIC \& Host E2 Agent (decoupled from silicon choice)};

\node[fig/group,fill=white,minimum width=56mm,minimum height=29.2mm] (c1) at (29,-71.4) {};
\node[font=\fontsize{6.3}{7.1}\selectfont\rmfamily\bfseries,text=FigNavy,anchor=north] at (29,-57.8)
  {1.\ Look-Aside Offload};
\node[hostlayer] at (29,-64.6) {Host CPU: RLC, MAC \& PHY chain};
\draw[cutline] (3.8,-69.5) -- (54.2,-69.5);
\node[font=\fontsize{5.5}{6.1}\selectfont\rmfamily,text=FigAmber,fill=white,inner sep=.6pt] at (29,-69.5)
  {Task offload and result return};
\node[applayer] at (29,-74.5) {\textbf{Offloaded:} FEC / LDPC kernels};
\node[sublayer] at (29,-81.6) {Discrete or processor-integrated accelerator};

\node[fig/group,draw=FigTeal,fill=FigTealFill!28!white,line width=.65pt,minimum width=56mm,minimum height=29.2mm] (c2) at (89,-71.4) {};
\node[font=\fontsize{6.3}{7.1}\selectfont\rmfamily\bfseries,text=FigTeal,anchor=north] at (89,-57.8)
  {2.\ Inline High-PHY \& AI-RAN};
\node[hostlayer] at (89,-64.6) {Host CPU: RLC \& MAC (FAPI)};
\draw[cutline] (63.8,-69.5) -- (114.2,-69.5);
\node[font=\fontsize{5.5}{6.1}\selectfont\rmfamily,text=FigAmber,fill=FigTealFill!28!white,inner sep=.6pt] at (89,-69.5)
  {Host--High-PHY interface};
\node[applayer] at (89,-74.5) {\textbf{In-stream:} Full High-PHY + AI for RAN};
\node[applayer,fill=FigTealFill] at (89,-81.6) {\textbf{Shared GPU Compute:} GPU + NIC + AI};

\node[fig/prop/group,fill=FigAmberFill!28!white,minimum width=56mm,minimum height=29.2mm] (c3) at (149,-71.4) {};
\node[font=\fontsize{6.3}{7.1}\selectfont\rmfamily\bfseries,text=FigAmber,anchor=north] at (149,-57.8)
  {3.\ Emerging 6G LPU Substrates};
\node[hostlayer] at (149,-64.6) {O-DU + In-node dApp host (E3)};
\draw[cutline] (123.8,-69.5) -- (174.2,-69.5);
\node[font=\fontsize{5.5}{6.1}\selectfont\rmfamily,text=FigAmber,fill=FigAmberFill!28!white,inner sep=.6pt] at (149,-69.5)
  {Abstracted AAL / LPU cut};
\node[proplayer,fill=white] at (149,-74.5) {\textbf{Offloaded:} Vector PHY \& neural MVM};
\node[proplayer,fill=FigAmberFill] at (149,-81.6) {\textbf{6G LPUs:} Vector processors, FPGA, Analog IMC};

\foreach \x in {29,89,149} \draw[fig/bidir] (\x,-54.2) -- (\x,-56.8);

\node[fig/node,fill=FigGreyFill!65!white,minimum width=176mm,minimum height=5.6mm,f58] (bench) at (89,-90.2)
  {\textbf{System-Level Evaluation Rule (Sec.~VI-C):} Compare complete cell sites under matched BW, MIMO, traffic, co-located AI load, slot misses, W, and \$/cell across AAL cuts.};
\end{tikzpicture}
\caption{AI for RAN reference implementations of O-RAN.
(a)~Multi-supplier inline-accelerated composition (X5G with OAI, Aerial, and
OSC RIC \cite{villa25x5g}) and a co-located AI-RAN compute pool orchestrated
via SMO. (b)~Accelerator placement for look-aside offload, inline GPU
processing, and prospective 6G LPUs based on RVV, FPGA, or analog IMC.
NIC: network interface card. NVIPC: NVIDIA inter-process communication.}
\label{fig:coverage}
\end{figure*}

%% file: sec6_airan.tex
\section{AI for RAN Reference Implementations of O-RAN}
\label{sec:airan}

The early O-RAN designs reviewed in Section~\ref{sec:beginning} placed
intelligence in a programmable RAN controller. The reference implementations
examined here show how the controller connects to a radio protocol stack
and where host software hands baseband processing to an accelerator. In
general usage, \emph{AI for RAN} refers to applying AI/ML to
radio control and physical-layer operation, whereas \emph{AI-RAN} specifically
designates the AI-RAN Alliance's architecture and implementation program uniting
AI for RAN, shared RAN/AI computing, and connectivity for AI workloads
\cite{airan26.architecture}.
O-RAN provides hierarchical RICs for radio control, O-Cloud and CCIN for
shared computing infrastructure, and RAIE over Y1 for read-only analytics
\cite{oran18whitepaper,oran.wg1.ccin,oran.wg3.y1gap,oran26.specupdate}.
Analytics exposure can inform differentiated connectivity for AI workloads.
Providing that connectivity also involves mechanisms such as uplink
optimization and network slicing \cite{airan26.wg3}. The reference integrations
examined here use selected O-RAN interfaces.

\begin{enumerate}
\item \textbf{Hardware acceleration boundaries.}
Look-aside and inline profiles place the boundary between host software and
accelerators at different points in the PHY pipeline
(Section~\ref{sec:acceleration-boundaries}).

\item \textbf{Multi-supplier integration.}
The X5G example shows how the host protocol stack, inline High-PHY accelerator,
and RIC connect, and which interfaces depend on system integration
(Section~\ref{sec:reference-integration}).

\item \textbf{Shared RAN and AI computing.}
Shared platforms run both workloads on the same infrastructure. Comparing
them requires system-level tests under matched operating conditions
(Section~\ref{sec:platform-benchmarking}).
\end{enumerate}

\subsection{Baseband Acceleration Paths and Hardware Implementations}
\label{sec:acceleration-boundaries}

Open RAN implementations combine protocol software with different types
of hardware accelerator. The AAL presents
ASICs, FPGAs, DSPs, and GPUs to the O-DU through a common abstraction
\cite{oran.wg6.aalganp}.

\subsubsection{Placement in the PHY data path}
In the look-aside profile, the host offloads a selected PHY function,
such as FEC, and uses the result to continue the
processing chain \cite{oran.wg6.aalfec}. In the inline High-PHY profile,
the accelerator executes High-PHY directly in the data path between
the upper protocol layers and Open Fronthaul \cite{oran.wg6.aalhiphy}.
Figure~\ref{fig:coverage}(b) compares accelerator placement in both profiles.

\subsubsection{Instruction sets and reconfigurable logic}
Baseband and AI kernels repeatedly apply the same operation to many samples
or matrix elements. Vector and matrix instructions let software express that
parallel work directly, as in x86 AVX-512/AMX and Arm Neon/SVE. The RISC-V
Vector extension (RVV) allows implementations with different numbers of
parallel lanes to execute the same vector instructions \cite{perotti24ara2}.
An FPGA offers another way to tailor the hardware. Its logic can implement a
vector or application-specific pipeline for a particular baseband kernel.

\subsubsection{Memory traffic and in-memory computing}
Massive-MIMO processing and neural inference both use matrix--vector
multiplication. Moving the matrix coefficients from memory to a separate
compute unit consumes energy each time they must be fetched. Analog and
mixed-signal in-memory accelerators store the coefficients in a memory array
and perform multiply--accumulate operations there. This reduces coefficient
transfers. The achievable accuracy and energy savings depend on device
non-idealities and the cost of data conversion \cite{sebastian20memory}.

For future AAL implementations, the LPU abstraction
could expose functions implemented with programmable processors,
reconfigurable logic, or analog memory arrays to O-DU software. The AAL
profile specifies where the function sits in the PHY pipeline and how it
exchanges data with the host.
GPUs, FPGAs, and ASICs underpin current AAL implementations in 5G-Advanced
and remain strong candidates for 6G, while RVV-based processors and analog
in-memory computing (IMC) are potential longer-term paths for LPU design
to address energy-efficiency bottlenecks in 6G.

\subsection{Multi-Supplier System Composition: Host Stack, Inline PHY, and RIC}
\label{sec:reference-integration}

X5G combines the OAI protocol stack, NVIDIA Aerial, and
the O-RAN Software Community's Near-RT RIC in one testbed
(Fig.~\ref{fig:coverage}(a)) \cite{villa25x5g}. Three aspects of this
integration connect the acceleration path to radio control.

\begin{enumerate}
\item \textbf{Baseband processing.} OAI runs the O-CU and the O-DU's RLC
and MAC layers on the host CPU. Aerial provides GPU-accelerated High-PHY
processing. The host and accelerator thus implement different parts of the
same logical O-DU.

\item \textbf{Host and radio interfaces.} X5G connects the OAI MAC to
Aerial's physical layer through the Small Cell Forum's functional application
platform interface (FAPI). The accelerated
O-DU connects to commercial O-RUs through Option~7.2x Open Fronthaul.
These interfaces sit on opposite sides of the High-PHY processing chain.

\item \textbf{RIC control.} An E2 agent in the OAI stack connects the O-DU
to the Near-RT RIC. An xApp can use the measurements and control actions
exposed by that agent. In the energy-saving workflow of
Section~\ref{sec:lifecycle}, this connection supports the radio control
step. Model registration, policy guidance, and performance monitoring also
involve the Non-RT RIC and SMO.
\end{enumerate}

\subsection{Converged AI-RAN Platforms and System-Level Benchmarking}
\label{sec:platform-benchmarking}

\input{figs/fig_rac_shell}
\input{figs/fig_6g_federated}

Inline GPU acceleration provides one implementation path for co-locating
virtualized RAN and AI workloads on shared accelerated servers. Following
technical contributions on GPU inline acceleration, the AAL API, and High-PHY
profiles in O-RAN working groups beginning in February 2020
\cite{nvidia.oranarchive}, operators, suppliers, and research institutions
formed the AI-RAN Alliance in February 2024 to advance AI for RAN, shared
infrastructure for AI and RAN, and edge AI applications served by the RAN
\cite{airan24.founding}. Operators have since trialed massive-MIMO baseband
processing and shared infrastructure for RAN and AI inference
\cite{softbank24.aitras,softbank25.mmimo,ericsson26.portable}.

Because look-aside and inline architectures partition the baseband stack at
different boundaries, isolated kernel microbenchmarks cannot rank hardware
classes fairly. Meaningful evaluation must benchmark the complete cell-site
configuration under matched bandwidth, antenna count, traffic profile, active
PHY/MAC features, and co-located AI load, reporting throughput, slot-deadline
misses, tail latency, power consumption, and cost per cell across AAL profile
boundaries \cite{oran.wg6.aalganp} (Fig.~\ref{fig:coverage}(b)).
These benchmarks show how much AI workload a site can sustain within its
radio deadlines, power budget, and cost constraints.
Figure~\ref{fig:rac_shell} previews the RAC
proposed in Section~\ref{sec:rac-concept} to coordinate applications
across hosts.

%% file: figs/fig_rac_shell.tex
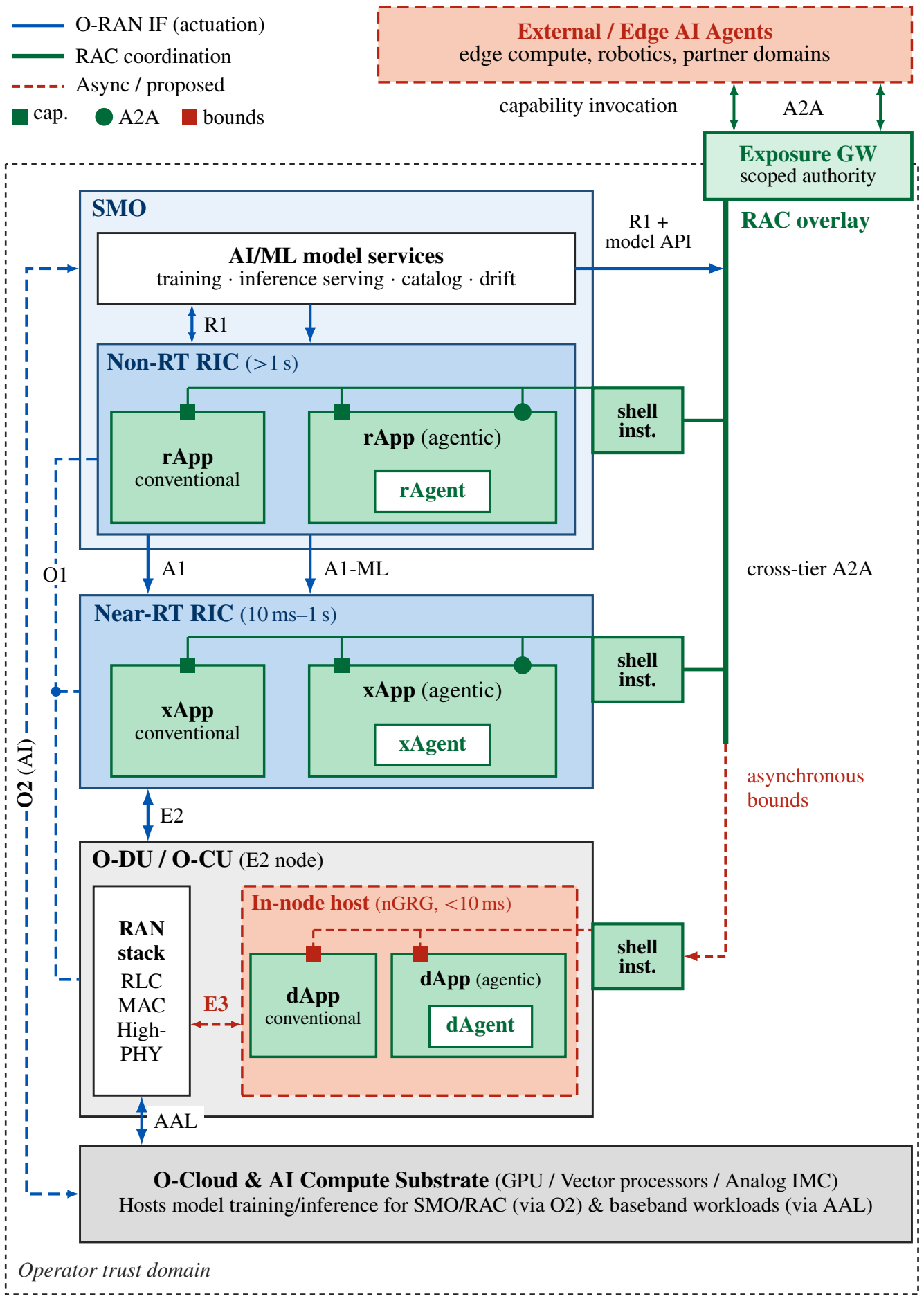
\begin{figure}[!t]
\centering
\begin{tikzpicture}[fig/base,
  f62/.style={font=\fontsize{6.2}{7}\selectfont\rmfamily},
  f58/.style={font=\fontsize{5.8}{6.6}\selectfont\rmfamily},
  hostbox/.style={fig/group,draw=FigHostLine,fill=FigBlueFill,line width=.6pt,inner sep=0pt,outer sep=0pt},
  htitle/.style={font=\fontsize{6.5}{7.4}\selectfont\rmfamily\bfseries,text=FigNavy,anchor=north west,inner sep=0pt},
  capp/.style={fig/app,minimum height=10.6mm,inner sep=.5mm,font=\fontsize{6.2}{7}\selectfont\rmfamily},
  agentbox/.style={fig/node,draw=FigTeal,fill=white,line width=.6pt,minimum width=11mm,minimum height=3.9mm,
    inner sep=0pt,font=\fontsize{6.0}{6.7}\selectfont\rmfamily\bfseries,text=FigTeal},
  capport/.style={fill=FigTeal,minimum size=1.4mm,inner sep=0pt,outer sep=0pt},
  aport/.style={circle,fill=FigTeal,minimum size=1.7mm,inner sep=0pt,outer sep=0pt},
  bport/.style={fill=FigAmber,minimum size=1.4mm,inner sep=0pt,outer sep=0pt},
  tab/.style={fill=FigTealFill,draw=FigTeal,line width=.8pt,minimum width=8.6mm,minimum height=6.2mm,
    inner sep=0pt,outer sep=0pt,align=center,font=\fontsize{5.6}{6.2}\selectfont\rmfamily\bfseries,text=FigInk},
  racline/.style={draw=FigTeal,line width=.6pt},
  spine/.style={draw=FigTeal,line width=1.5pt},
  async/.style={draw=FigAmber,line width=.75pt,dash pattern=on 2.2pt off 1.2pt}]
\path[use as bounding box] (0,2.0) rectangle (88,-121.5);

\draw[fig/interface] (1,0.4) -- (6,0.4);
\node[f58,anchor=west,inner sep=0pt] at (7,0.4) {O-RAN IF (actuation)};
\draw[spine,line width=1.2pt] (1,-2.5) -- (6,-2.5);
\node[f58,anchor=west,inner sep=0pt] at (7,-2.5) {RAC coordination};
\draw[async] (1,-5.4) -- (6,-5.4);
\node[f58,anchor=west,inner sep=0pt] at (7,-5.4) {Async / proposed};
\node[capport] at (1.8,-8.3) {};
\node[f58,anchor=west,inner sep=0pt] at (3,-8.3) {cap.};
\node[aport] at (9.8,-8.3) {};
\node[f58,anchor=west,inner sep=0pt] at (11.1,-8.3) {A2A};
\node[bport] at (18,-8.3) {};
\node[f58,anchor=west,inner sep=0pt] at (19.2,-8.3) {bounds};

\node[fig/prop,minimum width=51mm,minimum height=7.2mm,inner sep=.6mm,f62] (ext) at (61.5,-1.4)
  {\textbf{\textcolor{FigAmber}{External / Edge AI Agents}}\\[-1pt]edge compute, robotics, partner domains};

\draw[draw=FigLine,line width=.5pt,dash pattern=on 1.4pt off 1.2pt] (0.4,-12.8) rectangle (87.6,-120.8);

\node[fig/node,draw=FigTeal,line width=.9pt,fill=FigTealFill!55!white,minimum width=19.8mm,minimum height=6.4mm,
  inner sep=0pt,f62] (gw) at (77.1,-13.0)
  {\textbf{\textcolor{FigTeal}{Exposure GW}}\\[-1.5pt]{\fontsize{5.6}{6.2}\selectfont scoped authority}};
\draw[racline,line width=.75pt,<->,>={Latex[length=1.2mm,width=.85mm]}] (70,-5.0) -- (70,-9.8);
\draw[racline,line width=.75pt,<->,>={Latex[length=1.2mm,width=.85mm]}] (84,-5.0) -- (84,-9.8);
\node[f58,anchor=east,inner sep=0pt] at (64.7,-7.4) {capability invocation};
\node[f58,anchor=east,inner sep=0pt] at (78.7,-7.4) {A2A};

\draw[spine] (69.2,-16.2) -- (69.2,-68.2);
\node[font=\fontsize{6.4}{7}\selectfont\rmfamily\bfseries,text=FigTeal,anchor=north west,inner sep=0pt]
  at (70.6,-17.3) {RAC overlay};

\node[fig/group,draw=FigHostLine,fill=FigBlueFill!35!white,line width=.6pt,minimum width=49mm,
  minimum height=34.2mm,inner sep=0pt,outer sep=0pt] (smo) at (32.0,-32.5) {};
\node[htitle] at (8.6,-16.1) {SMO};
\node[fig/node,fill=white,minimum width=45.6mm,minimum height=6.8mm,inner sep=0pt,f62] (svc) at (32.0,-22.8)
  {\textbf{AI/ML model services}\\[-1pt]{\fontsize{5.7}{6.4}\selectfont training $\cdot$ inference serving $\cdot$ catalog $\cdot$ drift}};
\draw[fig/arrow] (svc.east) -- (69.2,-22.8);
\node[font=\fontsize{5.4}{6.0}\selectfont\rmfamily,align=center,fill=white,inner sep=.25mm]
  at (61.85,-19.2) {R1 +\\[-0.5pt]model API};

\node[hostbox,minimum width=45.6mm,minimum height=18.2mm] (nrt) at (32.0,-39.1) {};
\node[htitle] at (10.0,-30.8) {Non-RT RIC \textnormal{\fontsize{5.8}{6.6}\selectfont($>$1\,s)}};
\draw[fig/interface,<->,>={Latex[length=1.1mm,width=.8mm]}] (18.2,-26.2) -- (18.2,-30.0);
\node[f58,anchor=west,inner sep=0pt] at (19.3,-28.1) {R1};
\node[capp,minimum width=14.6mm] (rc) at (17.8,-41.8) {\textbf{rApp}\\[-1pt]{\fontsize{5.6}{6.4}\selectfont conventional}};
\node[capp,minimum width=23.6mm] (ra) at (41.2,-41.8) {};
\node[f62,anchor=north,inner sep=0pt] at (41.2,-38.0) {\textbf{rApp} (agentic)};
\node[agentbox] at (41.2,-44.1) {rAgent};

\draw[racline] (17.8,-34.2) -- (56.5,-34.2);
\foreach \x in {17.8,32.5,49.8} \draw[racline] (\x,-34.2) -- (\x,-36.5);
\node[capport] at (17.8,-36.5) {};
\node[capport] at (32.5,-36.5) {};
\node[aport] at (49.8,-36.5) {};

\node[tab,anchor=west] (tnrt) at (56.5,-37.3) {shell\\[-1pt]inst.};
\draw[racline,line width=.9pt] (tnrt.east) -- (69.2,-37.3);

\draw[fig/arrow] (14.0,-48.2) -- (14.0,-54.0);
\node[f58,anchor=west,fill=white,inner sep=.3mm] at (14.9,-51.4) {A1};
\draw[fig/arrow] (29.5,-26.2) -- (29.5,-30.0);
\draw[fig/arrow] (29.5,-48.2) -- (29.5,-54.0);
\node[f58,anchor=west,fill=white,inner sep=.3mm] at (30.4,-51.4) {A1-ML};

\node[f58,anchor=west,inner sep=0pt] at (71.2,-51.4) {cross-tier A2A};

\node[hostbox,minimum width=49mm,minimum height=18.4mm] (near) at (32.0,-63.2) {};
\node[htitle] at (8.8,-54.8) {Near-RT RIC \textnormal{\fontsize{5.8}{6.6}\selectfont(10\,ms--1\,s)}};
\node[capp,minimum width=14.6mm] (xc) at (17.8,-66.0) {\textbf{xApp}\\[-1pt]{\fontsize{5.6}{6.4}\selectfont conventional}};
\node[capp,minimum width=23.6mm] (xa) at (41.2,-66.0) {};
\node[f62,anchor=north,inner sep=0pt] at (41.2,-62.2) {\textbf{xApp} (agentic)};
\node[agentbox] at (41.2,-68.3) {xAgent};

\draw[racline] (17.8,-58.0) -- (56.5,-58.0);
\foreach \x in {17.8,32.5,49.8} \draw[racline] (\x,-58.0) -- (\x,-60.7);
\node[capport] at (17.8,-60.7) {};
\node[capport] at (32.5,-60.7) {};
\node[aport] at (49.8,-60.7) {};

\node[tab,anchor=west] (tnear) at (56.5,-61.1) {shell\\[-1pt]inst.};
\draw[racline,line width=.9pt] (tnear.east) -- (69.2,-61.1);

\draw[fig/bidir] (14.0,-72.4) -- (14.0,-77.6);
\node[f58,anchor=west,inner sep=0pt] at (15.1,-75.0) {E2};
\node[fig/group,fill=FigGreyFill!55!white,minimum width=49mm,minimum height=26.2mm,inner sep=0pt,outer sep=0pt]
  (odu) at (32.0,-90.7) {};
\node[font=\fontsize{6.5}{7.4}\selectfont\rmfamily\bfseries,anchor=north west,inner sep=0pt] at (8.6,-78.3)
  {O-DU / O-CU \textnormal{\fontsize{5.8}{6.6}\selectfont(E2 node)}};
\node[fig/node,fill=white,minimum width=9.2mm,minimum height=20.0mm,inner sep=0pt,f58] (stack) at (13.4,-91.8)
  {\textbf{RAN}\\\textbf{stack}\\[1pt]RLC\\MAC\\High-\\PHY};
\node[fig/prop/group,line width=.65pt,minimum width=32.0mm,minimum height=20.0mm,inner sep=0pt,outer sep=0pt]
  (inn) at (39.0,-91.8) {};
\node[font=\fontsize{6.0}{6.8}\selectfont\rmfamily\bfseries,text=FigAmber,anchor=north west,inner sep=0pt]
  at (23.8,-82.5) {In-node host \textnormal{\fontsize{5.4}{6.0}\selectfont(nGRG, $<$10\,ms)}};

\node[capp,minimum width=12.0mm,minimum height=9.8mm] (dc) at (29.8,-93.2) {\textbf{dApp}\\[-1pt]{\fontsize{5.1}{5.7}\selectfont conventional}};
\node[capp,minimum width=16.6mm,minimum height=9.8mm] (da) at (45.6,-93.2) {};
\node[font=\fontsize{5.8}{6.5}\selectfont\rmfamily,anchor=north,inner sep=0pt] at (45.8,-89.8) {\textbf{dApp} {\fontsize{5.0}{5.7}\selectfont(agentic)}};
\node[agentbox,minimum width=9.8mm] at (45.8,-95.2) {dAgent};

\draw[async,line width=.6pt] (29.8,-86.0) -- (56.5,-86.0);
\foreach \x in {29.8,40.0} \draw[async,line width=.6pt] (\x,-86.0) -- (\x,-88.3);
\node[bport] at (29.8,-88.3) {};
\node[bport] at (40.0,-88.3) {};

\node[tab,anchor=west] (tin) at (56.5,-88.5) {shell\\[-1pt]inst.};
\draw[async,<->,>={Latex[length=1.05mm,width=.75mm]}] (stack.east |- 18,-95.0) -- (inn.west |- 23,-95.0);
\node[f58,text=FigAmber,inner sep=0pt] at (20.5,-93.0) {\textbf{E3}};

\draw[async,-{Latex[length=1.4mm,width=.95mm]}] (69.2,-68.2) -- (69.2,-88.5) -- (tin.east);
\node[f58,text=FigAmber,anchor=north west,inner sep=0pt,align=left] at (71.2,-70.4)
  {asynchronous\\bounds};

\node[fig/cloud/group,minimum width=79.5mm,minimum height=9.2mm,inner sep=0pt,outer sep=0pt]
  (ocloud) at (47.25,-111.2) {};
\node[f62,align=center,inner sep=0pt] at (47.25,-111.2)
  {\textbf{O-Cloud \& AI Compute Substrate} {\fontsize{5.7}{6.4}\selectfont(GPU / Vector processors / Analog IMC)}\\[-0.5pt]
   {\fontsize{5.7}{6.4}\selectfont Hosts model training/inference for SMO/RAC (via O2) \& baseband workloads (via AAL)}};

\draw[fig/bidir] (13.4,-101.8) -- node[fig/label,right=.5mm,font=\fontsize{5.8}{6.5}\selectfont\rmfamily]{AAL} (13.4,-106.6);

\draw[fig/management] (9.2,-41.0) -- (5.2,-41.0) -- (5.2,-90.7) -- (7.5,-90.7);
\draw[fig/management] (5.2,-63.2) -- (7.5,-63.2);
\fill[FigBlue] (5.2,-63.2) circle (.48mm);
\node[fig/label,f58,inner sep=.3mm] at (5.2,-52.0) {O1};

\draw[fig/management,<->,>={Latex[length=1.2mm,width=.85mm]}]
  (7.5,-22.8) -- (2.4,-22.8) -- (2.4,-111.2) -- (ocloud.west);
\node[fig/label,f58,rotate=90,inner sep=.35mm] at (2.4,-71.0) {\textbf{O2} (AI)};

\node[font=\fontsize{5.8}{6.6}\selectfont\rmfamily\itshape,text=FigLine,anchor=west,inner sep=0pt]
  at (1.5,-118.6) {Operator trust domain};
\end{tikzpicture}
\caption{Proposed RAC architecture. Each host runs a local shell alongside its
applications. External agents access shells through the gateway at the operator
trust boundary. Agents embedded in applications exchange agent-to-agent (A2A)
messages. In-node hosts receive operating bounds asynchronously.}
\label{fig:rac_shell}
\end{figure}

%% file: figs/fig_6g_federated.tex
\begin{figure*}[!t]
\centering
\begin{tikzpicture}[fig/base,
  sbox/.style={fig/node,align=center,inner sep=1.2mm},
  rbox/.style={fig/app,align=center,inner sep=1.2mm},
  fbox/.style={fig/host,align=center,inner sep=1.2mm},
  pbox/.style={fig/prop,align=center,inner sep=1.2mm}]
\path[use as bounding box] (0,3.5) rectangle (178,-91.5);

\node[fig/smo/group,minimum width=176mm,minimum height=14.5mm] (sdo) at (89,-5.25) {};
\node[fig/heading] at (3,-0.6) {Standards \& Pre-Standardization Loop (O-RAN nGRG / Working Groups \& 3GPP Release 20+)};
\node[sbox,minimum width=52mm,minimum height=7.2mm] (cand) at (30,-7.5)
  {\textbf{Standard and research questions}\\Research reports, clauses, study items};
\node[sbox,minimum width=54mm,minimum height=7.2mm] (norm) at (90,-7.5)
  {\textbf{Candidate contributions}\\Information elements, profiles, tests};
\node[sbox,fill=FigGreyFill,minimum width=48mm,minimum height=7.2mm] (certspec) at (150,-7.5)
  {\textbf{Frozen normative releases}\\Post-release conformance specs};
\draw[ari] (norm.west) -- (cand.east);
\draw[fig/arrow] (norm.east) -- (certspec.west);

\node[fig/group,draw=FigTeal,fill=FigTealFill,line width=.7pt,
  minimum width=118mm,minimum height=32mm] (rec) at (60,-35.2) {};
\node[fig/heading,text=FigTeal] at (3,-21.6)
  {Proposed 6G open lab workflow (RIRDRT-RS)};

\node[sbox,minimum width=34mm,minimum height=10.8mm] (ri) at (21,-33.0)
  {\textbf{RI: Recursive Integration}\\Cross-site integration\\Semantics \& profiles};
\node[sbox,minimum width=35mm,minimum height=10.8mm] (rd) at (60,-33.0)
  {\textbf{RD: Recursive Development}\\Reference code \&\\lab workflows};
\node[sbox,minimum width=34mm,minimum height=10.8mm] (rt) at (99,-33.0)
  {\textbf{RT: Recursive Testing}\\Reference tests \&\\comparable results};
\draw[ari] (ri.east) -- (rd.west);
\draw[ari] (rd.east) -- (rt.west);
\draw[ari,dash pattern=on 2.5pt off 1.3pt] (rt.north) -- (99,-24.8) -- (21,-24.8) -- (ri.north);

\node[fig/node,fill=white,minimum width=112mm,minimum height=7.2mm,inner sep=1mm,font=\FigSmall] (pair) at (60,-45.4)
  {\textbf{AI harness:} Code revisions $\leftrightarrow$ programmable lab workflows $\leftrightarrow$ test results};

\node[fig/group,fill=FigGreyFill,minimum width=54mm,minimum height=32mm] (cat1) at (150,-35.2) {};
\node[fig/heading] at (125,-22.5) {Professional Testing Houses};
\node[font=\FigSmall,anchor=west,text=FigInk] at (125,-26.8)
  {\textbf{National \& operator OTICs}};
\node[font=\FigSmall,anchor=west,text=FigLine] at (125,-30.5)
  {Continuous integration and delivery};
\node[sbox,minimum width=48mm,minimum height=13.5mm] (c1role) at (150,-41.5)
  {6G intelligent-system testing,\\continuous certification, and\\multi-vendor assurance};

\draw[fig/arrow] (cand.south) -- node[fig/label,right=1mm] {Research question} (30,-19.2);
\draw[ari] (90,-19.2) -- node[fig/label,right=1mm] {Compared results} (norm.south);
\draw[fig/inherited,-{Latex[length=1.5mm,width=1mm]}]
  (141,-11.1) -- node[fig/label,left=.6mm,font=\fontsize{6.6}{7.4}\selectfont\rmfamily] {Specs} (141,-19.2);
\draw[fig/management,-{Latex[length=1.5mm,width=1mm]}]
  (159,-19.2) -- node[fig/label,right=.6mm,font=\fontsize{6.6}{7.4}\selectfont\rmfamily] {Assurance} (159,-11.1);

\node[fig/host/group,minimum width=176mm,minimum height=31mm] (fed) at (89,-74.2) {};
\node[fig/heading] at (3,-61.6)
  {Federated 6G Research Network (Academic-Led 6G Open Labs \& University Testbeds)};

\node[fig/node,fill=white,minimum width=170mm,minimum height=6.8mm] (gov) at (89,-68.3)
  {\textbf{Shared federation layer:} Capability catalogue $\cdot$ Code \& model registry $\cdot$ Scoped host authorization $\cdot$ Comparable run records};

\node[sbox,minimum width=39mm,minimum height=11.2mm] (tA) at (23.5,-81.5)
  {\textbf{Cloud simulation}\\Scale-out protocol \&\\RIC evaluation};
\node[sbox,minimum width=39mm,minimum height=11.2mm] (tB) at (67,-81.5)
  {\textbf{Digital twins}\\Site-specific RF \&\\workload replay};
\node[sbox,minimum width=40mm,minimum height=11.2mm] (tC) at (111,-81.5)
  {\textbf{HIL emulators}\\Repeatable channel \&\\accelerator benches};
\node[rbox,minimum width=40mm,minimum height=11.2mm] (tD) at (154.5,-81.5)
  {\textbf{Academic-Led Open Labs}\\Live OTA radios, open\\stacks \& multi-arch LPUs};

\draw[ari] (34,-51.2) -- node[fig/label,left=1.2mm] {Scoped actuation} (34,-58.7);
\draw[ari] (86,-58.7) -- node[fig/label,right=1.2mm] {Comparable run records \& failures} (86,-51.2);

\draw[draw=FigAmber,line width=.8pt,dash pattern=on 3pt off 1.5pt] (127,-55.0) -- (177,-55.0);
\node[fig/label,text=FigAmber,fill=white,font=\fontsize{6.2}{7.0}\selectfont\rmfamily] at (152,-55.0)
  {Mandate boundary (no badging overlap)};
\end{tikzpicture}
\caption{Recursive Integration, Development, and Testing for Recursive Standardization (RIRDRT-RS) in 6G. Federated open labs refine research code and lab workflows. Cross-site results inform candidate requirements, and revisions from standards review guide the next experiments. Professional Testing Houses provide continuous assurance after adoption.}
\label{fig:6g_federated}
\end{figure*}
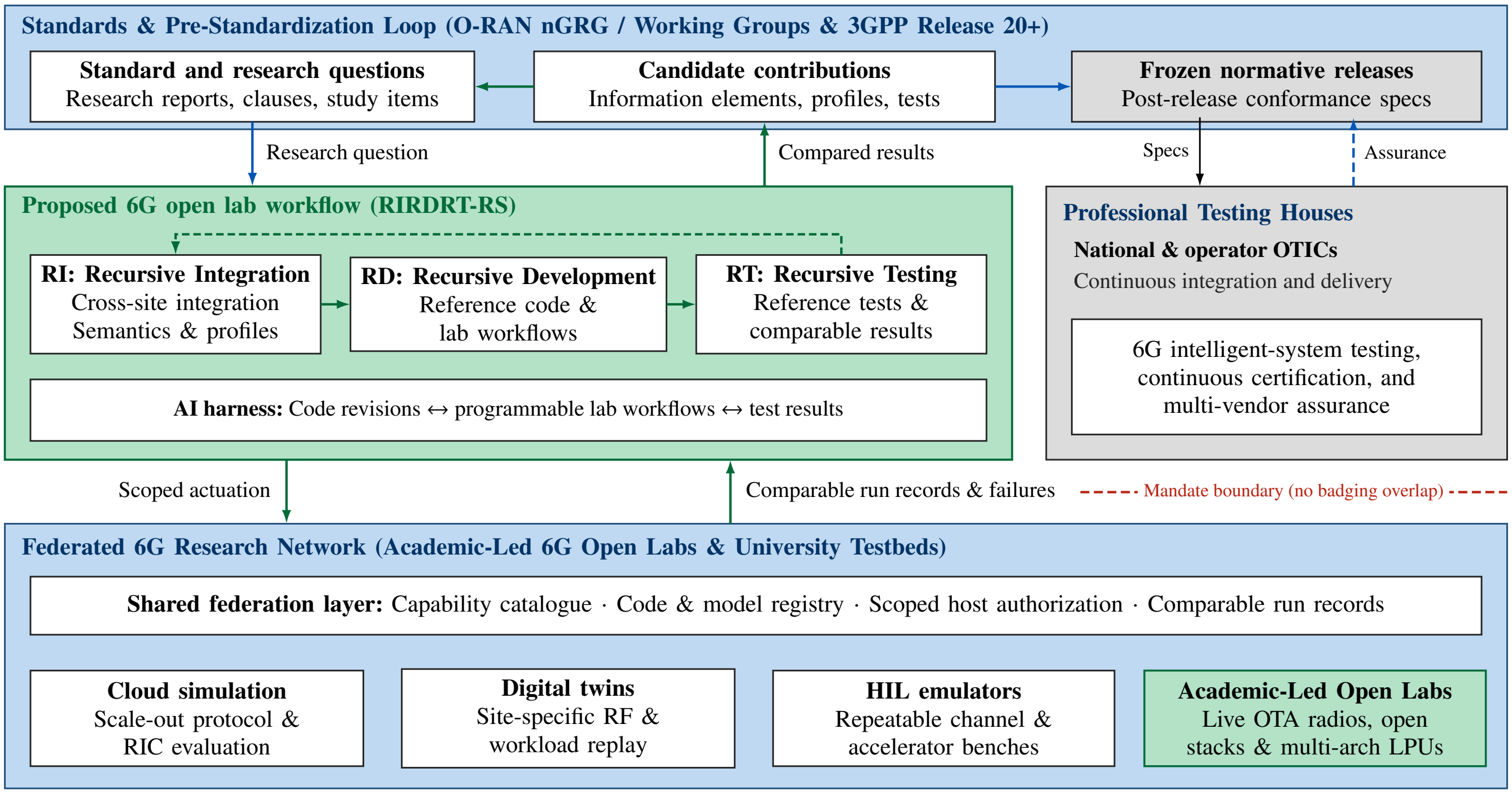

%% file: sec7_6g.tex
\section{Towards Open and Intelligent 6G RAN: Agent Coordination and Recursive Standardization}
\label{sec:6g}

Building on the industrial deployments, academic testbeds, and AI for RAN and
AI-RAN systems examined above, this section outlines how openness and
intelligence extend toward 6G. Next-generation frameworks such as IMT-2030
integrate AI, sensing, and distributed computing into mobile infrastructure
\cite{itu.m2160}.

\begin{enumerate}
\item \textbf{RAN control and edge inference.}
Section~\ref{sec:6g-theses} considers timing constraints inside radio nodes,
the resource demands of agent sessions, energy saving and
RAN/Edge---AI Silicon Co-design, and agent session mobility.

\item \textbf{RAN Agentic Controller architecture.}
The proposed RAC has a local software component
on each participating host and a shared gateway for external agents
(Section~\ref{sec:rac-concept}).

\item \textbf{Recursive standardization.}
Results from federated open labs could inform candidate 6G requirements.
The revised requirements and questions raised in standards review would
then guide the next round of development and experiments
(Section~\ref{sec:recursive-standardization}).
\end{enumerate}

\subsection{RAN Control and Edge Inference}
\label{sec:6g-theses}

An in-node dApp participating in radio scheduling
(Section~\ref{sec:academia}) must deliver its decision in time for the
slot it governs. An agent serving a mobile
user has a different execution pattern: successive model calls at the edge
share the history of an ongoing task. On the shared platforms reviewed in
Section~\ref{sec:airan}, those calls compete with baseband processing for
compute and memory. A handover can also change which site serves the next
call while the task is still in progress.

\subsubsection{\texorpdfstring{Real-Time Control Below the 10\,ms Boundary}{Real-Time Control Below the 10 ms Boundary}}

Standardized third-party application hosting currently reaches the Near-RT RIC,
whose operating range starts at 10\,ms. Vendor stacks already execute per-TTI
scheduling, link adaptation, and fast beam tracking inside radio nodes.
nGRG has studied dApps and proposed the E3 interface to give third-party
applications access to local measurements and control functions
\cite{ngrg2024.10,ngrg2025.05}. For such an application, a node needs to
bound execution time and isolate the real-time PHY/MAC threads. A concrete
design question is which action the scheduler takes when an application
delivers its result after the slot deadline. Independent labs could compare
how their implementations handle this case using the proposed workflow
in Fig.~\ref{fig:6g_federated}.

\subsubsection{Resource Profiles for AI Agent Sessions}

An edge-assisted agent may call a model several times to complete one task.
Each call uses uplink capacity to deliver its input, accelerator time for
inference, and downlink capacity to return its output. Session state retained
at the edge between calls also occupies memory.
Keeping the task history at the edge lets the device send only new input.
Retaining a compatible key--value (KV) cache can also avoid recomputing the
shared prefix. A scheduler can use the session profile to decide
which sessions to admit, which state to retain, and how to allocate radio
resources. For example, extra uplink capacity can help when requests wait
for transmission, but offers little relief when they already queue at the
accelerator.

\subsubsection{Energy Saving and RAN/Edge---AI Silicon Co-design}

Under light traffic, a cell can reduce the number of active RF chains while
maintaining its service target \cite{oran25.eswp}. The shared platforms
reviewed in Section~\ref{sec:platform-benchmarking} extend the energy
question to the silicon that runs baseband processing and AI inference.
As radio load varies across slots, inference can use spare compute cycles,
but contention for memory bandwidth can still delay the next radio task.

RAN/Edge---AI Silicon Co-design therefore needs to consider compute
capacity and memory bandwidth together, accounting for both radio slot
deadlines and inference response times. At a given radio load and site
power budget, the relevant measure is how much inference the platform can
complete on time while continuing to meet its radio deadlines.

\subsubsection{Agent Session Mobility During Radio Handover}

Vehicles and mobile robots that rely on edge inference need results while
their observations are still current \cite{ngrg2025.04}.
A vehicle may already be sending observations through its new serving cell
while the target edge site is still waiting for compute capacity or for the
state of the tracking task. For an agent making successive model calls, the
next call must also have access to the instructions and results accumulated
earlier in the session. ETSI multi-access edge computing supports application
relocation and context transfer. O-RAN describes compute-aware scheduling for
vehicle-to-everything service continuity \cite{etsi.mec021,oran26.v2xspecupdate}. For a session
spanning several model calls, relocation also requires deciding where to
keep the task history and whether to transfer the KV cache.

After radio handover, an agent session can continue at the source edge site
by relaying subsequent requests and responses over the backhaul. Alternatively,
an authorized target site can receive the session context and rebuild the
model's KV cache, or receive a compatible cache along with the context.
Continuing at the source adds backhaul delay to subsequent calls. Moving the
session incurs transfer or recomputation delay before the target can respond.
The choice depends on the context length, available backhaul throughput,
target compute capacity, and the next response deadline. A backhaul disruption
may require a local fallback. A cross-domain trial can
measure the interval between the last timely result received by the vehicle
before handover and the first one received afterward. This interval can be
compared with the maximum interruption the application tolerates whether
the session stays at the source or moves to the target.

These examples involve applications in two settings: radio nodes making
time-critical decisions and edge sites serving mobile AI sessions. This
raises an architectural question for the RAN: how can applications on
different hosts interact while the radio node retains responsibility for
slot-level execution?

\subsection{Architecture of the RAN Agentic Controller}
\label{sec:rac-concept}

The Non-RT RIC hosts rApps and the Near-RT RIC hosts xApps. nGRG has also
studied hosting dApps inside an O-DU or O-CU, close to the radio functions
they access \cite{ngrg2024.10,ngrg2025.05}. Building on this arrangement,
we propose the RAC as a distributed software layer.
It comprises a local software component, or shell, on each participating
host and a shared exposure gateway at the operator's trust boundary
(Fig.~\ref{fig:rac_shell}).

Through its local RAC shell, each host exposes selected interfaces of the
applications it deploys and manages so that applications on other hosts
can discover and invoke them. Agentic rApps, xApps, and dApps embed
reasoning components called \emph{rAgents}, \emph{xAgents}, and
\emph{dAgents}, respectively.
Agents on the RICs can communicate through the shells using agent-to-agent
(A2A) messages. An external edge or partner agent enters through the gateway,
which checks its identity and the scope of its access. Applications obtain
shared model services from the SMO, backed by O-Cloud compute.

The O-RAN interfaces carry the resulting radio policies and actions. A1
connects the two RICs, E2 connects the Near-RT RIC to radio nodes, and the
proposed E3 interface connects a dApp to local radio functions. The in-node
shell receives operating bounds asynchronously. The node executes each
slot-level action and handles late results locally, without waiting for
an external agent or a large language model. In the energy-saving example
of Section~\ref{sec:lifecycle}, an rApp and an xApp could use their shells
to coordinate a change in antenna configuration while the radio node
retains control of its timing.

\subsection{Recursive Standardization Across Federated Open Labs}
\label{sec:recursive-standardization}

3GPP is studying 6G radio requirements in Release~20, and O-RAN and 3GPP
are coordinating their respective work programmes and release schedules
\cite{3gpp.38914} (Section~\ref{sec:6g-coordination}). Within O-RAN, nGRG
studies candidate technologies, while the working groups develop normative
specifications. OTICs and PlugFests provide settings for testing
implementations from different suppliers \cite{oran.otic}.

One source of delay is the difficulty of comparing results from independent
implementations. An industry white paper on O-RAN testing reports that
differences in software versions, test setups, and how laboratories record
results can lead to repeat testing \cite{viavi25.orantesting}. During
specification drafting, two implementations may exhibit different behavior
under the same candidate clause. When their test conditions are unknown, a
working group may need another round of testing to determine whether to
revise the clause or the test procedure.

The proposed open lab initiative would allow academic-led OTICs and
university testbeds to begin with a research question about a prospective
6G feature, before its interface requirements have been settled
(Fig.~\ref{fig:6g_federated}). AI-assisted development could turn that
question into a working implementation and a laboratory workflow for
deploying it and measuring its behavior.
During pre-standardization, results from these independent sites could help
nGRG assess which candidate interface requirements hold across implementations
and recommend them to the relevant O-RAN working groups.
Drawing on experience from existing university-hosted OTICs and shared wireless testbeds
\cite{powder.oran}, federated open labs could link cloud simulators,
digital twins, HIL emulators, and OTA testbeds through shared registries and
comparable run records while preserving local spectrum and safety governance.

Recursive integration (RI), development (RD), and testing (RT) link two
versioned artifacts. One is
the research implementation, including its models and control logic. The
other is the laboratory implementation, including its adapters, deployment
scripts, instrumentation, and test scenarios. An integration run can
reveal a mismatch that calls for a change to a component or its adapter.
Testing can also identify a condition that the current setup cannot
reproduce or measure. Each run records both versions, the conditions,
and the outcome, including failures. These findings guide the next change
to the code, the laboratory, or both. The revised pair is compared with its
baseline under the same criteria before the change is adopted.

\emph{Recursive Standardization} (RS) is the outer process that connects
this work with standards development. Results from independent
implementations can help a research or working group formulate a candidate
requirement. The group may revise the requirement or identify an operating
condition it must cover. That revised requirement or research question then
guides what the labs implement, integrate, and test next. Their findings
return to the group for its next deliberation under the existing standards
procedures.

Consider the slot deadline discussed in Section~\ref{sec:6g-theses}. An accelerator
can complete inference before the radio deadline while the controller receives
the result after that deadline. The same result meets a completion-based
deadline and misses a delivery-based deadline. A lab that records only
accelerator completion cannot detect late delivery. In the next iteration
the lab could instrument receipt at the controller, while the controller
implementation could be revised to apply a fallback on late receipt. Two
sites could then replay the same load on the baseline and revised controllers,
using the new instrumentation for both, and observe the action each applied.
The resulting candidate contribution would specify
the event at which timeliness is judged and the observable fallback behavior,
leaving the accelerator and its internal scheduling to each supplier. The
working group could then ask whether the rule should also cover bursty load.
The labs would add a burst scenario and examine whether the controller still
applies the candidate rule. Those results would inform the scope of the next
draft. After adoption, professional testing houses could use the agreed
event definitions to detect regressions as components and software versions
change (Section~\ref{sec:otic-testing}) \cite{oran.otic}.

The proposed process could be used to test service semantics, information
elements, and fallback profiles across implementations in the context of
3GPP Release~20 6G studies (TR~38.914 and TR~38.960)
\cite{3gpp.38914,3gpp.38960} and in related 3GPP--O-RAN coordination
(Section~\ref{sec:6g-coordination}).
Working implementations would thus inform a requirement while it is still
being formulated, and each substantive revision could guide the next round
of implementation and experiments.

%% file: sec8_conclusion.tex
\section{Conclusion}
\label{sec:conclusion}

The 2018 O-RAN design paired open fronthaul between radio and baseband
equipment with a hierarchical RIC that could host independently developed
control applications. The AAL extends this modularity to hardware
acceleration by defining how baseband software accesses accelerated
functions. The AI-RAN reference implementations examined here show how GPU
acceleration can realize the host--accelerator split described by the O-RAN
AAL inline High-PHY profile.

The proposed RAC extends open application hosting to cross-host
coordination through local shells and a shared exposure gateway. Explicit
timing boundaries govern how agent coordination interacts with radio
control.

Through Recursive Standardization, federated open labs would compare
independent implementations to inform candidate requirements. Standards
review would guide the next round of implementation and testing. This
process gives working groups a basis for defining timing requirements
that can be verified across suppliers.

%% file: paper.bbl
\begin{thebibliography}{100}
\providecommand{\url}[1]{#1}
\csname url@samestyle\endcsname
\providecommand{\newblock}{\relax}
\providecommand{\bibinfo}[2]{#2}
\providecommand{\BIBentrySTDinterwordspacing}{\spaceskip=0pt\relax}
\providecommand{\BIBentryALTinterwordstretchfactor}{4}
\providecommand{\BIBentryALTinterwordspacing}{\spaceskip=\fontdimen2\font plus
\BIBentryALTinterwordstretchfactor\fontdimen3\font minus
  \fontdimen4\font\relax}
\providecommand{\BIBforeignlanguage}[2]{{%
\expandafter\ifx\csname l@#1\endcsname\relax
\typeout{** WARNING: IEEEtran.bst: No hyphenation pattern has been}%
\typeout{** loaded for the language `#1'. Using the pattern for}%
\typeout{** the default language instead.}%
\else
\language=\csname l@#1\endcsname
\fi
#2}}
\providecommand{\BIBdecl}{\relax}
\BIBdecl

\bibitem{oran18launch}
\BIBentryALTinterwordspacing
{O-RAN ALLIANCE}, ``Industry leaders launch {ORAN} alliance,'' Media release,
  Feb. 2018, {F}or {MWC} 2018, dated February 27, 2018. [Online]. Available:
  \url{https://cdn.prod.website-files.com/60b4ffd4ca081979751b5ed2/61197a5bf792d7a65c3073aa_O-RAN%2BAlliance%2BLaunch%2BFebruary%2B2018.pdf}
\BIBentrySTDinterwordspacing

\bibitem{oran18whitepaper}
\BIBentryALTinterwordspacing
------, ``{O-RAN}: Towards an open and smart {RAN},'' O-RAN ALLIANCE, White
  Paper, Oct. 2018. [Online]. Available:
  \url{https://mediastorage.o-ran.org/white-papers/O-RAN.White-Paper-2018-10.pdf}
\BIBentrySTDinterwordspacing

\bibitem{oran.wg1.arch}
------, ``{O-RAN} architecture description,'' ETSI, Technical Specification TS
  103 982 V14.0.0, Mar. 2026, {O-RAN.WG1.TS.OAD-R004-v14.00}.

\bibitem{airan26.architecture}
\BIBentryALTinterwordspacing
{AI-RAN Alliance}, ``{AI-RAN} architecture overview and component definitions,
  version 1.2,'' AI-RAN Alliance, Tech. Rep., Feb. 2026. [Online]. Available:
  \url{https://ai-ran.org/documents/AI-RAN-Architecture-v1.2.pdf}
\BIBentrySTDinterwordspacing

\bibitem{oran.wg6.aalhiphy}
{O-RAN ALLIANCE}, ``Acceleration abstraction layer: High-{PHY} profiles,''
  O-RAN ALLIANCE, Technical Specification O-RAN.WG6.AAL-HighPHY, 2023.

\bibitem{villa25x5g}
D.~Villa, I.~Khan, F.~Kaltenberger, N.~Hedberg, R.~{Soares da Silva},
  S.~Maxenti, L.~Bonati, A.~Kelkar, C.~Dick, E.~Baena, J.~M. Jornet,
  T.~Melodia, M.~Polese, and D.~Koutsonikolas, ``{X5G}: An open, programmable,
  multi-vendor, end-to-end, private {5G} {O-RAN} testbed with {NVIDIA ARC} and
  {OpenAirInterface},'' \emph{IEEE Trans. Mobile Comput.}, vol.~24, no.~11, pp.
  11\,305--11\,322, Nov. 2025.

\bibitem{oran22.ngrg}
\BIBentryALTinterwordspacing
{O-RAN ALLIANCE}, ``{O-RAN} announces certification and badging program, next
  generation research group, open {RAN} summit, and results of spring 2022
  {PlugFest},'' Press release, Jun. 2022. [Online]. Available:
  \url{https://www.businesswire.com/news/home/20220630005362/en/}
\BIBentrySTDinterwordspacing

\bibitem{dt23.nonrtric}
\BIBentryALTinterwordspacing
{Deutsche Telekom}, {AirHop}, {Juniper Networks}, {VIAVI Solutions}, and
  {VMware}, ``Demonstration of non-real time {RAN} optimization in a
  multi-vendor environment,'' O-RAN ALLIANCE, White Paper, Sep. 2023. [Online].
  Available:
  \url{https://www.o-ran.org/best-practices/deutsche-telekom-and-partners-demonstrate-non-real-time-ran-optimization-in-a-multi-vendor-environment}
\BIBentrySTDinterwordspacing

\bibitem{xran18merger}
\BIBentryALTinterwordspacing
{xRAN Forum}, ``{xRAN} forum merges with {C-RAN} alliance to form {O-RAN}
  alliance,'' Press release, Barcelona, Feb. 2018. [Online]. Available:
  \url{https://www.o-ran.org/press-releases/xran-forum-merges-with-c-ran-alliance-to-form-o-ran-alliance}
\BIBentrySTDinterwordspacing

\bibitem{ieee1914.1}
{IEEE}, ``{IEEE} standard for packet-based fronthaul transport networks,''
  IEEE, Tech. Rep. 1914.1-2019, 2020.

\bibitem{ieee1914.3}
------, ``{IEEE} standard for radio over {Ethernet} encapsulations and
  mappings,'' IEEE, Tech. Rep. 1914.3-2018, 2018.

\bibitem{oran.wg4.cus}
{O-RAN ALLIANCE}, ``Control, user and synchronization plane specification,''
  O-RAN ALLIANCE, Technical Specification O-RAN.WG4.CUS.0, 2026.

\bibitem{softran13}
A.~Gudipati, D.~Perry, L.~E. Li, and S.~Katti, ``{SoftRAN}: Software defined
  radio access network,'' in \emph{Proc. ACM SIGCOMM Workshop on Hot Topics in
  Software Defined Networking (HotSDN)}, Aug. 2013, pp. 25--30.

\bibitem{flexran16}
X.~Foukas, N.~Nikaein, M.~M. Kassem, M.~K. Marina, and K.~Kontovasilis,
  ``{FlexRAN}: A flexible and programmable platform for software-defined radio
  access networks,'' in \emph{Proc. 12th International Conference on Emerging
  Networking Experiments and Technologies (CoNEXT)}, Dec. 2016, pp. 427--441.

\bibitem{etsi17.eni}
\BIBentryALTinterwordspacing
{ETSI}, ``New {ETSI} group on improving operator experience using artificial
  intelligence,'' ETSI news, Sophia Antipolis, Feb. 2017. [Online]. Available:
  \url{https://www.etsi.org/newsroom/news/1171-2017-02-new-etsi-group-on-improving-operator-experience-using-artificial-intelligence/}
\BIBentrySTDinterwordspacing

\bibitem{vodafone17.tip}
\BIBentryALTinterwordspacing
{Vodafone}, ``Vodafone to join the {Telecom Infra Project} board and co-chair a
  new {Open RAN} project group with {Intel},'' Newsroom, Nov. 2017. [Online].
  Available:
  \url{https://www.vodafone.com/news/newsroom/technology/telecom-infra-project}
\BIBentrySTDinterwordspacing

\bibitem{ieee5gtf18.levels}
\BIBentryALTinterwordspacing
Y.~Wang, H.~Huang, Y.~Li, W.~Zhou, C.-L. I, Q.~Sun, and S.~Zhang, ``The levels
  of intelligence of mobile networks and consideration of architecture
  evolution,'' \emph{IEEE Future Networks Tech Focus}, vol.~2, no.~3, Dec.
  2018. [Online]. Available:
  \url{https://futurenetworks.ieee.org/tech-focus/december-2018/levels-of-intelligence}
\BIBentrySTDinterwordspacing

\bibitem{wwrf.wgwai}
\BIBentryALTinterwordspacing
{WWRF Working Group on Wireless Artificial Intelligence}, ``Artificial
  intelligence in future wireless communication systems,'' Wireless World
  Research Forum, White Paper Outlook 24, 2019. [Online]. Available:
  \url{https://wwrf.ch/wp-content/publications/outlook/Outlook24.pdf}
\BIBentrySTDinterwordspacing

\bibitem{3gpp.ts23501}
{3GPP}, ``System architecture for the {5G} system; stage 2,'' 3rd Generation
  Partnership Project, Tech. Rep. TS 23.501, 2017, v1.0.0, June 2017.

\bibitem{i17bigdata}
C.-L. I, Q.~Sun, Z.~Liu, S.~Zhang, and S.~Han, ``The big-data-driven
  intelligent wireless network: Architecture, use cases, solutions, and future
  trends,'' \emph{IEEE Veh. Technol. Mag.}, vol.~12, no.~4, pp. 20--29, Dec.
  2017.

\bibitem{ieee5gtf18.cudaduda}
\BIBentryALTinterwordspacing
C.-L. I, J.~Sun, X.~Han, Y.~Wang, and X.~Huang, ``A functional overview on
  {CUDA} and {DUDA},'' \emph{IEEE 5G Tech Focus}, vol.~2, no.~1, Mar. 2018.
  [Online]. Available:
  \url{https://futurenetworks.ieee.org/tech-focus/march-2018/functional-overview-of-cuda-and-duda}
\BIBentrySTDinterwordspacing

\bibitem{itu19.ml5garc}
\BIBentryALTinterwordspacing
{ITU-T Focus Group on Machine Learning for Future Networks including 5G},
  ``Unified architecture for machine learning in {5G} and future networks,''
  International Telecommunication Union, Technical Specification FG-ML5G-ARC5G,
  Jan. 2019. [Online]. Available:
  \url{https://www.itu.int/en/ITU-T/focusgroups/ml5g/Documents/ML5G-delievrables.pdf}
\BIBentrySTDinterwordspacing

\bibitem{itu19.y3172}
{ITU-T}, ``Architectural framework for machine learning in future networks
  including {IMT-2020},'' International Telecommunication Union, Recommendation
  Y.3172, Jun. 2019.

\bibitem{3gpp.tr38801}
{3GPP}, ``Study on new radio access technology: Radio access architecture and
  interfaces,'' 3rd Generation Partnership Project, Tech. Rep. TR 38.801, 2017,
  v14.0.0, March 2017.

\bibitem{oran.wg6.o2}
{O-RAN ALLIANCE}, ``{O2} interface: General aspects and principles,'' O-RAN
  ALLIANCE, Technical Specification O-RAN.WG6.O2GAP, 2026.

\bibitem{3gpp.ts32500}
{3GPP}, ``Telecommunication management; self-organizing networks {(SON)};
  concepts and requirements,'' 3rd Generation Partnership Project, Technical
  Specification TS 32.500, 2008, {R}elease 8.

\bibitem{3gpp.ts28313}
------, ``Management and orchestration; self-organizing networks {(SON)} for
  {5G} networks,'' 3rd Generation Partnership Project, Technical Specification
  TS 28.313, 2026, {R}elease 19, {ETSI} {TS} 128 313 V19.2.0.

\bibitem{3gpp.tr36902}
------, ``Evolved universal terrestrial radio access network {(E-UTRAN)};
  self-configuring and self-optimizing network {(SON)} use cases and
  solutions,'' 3rd Generation Partnership Project, Technical Report TR 36.902,
  2011, {R}elease 9, {ETSI} {TR} 136 902 V9.3.1.

\bibitem{oran.uc036}
{O-RAN ALLIANCE}, ``Publicly available specification {(PAS)}: {O-RAN} use cases
  detailed specification,'' European Telecommunications Standards Institute,
  Technical Specification ETSI TS 104 036 V12.0.0, 2025.

\bibitem{ngrg2024.10}
\BIBentryALTinterwordspacing
{O-RAN ALLIANCE next Generation Research Group}, ``{dApps} for real-time {RAN}
  control: Use cases and requirements,'' O-RAN ALLIANCE, Research Report
  nGRG-RR-2024-10, Oct. 2024. [Online]. Available:
  \url{https://mediastorage.o-ran.org/ngrg-rr/nGRG-RR-2024-10-dApp%20use%20cases%20and%20requirements.pdf}
\BIBentrySTDinterwordspacing

\bibitem{doro22dapps}
S.~D'Oro, M.~Polese, L.~Bonati, H.~Cheng, and T.~Melodia, ``{dApps}:
  Distributed applications for real-time inference and control in {O-RAN},''
  \emph{IEEE Commun. Mag.}, vol.~60, no.~11, pp. 52--58, Nov. 2022.

\bibitem{lacava25dapps}
A.~Lacava, L.~Bonati, N.~Mohamadi, R.~Gangula, F.~Kaltenberger, P.~Johari,
  S.~D'Oro, F.~Cuomo, M.~Polese, and T.~Melodia, ``{dApps}: Enabling real-time
  {AI}-based open {RAN} control,'' \emph{Comput. Netw.}, vol. 269, p. 111342,
  Sep. 2025.

\bibitem{ngrg2025.05}
\BIBentryALTinterwordspacing
{O-RAN ALLIANCE next Generation Research Group}, ``{dApps} architecture and
  interfaces,'' O-RAN ALLIANCE, Research Report nGRG-RR-2025-05, Feb. 2026.
  [Online]. Available:
  \url{https://mediastorage.o-ran.org/ngrg-rr/nGRG-RR-2025-05-dApps%20Architecture%20and%20Interfaces-v2.0.pdf}
\BIBentrySTDinterwordspacing

\bibitem{oran22.release002}
\BIBentryALTinterwordspacing
{O-RAN ALLIANCE}, ``{O-RAN ALLIANCE} introduces 40 new specifications released
  since {November} 2021,'' Technical release announcement, Apr. 2022. [Online].
  Available:
  \url{https://www.o-ran.org/blog/o-ran-alliance-introduces-40-new-specifications-released-since-november-2021}
\BIBentrySTDinterwordspacing

\bibitem{oran26.specupdate}
\BIBentryALTinterwordspacing
------, ``71 new or updated {O-RAN} technical documents released since
  {November} 2025,'' Specification release announcement, Feb. 2026, {A}ccessed:
  Nov. 15, 2026. [Online]. Available:
  \url{https://www.o-ran.org/blog/71-new-or-updated-o-ran-technical-documents-released-since-november-2025}
\BIBentrySTDinterwordspacing

\bibitem{oran25.eswp}
\BIBentryALTinterwordspacing
{O-RAN ALLIANCE Sustainability Focus Group}, ``Potential energy savings
  features in {O-RAN},'' O-RAN ALLIANCE, White Paper O-RAN.WP.Potential Energy
  Savings Features-v01.00, Jan. 2025. [Online]. Available:
  \url{https://mediastorage.o-ran.org/white-papers/O-RAN.SuFG.Potential%20Energy%20Savings%20Features%20in%20O-RAN%20white%20paper%202025-01.pdf}
\BIBentrySTDinterwordspacing

\bibitem{oran.ts104226}
{O-RAN ALLIANCE}, ``Publicly available specification {(PAS)}: Non-{RT} {RIC}
  use cases and requirements,'' European Telecommunications Standards
  Institute, Technical Specification ETSI TS 104 226 V10.1.0, 2025.

\bibitem{oran.wg2.r1ap}
------, ``{R1} interface: Application protocols for {R1} services,'' ETSI,
  Technical Specification TS 104 231 V8.0.0, Mar. 2026,
  {O-RAN.WG2.TS.R1AP-R004-v08.00}.

\bibitem{oran.wg3.y1gap}
------, ``{Y1} interface: General aspects and principles,'' O-RAN ALLIANCE,
  Technical Specification O-RAN.WG3.Y1GAP-v01.00, Mar. 2024.

\bibitem{oran.wg3.e2gap}
------, ``Near-{RT} {RIC} architecture and {E2} general aspects and
  principles,'' O-RAN ALLIANCE, Technical Specification O-RAN.WG3.E2GAP, 2026.

\bibitem{ngrg2024.05}
\BIBentryALTinterwordspacing
{O-RAN ALLIANCE next Generation Research Group}, ``Principles and methodologies
  for {AI/ML} testing in next generation networks,'' O-RAN ALLIANCE, Research
  Report nGRG-RR-2024-05, Apr. 2024. [Online]. Available:
  \url{https://mediastorage.o-ran.org/ngrg-rr/nGRG-RR-2024-05-Principals%20Methodologies%20AIML%20testing%20next%20Generation%20Networks-v1.9.pdf}
\BIBentrySTDinterwordspacing

\bibitem{oran.security.aiml}
\BIBentryALTinterwordspacing
{O-RAN ALLIANCE}, ``Study on security for {AI} and {ML} in {O-RAN},'' O-RAN
  ALLIANCE, Technical Report v02.00, Oct. 2024. [Online]. Available:
  \url{https://www.o-ran.org/blog/74-new-or-updated-o-ran-technical-documents-released-since-july-2024}
\BIBentrySTDinterwordspacing

\bibitem{oran.wg1.ccin}
\BIBentryALTinterwordspacing
------, ``{O-RAN} communication and computing integrated networks,'' O-RAN
  ALLIANCE, Technical Report O-RAN.WG1.CCIN-TR, 2026, {V}ersion 2.00. [Online].
  Available:
  \url{https://www.o-ran.org/blog/71-new-or-updated-o-ran-technical-documents-released-since-november-2025}
\BIBentrySTDinterwordspacing

\bibitem{oran.wg6.aalganp}
------, ``Acceleration abstraction layer: General aspects and principles,''
  O-RAN ALLIANCE, Technical Specification O-RAN.WG6.AAL-GAnP-v01.00, Nov. 2020.

\bibitem{oran.wg6.aalapi}
------, ``Acceleration abstraction layer: Common {API},'' O-RAN ALLIANCE,
  Technical Specification O-RAN.WG6.AAL-CommonAPI-v06.00, Mar. 2024.

\bibitem{oran.wg6.aalfec}
------, ``Acceleration abstraction layer: {FEC} profiles,'' O-RAN ALLIANCE,
  Technical Specification O-RAN.WG6.AAL-FEC-v01.00, 2021.

\bibitem{oran.wg6.aalmumimo}
------, ``{O-RAN} acceleration abstraction layer multi-user {MIMO} beamforming
  profile,'' O-RAN ALLIANCE, Technical Specification O-RAN.WG6.AAL-MU-MIMO-BF,
  2026.

\bibitem{3gpp.38843}
{3GPP}, ``Study on artificial intelligence {(AI)}/machine learning {(ML)} for
  {NR} air interface,'' 3rd Generation Partnership Project, Technical Report TR
  38.843, 2023, {R}elease 18.

\bibitem{3gpp.38744}
------, ``Study on {AI/ML} for mobility in {NR},'' 3rd Generation Partnership
  Project, Technical Report TR 38.744, 2025, {R}elease 19.

\bibitem{3gpp.ts38300}
------, ``{NR}; {NR} and {NG-RAN} overall description; stage 2,'' 3rd
  Generation Partnership Project, Technical Specification 38.300 V18.3.0, 2024.

\bibitem{3gpp.ts28105}
------, ``Management and orchestration; artificial intelligence / machine
  learning {(AI/ML)} management,'' 3rd Generation Partnership Project,
  Technical Specification TS 28.105, 2026, {R}elease 19, {ETSI} {TS} 128 105
  V19.5.0.

\bibitem{oran.otic}
\BIBentryALTinterwordspacing
{O-RAN ALLIANCE}, ``Open testing and integration centres,'' Directory, 2026.
  [Online]. Available: \url{https://www.o-ran.org/otic}
\BIBentrySTDinterwordspacing

\bibitem{oran25.mwcsummit}
\BIBentryALTinterwordspacing
------, ``{O-RAN} {ALLIANCE} advances open and {AI}-driven {RAN}
  standardization,'' Press release, {MWC} Barcelona, Mar. 2025. [Online].
  Available:
  \url{https://mediastorage.o-ran.org/press-release/O-RAN.EC.2025-03-28.PR%20MWC25%20summit%20messages.pdf}
\BIBentrySTDinterwordspacing

\bibitem{oran25.plugfestfall}
\BIBentryALTinterwordspacing
------, ``{O-RAN} global {PlugFest} fall 2025 results,'' Press release, Dec.
  2025. [Online]. Available:
  \url{https://mediastorage.o-ran.org/press-release/O-RAN.EC.2025-12-11.PR%20O-RAN%20Global%20PlugFest%20Fall%202025%20results-v01.pdf}
\BIBentrySTDinterwordspacing

\bibitem{oran26.plugfestspring}
\BIBentryALTinterwordspacing
------, ``{O-RAN} global {PlugFest} spring 2026 results,'' Press release, Jun.
  2026. [Online]. Available:
  \url{https://mediastorage.o-ran.org/o-ran-pr/pr-on-o-ran-global-plugfest-spring-2026-results/o-ran-ec-2026-06-04-pr-o-ran-global-plugfest-spring-2026-results-v01.pdf}
\BIBentrySTDinterwordspacing

\bibitem{ukgov22.sonic}
\BIBentryALTinterwordspacing
{Department for Digital, Culture, Media and Sport}, ``{SmartRAN} open network
  interoperability centre ({SONIC Labs}),'' UK government case study, Aug.
  2022. [Online]. Available:
  \url{https://www.gov.uk/government/case-studies/smartran-open-network-interoperability-centre-sonic-labs}
\BIBentrySTDinterwordspacing

\bibitem{powder.oran}
\BIBentryALTinterwordspacing
{POWDER}, ``{O-RAN} on {POWDER},'' Platform documentation, 2026. [Online].
  Available: \url{https://powderwireless.net/oran}
\BIBentrySTDinterwordspacing

\bibitem{rakuten20.launch}
\BIBentryALTinterwordspacing
{Rakuten Mobile}, ``{Rakuten} mobile launches full commercial service on a
  fully virtualized mobile network,'' Press release, Apr. 2020. [Online].
  Available: \url{https://global.rakuten.com/corp/news/press/2020/0408_01.html}
\BIBentrySTDinterwordspacing

\bibitem{mou24.priorities}
\BIBentryALTinterwordspacing
{Deutsche Telekom}, {Orange}, {Telef\'onica}, {TIM}, and {Vodafone}, ``Open
  {RAN} technical priorities, release 4,'' Industry report, Jun. 2024.
  [Online]. Available:
  \url{https://www.telefonica.com/en/communication-room/reports/open-ran-mou/}
\BIBentrySTDinterwordspacing

\bibitem{dish23.buildout}
\BIBentryALTinterwordspacing
{DISH Wireless}, ``The {DISH} {5G} network is now available to over 70 percent
  of the {US} population,'' Press release, Jun. 2023. [Online]. Available:
  \url{https://www.prnewswire.com/news-releases/the-dish-5g-network-is-now-available-to-over-70-percent-of-the-us-population-301851576.html}
\BIBentrySTDinterwordspacing

\bibitem{vodafone22.bath}
\BIBentryALTinterwordspacing
{Vodafone}, ``{Vodafone} switches on its first {5G} {OpenRAN} site,'' Press
  release, Jan. 2022. [Online]. Available:
  \url{https://www.vodafone.co.uk/newscentre/press-release/switches-on-first-5g-openran-site/}
\BIBentrySTDinterwordspacing

\bibitem{docomo24.orexsai}
\BIBentryALTinterwordspacing
{NTT DOCOMO}, ``{OREX} {SAI} established to accelerate open {RAN} deployment,''
  Press release, Feb. 2024. [Online]. Available:
  \url{https://www.docomo.ne.jp/english/info/media_center/pr/2024/0226_01.html}
\BIBentrySTDinterwordspacing

\bibitem{etsi.ts103859}
{ETSI}, ``Publicly available specification {(PAS)}: {O-RAN} fronthaul control,
  user and synchronization plane specification,'' European Telecommunications
  Standards Institute, Technical Specification ETSI TS 103 859 V7.0.2, Sep.
  2022.

\bibitem{oran26.atisdone}
\BIBentryALTinterwordspacing
{ATIS}, ``Transposition of {O-RAN} {ALLIANCE} specifications to {ATIS}
  standards,'' Press release, Feb. 2026. [Online]. Available:
  \url{https://mediastorage.o-ran.org/o-ran-pr/O-RAN.EC.2026-02-17.PR%20initial%20transposition%20of%20O-RAN%20specs%20to%20ATIS%20standards-v01.pdf}
\BIBentrySTDinterwordspacing

\bibitem{oran24.tta}
\BIBentryALTinterwordspacing
{TTA} and {O-RAN ALLIANCE}, ``{TTA} and {O-RAN} {ALLIANCE} signed agreement to
  cooperate on standardization of open {RAN} technology,'' Press release, Nov.
  2024. [Online]. Available:
  \url{https://www.o-ran.org/press-releases/tta-and-o-ran-alliance-signed-agreement-to-cooperate-on-standardization-of-open-ran-technology}
\BIBentrySTDinterwordspacing

\bibitem{oran21.etsitsdsi}
\BIBentryALTinterwordspacing
{O-RAN ALLIANCE}, ``{O-RAN} {ALLIANCE} broadens its industry collaboration with
  {ETSI}, announces its 3rd whitepaper, and 4th release of open software,''
  Press release, Aug. 2021. [Online]. Available:
  \url{https://www.o-ran.org/press-releases/o-ran-alliance-broadens-its-industry-collaboration-with-etsi-announces-its-3rd-whitepaper-and-4th-release-of-open-software}
\BIBentrySTDinterwordspacing

\bibitem{ntia.pwscif.r1}
\BIBentryALTinterwordspacing
{National Telecommunications and Information Administration}, ``Public wireless
  supply chain innovation fund, round 1: Research and development, testing and
  evaluation,'' Award announcement, 2023. [Online]. Available:
  \url{https://www.ntia.gov/funding-programs/public-wireless-supply-chain-innovation-fund/round-1-2023-research-and-development-testing-and-evaluation}
\BIBentrySTDinterwordspacing

\bibitem{bmvi21.openran}
\BIBentryALTinterwordspacing
{Bundesministerium f\"ur Verkehr und digitale Infrastruktur}, ``{BMVI} startet
  open {RAN} f\"orderung,'' Information paper, Germany, Nov. 2021. [Online].
  Available:
  \url{https://www.bmv.de/SharedDocs/DE/Anlage/K/presse/126-bmvi-startet-open-ran-foerderung-infopapier.pdf}
\BIBentrySTDinterwordspacing

\bibitem{oran25.6gworkshop}
\BIBentryALTinterwordspacing
{O-RAN ALLIANCE}, ``Road to open, intelligent {6G}: {O-RAN}'s first {6G}
  workshop,'' Blog post, Dec. 2025. [Online]. Available:
  \url{https://www.o-ran.org/blog/road-to-open-intelligent-6g-o-rans-first-6g-workshop-toward-a-structured-and-collaborative-6g-journey}
\BIBentrySTDinterwordspacing

\bibitem{lfn26.oscmigration}
\BIBentryALTinterwordspacing
{LF Networking}, ``{LF Networking} welcomes the {O-RAN Software Community} as a
  formal project,'' Press release, Apr. 2026. [Online]. Available:
  \url{https://www.linuxfoundation.org/press/lf-networking-welcomes-o-ran-software-community-as-formal-project-deepening-open-ran-ecosystem-collaboration}
\BIBentrySTDinterwordspacing

\bibitem{osc.releases}
\BIBentryALTinterwordspacing
{O-RAN Software Community}, ``{O-RAN} {SC} release notes,'' Project
  documentation, 2026. [Online]. Available:
  \url{https://docs.o-ran-sc.org/en/latest/release-notes.html}
\BIBentrySTDinterwordspacing

\bibitem{schmidt21flexric}
R.~Schmidt, M.~Irazabal, and N.~Nikaein, ``{FlexRIC}: An {SDK} for
  next-generation {SD-RANs},'' in \emph{Proc. 17th Int. Conf. on Emerging
  Networking Experiments and Technologies (CoNEXT)}.\hskip 1em plus 0.5em minus
  0.4em\relax ACM, Dec. 2021, pp. 411--425.

\bibitem{johnson21nexran}
D.~Johnson, D.~Maas, and J.~{Van der Merwe}, ``{NexRAN}: Closed-loop {RAN}
  slicing in {POWDER}, a top-to-bottom open-source open-{RAN} use case,'' in
  \emph{Proc. 15th ACM Workshop on Wireless Network Testbeds, Experimental
  Evaluation and Characterization (WiNTECH)}.\hskip 1em plus 0.5em minus
  0.4em\relax ACM, Oct. 2021, pp. 17--23.

\bibitem{oai.fhi72}
\BIBentryALTinterwordspacing
{OpenAirInterface}, ``{O-RAN} 7.2 fronthaul interface driver,'' Source
  repository documentation, radio/fhi\_72/README.md, 2026. [Online]. Available:
  \url{https://gitlab.eurecom.fr/oai/openairinterface5g/-/blob/develop/radio/fhi_72/README.md}
\BIBentrySTDinterwordspacing

\bibitem{lfn25.duranta}
\BIBentryALTinterwordspacing
{LF Networking} and {OpenAirInterface Software Alliance}, ``{LF Networking} and
  {OSA} announce {Duranta} to advance open-source {RAN} innovation,'' Press
  release, Aug. 2025. [Online]. Available:
  \url{https://www.linuxfoundation.org/press/lf-networking-and-openairinterface-software-alliance-osa-collaborate-to-announce-duranta-to-advance-open-source-ran-innovation}
\BIBentrySTDinterwordspacing

\bibitem{srsran.release}
\BIBentryALTinterwordspacing
{Software Radio Systems}, ``{srsRAN} project release 23.10,'' Release notes,
  Oct. 2023. [Online]. Available:
  \url{https://github.com/srsran/srsRAN_Project/releases}
\BIBentrySTDinterwordspacing

\bibitem{ocudu26}
\BIBentryALTinterwordspacing
{OCUDU Ecosystem Foundation}, ``{OCUDU Ecosystem Foundation} launch,'' Press
  release, {MWC} Barcelona, Mar. 2026. [Online]. Available:
  \url{https://www.linuxfoundation.org/press}
\BIBentrySTDinterwordspacing

\bibitem{bonati21colosseum}
L.~Bonati, P.~Johari, M.~Polese, S.~D'Oro, S.~Mohanti, M.~Tehrani-Moayyed,
  D.~Villa, S.~Shrivastava, C.~Tassie, K.~Yoder, A.~Bagga, P.~Patel, V.~Petkov,
  M.~Seltser, F.~Restuccia, A.~Gosain, K.~R. Chowdhury, S.~Basagni, and
  T.~Melodia, ``Colosseum: Large-scale wireless experimentation through
  hardware-in-the-loop network emulation,'' in \emph{Proc. IEEE Int. Symp. on
  Dynamic Spectrum Access Networks (DySPAN)}, Dec. 2021, pp. 105--113.

\bibitem{nguyen24crosslab}
T.-T. Nguyen, M.~V. Ngo, B.~Chen, M.~Kuchitsu, S.~Wai, S.~Kawai, K.~Suzuki,
  E.~W. Koo, and T.~Quek, ``Consistent and repeatable testing of {mMIMO O-RU}
  across labs: A {Japan--Singapore} experience,'' in \emph{Proc. IEEE 100th
  Vehicular Technology Conference (VTC2024-Fall)}, 2024.

\bibitem{oran.otic.utdallas}
\BIBentryALTinterwordspacing
{O-RAN ALLIANCE}, ``North american {OTIC} in the central {US} telecom corridor
  (university of {Texas Dallas}),'' OTIC directory entry. [Online]. Available:
  \url{https://www.o-ran.org/otics/north-american-otic-in-the-central-us-telecom-corridor-university-of-texas-dallas}
\BIBentrySTDinterwordspacing

\bibitem{bonati23openrangym}
L.~Bonati, M.~Polese, S.~D'Oro, S.~Basagni, and T.~Melodia, ``{OpenRAN Gym}:
  {AI/ML} development, data collection, and testing for {O-RAN} on {PAWR}
  platforms,'' \emph{Comput. Netw.}, vol. 220, p. 109502, Jan. 2023.

\bibitem{lacava23nsoran}
A.~Lacava, M.~Bordin, M.~Polese, R.~Sivaraj, T.~Zugno, F.~Cuomo, and
  T.~Melodia, ``{ns-O-RAN}: Simulating {O-RAN} {5G} systems in {ns-3},'' in
  \emph{Proc. Workshop on ns-3 (WNS3)}.\hskip 1em plus 0.5em minus 0.4em\relax
  ACM, Jun. 2023, pp. 35--44.

\bibitem{cammerer25sionnark}
\BIBentryALTinterwordspacing
S.~Cammerer, G.~Marcus, T.~Zirr, F.~{A\"{i}t Aoudia}, L.~Maggi, J.~Hoydis, and
  A.~Keller, ``{Sionna Research Kit}: A {GPU}-accelerated research platform for
  {AI-RAN},'' arXiv:2505.15848v1, May 2025, {A}ccepted as a demonstration at
  IEEE ICMLCN 2025. [Online]. Available:
  \url{https://arxiv.org/abs/2505.15848v1}
\BIBentrySTDinterwordspacing

\bibitem{i14y.lab}
\BIBentryALTinterwordspacing
{i14y Lab}, ``{i14y Lab},'' Laboratory site, Berlin, 2026. [Online]. Available:
  \url{https://www.i14y-lab.com/}
\BIBentrySTDinterwordspacing

\bibitem{pawr.program}
\BIBentryALTinterwordspacing
{Platforms for Advanced Wireless Research}, ``{PAWR} program overview,''
  Program site, 2026. [Online]. Available:
  \url{https://advancedwireless.org/about-pawr/}
\BIBentrySTDinterwordspacing

\bibitem{polese24empowering}
M.~Polese, M.~Dohler, F.~Dressler, M.~Erol-Kantarci, R.~Jana, R.~Knopp, and
  T.~Melodia, ``Empowering the {6G} cellular architecture with open {RAN},''
  \emph{IEEE J. Sel. Areas Commun.}, vol.~42, no.~2, pp. 245--262, Feb. 2024.

\bibitem{chen25oranrole}
T.~Chen, C.-L. I, and T.~Melodia, ``{O-RAN's} role in shaping {6G}: Industry
  perspectives on open and smart {RAN},'' \emph{IEEE Wireless Commun.},
  vol.~32, no.~1, pp. 10--12, Feb. 2025.

\bibitem{airan26.wg3}
\BIBentryALTinterwordspacing
{AI-RAN Alliance Working Group 3}, ``{AI-on-RAN}: Enabling monetizable
  differentiated connectivity for {AI},'' AI-RAN Alliance, Tech. Rep., Feb.
  2026. [Online]. Available:
  \url{https://ai-ran.org/documents/AI-RAN-WG3-AI-on-RAN-Whitepaper.pdf}
\BIBentrySTDinterwordspacing

\bibitem{perotti24ara2}
M.~Perotti, M.~Cavalcante, R.~Andri, L.~Cavigelli, and L.~Benini, ``Ara2:
  Exploring single- and multi-core vector processing with an efficient {RVV}
  1.0 compliant open-source processor,'' \emph{IEEE Transactions on Computers},
  vol.~73, no.~7, pp. 1822--1836, Jul. 2024.

\bibitem{sebastian20memory}
A.~Sebastian, M.~Le~Gallo, R.~Khaddam-Aljameh, and E.~Eleftheriou, ``Memory
  devices and applications for in-memory computing,'' \emph{Nature
  Nanotechnology}, vol.~15, no.~7, pp. 529--544, 2020.

\bibitem{nvidia.oranarchive}
\BIBentryALTinterwordspacing
{NVIDIA}, ``Technical contributions submitted to the {O-RAN} alliance by
  {NVIDIA},'' Public archive, 2021. [Online]. Available:
  \url{https://github.com/NVIDIA/O-RAN-Archive}
\BIBentrySTDinterwordspacing

\bibitem{airan24.founding}
\BIBentryALTinterwordspacing
{AI-RAN Alliance}, ``Industry leaders in {AI} and wireless form the {AI-RAN}
  alliance,'' Press release, {MWC} Barcelona, Feb. 2024. [Online]. Available:
  \url{https://ai-ran.org/press-releases}
\BIBentrySTDinterwordspacing

\bibitem{softbank24.aitras}
\BIBentryALTinterwordspacing
{SoftBank Corp.}, ``{SoftBank} develops {AITRAS}, a converged {AI-RAN}
  solution,'' Press release, Nov. 2024. [Online]. Available:
  \url{https://www.softbank.jp/en/corp/news/press/sbkk/2024/20241113_06/}
\BIBentrySTDinterwordspacing

\bibitem{softbank25.mmimo}
\BIBentryALTinterwordspacing
------, ``{SoftBank} achieves software-only massive {MIMO} on {GPU},'' Press
  release, Oct. 2025. [Online]. Available:
  \url{https://www.softbank.jp/en/corp/news/press/sbkk/2025/20251029_02/}
\BIBentrySTDinterwordspacing

\bibitem{ericsson26.portable}
\BIBentryALTinterwordspacing
{Ericsson}, ``{Ericsson} and {T-Mobile} boost portable {AI-RAN} on the {NVIDIA}
  platform,'' Press release, Mar. 2026. [Online]. Available:
  \url{https://www.ericsson.com/en/news/2026/3/ericsson-t-mobile-boost-portable-ai-ran-on-nvidia-platform}
\BIBentrySTDinterwordspacing

\bibitem{itu.m2160}
{ITU-R}, ``Framework and overall objectives of the future development of {IMT}
  for 2030 and beyond,'' International Telecommunication Union, Recommendation
  M.2160-0, Nov. 2023.

\bibitem{ngrg2025.04}
\BIBentryALTinterwordspacing
{O-RAN ALLIANCE next Generation Research Group}, ``Evolution of {O-RAN}
  near-{RT} {RIC} toward {6G},'' O-RAN ALLIANCE, Research Report
  nGRG-RR-2025-04, Oct. 2025. [Online]. Available:
  \url{https://mediastorage.o-ran.org/ngrg-rr/nGRG-RR-2025-04-Evolution%20of%20O-RAN%20Near-RT%20RIC%20toward%206G-v1.0.pdf}
\BIBentrySTDinterwordspacing

\bibitem{etsi.mec021}
{ETSI}, ``Multi-access edge computing ({MEC}); application mobility service
  {API},'' European Telecommunications Standards Institute, Group Specification
  ETSI GS MEC 021 V3.1.1, Feb. 2024.

\bibitem{oran26.v2xspecupdate}
\BIBentryALTinterwordspacing
{O-RAN ALLIANCE}, ``66 new or updated {O-RAN} technical documents released
  since {July} 2026,'' Specification release announcement, Oct. 2026. [Online].
  Available:
  \url{https://www.o-ran.org/blog/66-new-or-updated-o-ran-technical-documents-released-since-july-2026}
\BIBentrySTDinterwordspacing

\bibitem{3gpp.38914}
{3GPP}, ``Study on {6G} scenarios and requirements,'' 3rd Generation
  Partnership Project, Technical Report TR 38.914, 2026, {R}elease 20.

\bibitem{viavi25.orantesting}
\BIBentryALTinterwordspacing
{VIAVI Solutions}, {Tata Consultancy Services}, {Spirent} \emph{et~al.},
  ``{O-RAN} testing: Challenges, and recommendations,'' {O-RAN} ALLIANCE,
  Ecosystem white paper O-RAN.WP.IEFG.O-RAN.Testing, Feb. 2025. [Online].
  Available:
  \url{https://mediastorage.o-ran.org/ecosystem-resources/O-RAN-Ecosystem-2025.04.02.WP.O-RAN_Testing-v07.3.pdf}
\BIBentrySTDinterwordspacing

\bibitem{3gpp.38960}
{3GPP}, ``Study on {6G} radio,'' 3rd Generation Partnership Project, Technical
  Report TR 38.960, 2026, {R}elease 20.

\end{thebibliography}
